\pdfoutput=1  % Instructs arXiv to run this with pdflatex rather than latex.

\documentclass[aps,twocolumn,groupedaddress,floatfix,preprintnumbers,nofootinbib,eqsecnum]{revtex4}

\usepackage[version=3]{mhchem}
\usepackage{graphicx,amssymb,epstopdf}
\usepackage{relsize,array}
\usepackage{xspace}
\usepackage{multirow,placeins}

\def\be{\begin{equation}}
\def\ee{\end{equation}}
\def\bea{\begin{eqnarray}}
\def\eea{\end{eqnarray}}

\def\ie{{\it i.e.}\/}
\def\eg{{\it e.g.}\/}
\def\etc{{\it etc}.\/}

\def\gev{\, {\rm GeV}}
\def\mev{\, {\rm MeV}}

\newcommand{\gsim}{\lower.7ex\hbox{$\;\stackrel{\textstyle>}{\sim}\;$}}
\newcommand{\lsim}{\lower.7ex\hbox{$\;\stackrel{\textstyle<}{\sim}\;$}}

\newcommand {\s}{\rm s}

\def\tnow{t_{\mathrm{now}}}
\def\Tnow{T_{\mathrm{now}}}
\def\tinj{t_{\mathrm{inj}}}
\def\Tinj{T_{\mathrm{inj}}}
\def\tMRE{t_{\mathrm{MRE}}}
\def\tLS{t_{\mathrm{LS}}}

\def\OmegaDM{\Omega_{\mathrm{DM}}}
\def\Omegatot{\Omega_{\mathrm{tot}}}
\def\rhocrit{\rho_{\mathrm{crit}}}

\begin{document}%\special{papersize=8.5in,11in}

\setcounter{footnote}{0}
\setcounter{page}{1}
\setcounter{figure}{0}
\setcounter{table}{0}

%\tableofcontents

%===========================================
\title{
Cosmological Constraints on Unstable Particles:\\
  Numerical Bounds and Analytic Approximations}
\author{Keith R.~Dienes$^{1,2}$\footnote{E-mail address: {\tt dienes@email.arizona.edu}},
  Jason Kumar$^3$\footnote{E-mail address: {\tt jkumar@hawaii.edu}},
  Patrick Stengel$^{4,5}$\footnote{E-mail address: {\tt patrick.stengel@fysik.su.se}},
  Brooks Thomas$^6$\footnote{E-mail address: {\tt thomasbd@lafayette.edu}}
  }
\affiliation{
    $^1$Department of Physics, University of Arizona, Tucson, AZ  85721  USA\\
	$^2$Department of Physics, University of Maryland, College Park, MD  20742  USA\\
    $^3$Department of Physics and Astronomy, University of Hawaii, Honolulu, HI 96822 USA\\
    $^4$Leinweber Center for Theoretical Physics, Department of Physics, 
        University of Michigan, Ann Arbor, MI 48109  USA\\
    $^5$Oskar Klein Centre for Cosmoparticle Physics, Department of Physics,  
        Stockholm University, Alba Nova, 10691 Stockholm, Sweden\\
    $^6$Department of Physics, Lafayette College, Easton, PA 18042  USA}

\begin{abstract}
Many extensions of the Standard Model predict large numbers of additional unstable 
particles whose decays in the early universe are tightly constrained by observational 
data.  For example, the decays of such particles can alter the ratios of light-element 
abundances, give rise to distortions in the cosmic microwave background, alter the 
ionization history of the universe, and contribute to the diffuse photon flux.  Constraints 
on new physics from such considerations are typically derived for a single 
unstable particle species with a single well-defined mass and characteristic lifetime. 
In this paper, by contrast, we investigate the cosmological constraints on theories 
involving entire ensembles of decaying particles --- ensembles which span potentially broad 
ranges of masses and lifetimes.  In addition to providing a detailed numerical analysis of 
these constraints, we also formulate a set of simple analytic approximations for these 
constraints which may be applied to generic ensembles of unstable particles which decay 
into electromagnetically-interacting final states.  We then illustrate how these analytic 
approximations can be used to constrain a variety of toy scenarios for physics beyond the 
Standard Model.  For ease of reference, we also compile our results in the form 
of a table which can be consulted independently of the rest of the paper.  It is thus our 
hope that this work might serve as a useful reference for future model-builders concerned 
with cosmological constraints on decaying particles, regardless of the particular model 
under study. 
\end{abstract}


\maketitle

%\newpage
  \tableofcontents
%\newpage

%%%%%%%%%%%%%%%%%%%%%%%%%%%%%%%%%%%%%%%%%%%%%%%%%%%%%%%%%%%%%%%%%%%%%%%%%%%%%%%%%%%%%%
\FloatBarrier
\section{Introduction\label{sec:Intro}}

%%%%%%%%%%%%%%%%%%%%%%%%%%%%%%%%%%%%%%%%%%%%%%%%%%%%%%%%%%%%%%%%%%%%%%%%%%%%%%%%%%%%%%

Many proposals for physics beyond the Standard Model (SM) predict the existence of 
additional unstable particles.
The decays of such particles can have a variety of observable 
consequences --- especially if the final states into which these particles decay involve 
visible-sector particles.  Indeed, electromagnetic or hadronic showers precipitated 
by unstable-particle decays within the recent cosmological past
can alter the primordial abundances of light nuclei both during
and after Big-Bang nucleosynthesis (BBN)~\cite{KawasakiMoroi,Sarkar:1995dd,CyburtEllisUpdated,
CyburtEllisGravitino,Kawasaki:2017bqm}, give rise to spectral distortions in the 
cosmic microwave background (CMB)~\cite{HuAndSilk,HuAndSilk2}, alter the 
ionization history of the universe~\cite{Adams:1998nr,ChenKamionkowski,SlatyerFinkbeiner,
Finkbeiner:2011dx,SlatyerInjectionHistory},
and give rise to characteristic features in the diffuse photon background.  
These considerations therefore place stringent constraints on models for new physics
involving unstable particles.  

Much previous work has focused on 
examining the cosmological consequences of a single particle species decaying in isolation, 
and the corresponding limits on the 
properties of such a particle species are now well established.  Indeed, simple analytic
approximations can be derived which accurately model the effects that the decays of such 
a particle can have on many of the relevant observables~\cite{HuAndSilk,CyburtEllisUpdated,KhatriSunyaev}.
However, many theories for new physics involve not merely one or a few unstable particles,
but rather a large --- and potentially vast --- number of such particles with a 
broad spectrum of masses, lifetimes, and cosmological abundances.   
For example, 
theories with additional spacetime dimensions
give rise to infinite towers of Kaluza-Klein (KK) excitations for any field which propagates in the 
higher-dimensional bulk.  Likewise, string theories generally predict
large numbers of light moduli~\cite{BanksModuliProblem1,deCarlosModuliProblem,BanksModuliProblem2} 
or axion-like particles~\cite{WittenStringAxion,SvrcekAndWitten,Axiverse}.
Collections of similar light fields also arise in supergravity theories~\cite{SUGRAModuliProblem},
as well as in other scenarios for new physics~\cite{NNaturalness,TimRaffaeleMatt}.  
There even exist approaches to the dark-matter problem such as the
Dynamical Dark Matter (DDM) framework~\cite{DDM1,DDM2} 
which posit the existence of potentially vast ensembles of unstable dark-sector particles.
It is therefore crucial to understand the cosmological consequences of entire ensembles
of decaying particles in the early universe and, if possible, to formulate a 
corresponding set of analytic approximations which model the effects of these decays.   

For a variety of reasons, 
assessing the effects of an
entire ensemble of decaying particles with a broad range of 
masses, lifetimes, and cosmological abundances is not merely a matter of trivially 
generalizing the results obtained in the single-particle case.
The decay of a given unstable particle amounts to an injection of 
additional electromagnetic radiation and/or other energetic particles
into the evolution of the universe,
and injection at different characteristic timescales during this evolution can have 
markedly different effects on the same observable.  Moreover, since many of
these observables evolve in time according to a complicated system of
coupled equations, the effects of injection at any particular time $\tinj$
depend in a non-trivial way on the entire injection history prior to $\tinj$ through
feedback effects.

In principle, the cosmological constraints on ensembles of decaying particles in the 
early universe can be evaluated numerically.  Indeed, there are several publicly 
available codes~\cite{ChlubaGreensFns1,SlatyerInjectionHistory,PPPC4DMID} which can readily 
be modified in order to assess the effects of an arbitrary additional injection history 
on the relevant observables.  Computational methods can certainly yield useful results 
in any particular individual case.  However, another complementary approach which
can provide additional physical insight into the underlying dynamics involves the
formulation of approximate analytic expressions for the relevant observables --- 
expressions analogous to those which already exist for a single particle species decaying 
in isolation.  Our aim in this paper is to derive such a set of analytic expressions --- 
expressions which are applicable to generic theories involving large numbers of unstable 
particles, but which nevertheless provide accurate approximations for the relevant 
cosmological observables.  
Thus, our results can 
serve as a useful reference for future model-builders 
concerned with cosmological constraints on decaying particles, regardless of the particular model 
under study.

In this paper, we shall focus primarily on the case in which electromagnetic 
injection dominates --- {\it i.e.}\/, the case in which the energy liberated by the decays
of these particles is released primarily in the form of photons, electrons, and 
positrons rather than hadrons.
We also emphasize that the approximations we shall derive in this paper are not {\it ad hoc}\/ in nature;
in particular, they are not the results of empirical fits.
Rather, as we shall see, they emerge organically 
from the underlying physics   
and thus carry direct information about the underlying processes involved.

This paper is organized as follows. In Sect.~\ref{sec:InjectionConstraints}, we
begin by establishing the notation and conventions that we shall use throughout this paper.  
We also review the various scattering processes through which energetic photons injected
by particle decay interact with other particles present in the radiation bath.   
In subsequent sections, we then turn our attention to entire 
ensembles of unstable particles, focusing
on the electromagnetic injections arising from decays occurring after the BBN epoch.
Each section is devoted to a different cosmological consideration arising from
such injection, and in each case
we ultimately obtain a simple, 
analytic approximation for the corresponding constraint.  
For example, in Sect.~\ref{sec:LightElemAbundances} we consider the constraints associated with 
the modification of the abundances of light nuclei after BBN, 
and in Sect.~\ref{sec:CMBDistortions} we consider limits on distortions of the 
CMB-photon spectrum.  Likewise, in Sect.~\ref{sec:Ionization} we consider the constraints 
associated with the ionization history of the universe and its impact on the CMB,
and in Sect.~\ref{sec:DiffusePhoton} we consider the constraints associated with additional 
contributions from unstable-particle decays to the diffuse photon background.
Ultimately, the results from these sections furnish us with the tools needed to constrain
decaying ensembles of various types.
This is then illustrated in Sect.~\ref{sec:MultiCompResults}, where we consider 
how our results may be applied 
to two classes of ensembles 
whose constituents exhibit 
different representative mass spectra.
Finally, in Sect.~\ref{sec:Conclusions}, we conclude with a discussion of 
our main results and avenues for future work.  For future reference,
we also provide (in Table~\ref{tab:SummaryTable}) a summary/compilation 
of our main results.

%%%%%%%%%%%%%%%%%%%%%%%%%%%%%%%%%%%%%%%%%%%%%%%%%%%%%%%%%%%%%%%%%%%%%%%%%%%%%%%%%%%%%%

\FloatBarrier
\section{Electromagnetic Injection:   
 ~Overview and Classification of Relevant Processes\label{sec:InjectionConstraints}}

%%%%%%%%%%%%%%%%%%%%%%%%%%%%%%%%%%%%%%%%%%%%%%%%%%%%%%%%%%%%%%%%%%%%%%%%%%%%%%%%%%%%%%

Our aim in this paper is to assess the cosmological constraints on an ensemble consisting 
of a potentially large number $N$ of unstable particle species $\chi_i$ with 
masses $m_i$ and decay widths $\Gamma_i$ (or, equivalently, lifetimes 
$\tau_i \equiv \Gamma_i^{-1}$), where the index $i = 0,1,\ldots,N-1$ labels these 
particle species in order of increasing mass $m_i$.  We shall characterize the cosmological 
abundance of each of the $\chi_i$ in terms of a quantity $\Omega_i$ which we call the 
``extrapolated abundance.''  This quantity represents the abundance that the species 
$\chi_i$ {\it would}\/ have had at present time, had it been absolutely stable.
We shall assume that the total abundance $\Omegatot \equiv \sum_i \Omega_i$ of the ensemble is sufficiently small that the universe remains radiation-dominated until the time of matter-radiation equality $\tMRE \approx 10^{12}$~s.
Moreover, we shall focus on the regime in which $m_i \gtrsim 1$~GeV for all $\chi_i$ and 
all of the ensemble constituents are non-relativistic by end of the BBN
epoch.  Within this regime, as we shall discuss in further detail below, 
the spectrum of energetic photons produced by electromagnetic injection takes a 
characteristic form which to a very good approximation depends only on the overall 
energy density injected~\cite{KawasakiMoroi}.  By contrast, for much lighter
decaying particles, the form of the resulting photon spectrum can differ from
this characteristic form as a result of the immediate decay products lacking  
sufficient energy to induce the production of $e^+ e^-$ pairs by scattering off
background photons~\cite{PoulinSerpico1,PoulinSerpico2}.
The cosmological constraints on a single electromagnetically-decaying particle species  
with a mass below $1$~GeV 
--- investigated earlier to constrain neutrinos (see, {\it e.g.}\/, Ref.~\cite{Sarkar:1984tt}) ---
have recently been investigated in Refs.~\cite{Hufnagel,ForestellMorrisseyWhite}.
    
The considerations which place the most stringent constraints on the ensemble depend
on the values of $\Omega_i$ and $\tau_i$ for the individual ensemble constituents.
For ensembles of particles with lifetimes in the range 
$1\mbox{~s} \lesssim \tau_i \lesssim \tnow$, where $\tnow$ denotes the present age of 
the universe, the dominant constraints are those related to the abundances of light 
elements, to spectral distortions of the CMB, and to the ionization history of the 
universe.  The effect of electromagnetic injection on the corresponding observables 
is sensitive to the overall energy density injected and to the timescales over which 
that energy is injected, but not to the details of the decay kinematics or the particular
channels through which the $\chi_i$ decay.  Thus, in order to retain as much generality as 
possible in our analysis, we shall focus on ensembles for which 
all constituents with non-negligible $\Omega_i$ have lifetimes within this range;
moreover, we shall refrain from specifying any particular decay channel for the 
$\chi_i$ when assessing the bounds on these ensembles due to these considerations.  
By contrast, the constraints on decaying ensembles that follow from limits on features 
in the diffuse photon background {\it do}\/ depend on the particulars of the decay 
kinematics.  Thus, when analyzing these constraints in Sect.~\ref{sec:DiffusePhoton},
we not only present a general expression for the relevant observable --- namely the 
contribution to the differential photon flux from the decaying ensemble --- but also 
apply this result to a concrete example involving a particular decay topology. 

Many of the constraints on electromagnetic injection are 
insensitive to the details of the decay kinematics because the injection of 
photons and other electromagnetically-interacting particles prior to CMB decoupling
sets into motion a complicated chain of interactions which serve to redistribute the 
energies of these particles.  In particular, the effects of electromagnetic 
injection on cosmological observables ultimately depend on the interplay between 
three broad classes of processes through which these photons interact with other 
particles in the background plasma.  
These are:   
\begin{itemize}
  \item \underbar{Class~I}: Cascade and cooling processes which rapidly redistribute the energy 
    of the injected photons.  Processes in this class include 
    $\gamma + \gamma_{\mathrm{BG}} \rightarrow e^+ + e^-$ and 
    $\gamma + \gamma_{\mathrm{BG}} \rightarrow \gamma + \gamma$, where 
    $\gamma_{\mathrm{BG}}$ denotes a background photon, as well as inverse-Compton 
    scattering and $e^+ e^-$ pair production off nuclei.  These processes occur on     
    timescales far shorter than the timescales associated with other
    relevant processes, and thus may be considered to be effectively instantaneous.
    As we shall see, these processes serve to establish a non-thermal population
    of photons with a characteristic spectrum.     
  \item \underbar{Class~II}: Processes through which the non-thermal population of photons
    established by Class-I processes can have a direct effect on cosmological
    observables.  These include the photoproduction and photodisintegration of light 
    elements during or after BBN, as well as the photoionization of neutral hydrogen 
    and helium after recombination.
  \item \underbar{Class~III}: Processes which serve to bring the non-thermal population of 
    photons established by Class-I processes into kinetic and/or thermal equilibrium 
    with the radiation bath.  Processes in this class include 
    Compton scattering, bremsstrahlung, and $e^+ e^-$ pair production off nuclei.
\end{itemize}
We emphasize that these classes are not necessarily mutually exclusive, and that
certain processes play different roles during different cosmological epochs. 

Any energy injected in the form of photons prior to last scattering is rapidly 
redistributed to lower energies due to the Class-I processes discussed above.  
The result is a non-thermal contribution to the photon spectrum at high energies 
with a normalization that depends on the total injected power and a generic shape 
which is essentially independent of the shape of the initial injection spectrum 
directly produced by $\chi_i$ decays.  This ``reprocessed'' photon spectrum   
serves as a source of for Class-II processes --- processes
which include, for example, reactions that alter the abundances of light nuclei and 
scattering processes which contribute to the ionization of neutral hydrogen and helium 
after recombination.  Since all information about the detailed shape of the initial injection 
spectrum from decays is effectively washed out by Class-I processes in establishing this 
reprocessed photon spectrum, the results of our analysis are largely independent of the 
kinematics of $\chi_i$ decay.  This is ultimately why many of our results --- 
including those pertaining to the alteration of light abundances after BBN, distortions 
in the CMB, and the ionization history of the universe --- are likewise largely insensitive to the 
decay kinematics of the $\chi_i$.
   
The timescale over which injected photons can cause these alterations is controlled by 
the Class-III processes.  Prior to CMB decoupling, these processes serve to ``degrade'' the 
reprocessed photon spectrum established by Class-I processes by bringing this non-thermal 
population of photons into kinetic or thermal equilibrium with the photons in the radiation
bath.  As this occurs, these Class-III interactions reduce the energies of the photons below 
the threshold for Class-II processes while also potentially altering the shape of the 
CMB-photon spectrum.  These Class-III processes eventually freeze out as well, after which 
point any photons injected by particle decays 
simply contribute to the diffuse extra-galactic photon background.

%%%%%%%%%%%%%%%%%%%%%%%%%%%%%%%%%%%%%%%%%%%%%%%%%%%%%%%%%%%%%%%%%%%%%%%%%%%%%%%%%%%%%%

\FloatBarrier
\section{Impact on Light-Element Abundances\label{sec:LightElemAbundances}}

%%%%%%%%%%%%%%%%%%%%%%%%%%%%%%%%%%%%%%%%%%%%%%%%%%%%%%%%%%%%%%%%%%%%%%%%%%%%%%%%%%%%%%

We begin our analysis of the cosmological constraints on ensembles of unstable particles 
by considering the effect that the decays of these particles have on the abundances of
light nuclei generated during BBN.~  
We shall assume that these decays occur after BBN has concluded, {\it i.e.}\/, after 
initial abundances for these nuclei are already established.
We begin by reviewing the properties of the 
non-thermal photon spectrum which is established by the rapid reprocessing of injected photons
from these decays by Class-I processes.  We then review the corresponding constraints 
on a single unstable particle species~\cite{CyburtEllisUpdated} --- constraints derived
from a numerical analysis of the coupled system of Boltzmann equations which govern
the evolution of these abundances.  We then set the stage for our eventual analysis 
by deriving a set of analytic approximations
for the above constraints and demonstrating that the results obtained from these approximations
are in excellent agreement with the results of a full numerical computation within
our regime of interest.  Finally, we apply our analytic approximations in order to constrain 
scenarios involving an entire ensemble of multiple decaying particles exhibiting 
a range of masses and lifetimes.

%%%%%%%%%%%%%%%%%%%%%%%%%%%%%%%%%%%%%%%%%%%%%%%%%%%%%%%%%%%%%%%%%%%%%%%%%%%%%%%%%%%%%%
\FloatBarrier
\subsection{Reprocessed Injection Spectrum\label{sec:InjectionSpectrum}}
%%%%%%%%%%%%%%%%%%%%%%%%%%%%%%%%%%%%%%%%%%%%%%%%%%%%%%%%%%%%%%%%%%%%%%%%%%%%%%%%%%%%%%

As discussed in Sect.~\ref{sec:InjectionConstraints}, the initial spectrum of photons
injected at time $\tinj$ is redistributed effectively instantaneously by Class-I 
processes.  A detailed treatment of the Boltzmann equations governing these processes 
in a radiation-dominated epoch can be found, \eg, in Ref.~\cite{KawasakiMoroi}.
For injection at times $\tinj \lesssim 10^{12}$~s, the resulting reprocessed photon 
spectrum turns out to take a characteristic form which we may parametrize
as follows:
\begin{equation}
  \frac{dn_{\gamma}(E,\tinj)}{dE d\tinj} ~=~
    \frac{d\rho(\tinj)}{d\tinj} K(E,\tinj)~.
  \label{eq:spectrum}
\end{equation}
The quantity $d\rho(\tinj)/d\tinj$ appearing in this expression, which specifies
the overall normalization of the contribution to the reprocessed photon spectrum, 
represents the energy density injected by particle decays during the infinitesimal 
time interval from $\tinj$ to $\tinj + d\tinj$.  The function $K(E,\tinj)$, on the 
other hand, specifies the shape of the spectrum as a function of the photon energy 
$E$.  This function is normalized such that 
\begin{equation}
  \int_0^\infty K(E,\tinj) E dE ~=~ 1~.
  \label{eq:NormalizeK}
\end{equation} 
It can be shown that for 
any process that injects energy primarily through 
electromagnetic (rather than hadronic) channels, the function $K(E,\tinj)$ takes 
the universal form~\cite{EllisGelmini,ProtheroeBerezinsky,KawasakiMoroi}
\begin{equation}
  K(E,\tinj) ~=~
  \begin{cases}
    K_0 (E_X / E)^{3/2} & E < E_X \\
    K_0 (E_X / E)^{2} & E_X < E < E_C ~~~\\
    0 & E > E_C~,
  \end{cases}\\
  \label{eq:Kfndef}
\end{equation}
where $K_0$ is an overall normalization constant and where $E_C$ and $E_X$ are energy scales 
associated with specific Class-I processes whose interplay determines the shape of the 
reprocessed photon spectrum.  The normalization convention in 
Eq.~(\ref{eq:NormalizeK}) implies that $K_0$ is given by 
\begin{equation}
  K_0 ~=~ \frac{1}{E_X^2 \big[2+\ln (E_C / E_X)\big]}~.
  \label{eq:K0def}
\end{equation}  

Physically, the energy scales $E_C$ and $E_X$ appearing in Eq.~(\ref{eq:Kfndef})
can be understood as follows.  The scale $E_C$ represents the energy above which 
the photon spectrum is effectively extinguished by the pair-production process 
$\gamma + \gamma_{\mathrm{BG}} \rightarrow e^+ + e^-$,
in conjunction with interactions between the resulting electron and 
positron and other particles in the thermal bath.  The energy scale $E_X$ represents the 
threshold above which $\gamma + \gamma_{\mathrm{BG}} \rightarrow \gamma + \gamma$ is   
the dominant process through which photons lose energy.  By contrast, below this energy
threshold, the dominant processes are Compton scattering and $e^+e^-$ pair-production 
off nuclei.  Note that while the normalization of the reprocessed photon spectrum   
is set by $d\rho(\tinj)/d\tinj$, the shape of this spectrum is entirely controlled by the temperature
$\Tinj$ at injection.  This temperature behaves like $\Tinj \propto \tinj^{-1/2}$ in a 
radiation-dominated epoch.  This implies that as $\tinj$ increases, the value of $E_C$ 
also increases.  This reflects the fact that the thermal bath is colder at later injection times,
and thus an injected photon must be more energetic in order for the pair-production
process $\gamma + \gamma_{\mathrm{BG}} \rightarrow e^+ + e^-$ to be effective.
Numerically, the values of $E_C$ and $E_X$ at a given injection time $\tinj$ 
are estimated to be~\cite{KawasakiMoroi,CyburtEllisUpdated}
\begin{eqnarray}
  E_C &=& \frac{m_e^2}{22\Tinj} \approx  \big(103 \,\mev\big) \times 
    \left(\frac{\tinj}{10^8\, \s}\right)^{1/2}\nonumber\\
  E_X &=& \frac{m_e^2}{80\Tinj} \approx \big(28 \,\mev\big) \times  
    \left(\frac{\tinj}{10^8\, \s}\right)^{1/2}~,~~
  \label{eq:ECandEX}
\end{eqnarray}
where $m_e$ is the electron mass and $\Tinj$ is the temperature of the thermal bath 
at $\tinj$.

\begin{center}
\begin{figure}[t]
\includegraphics[width=0.45\textwidth]{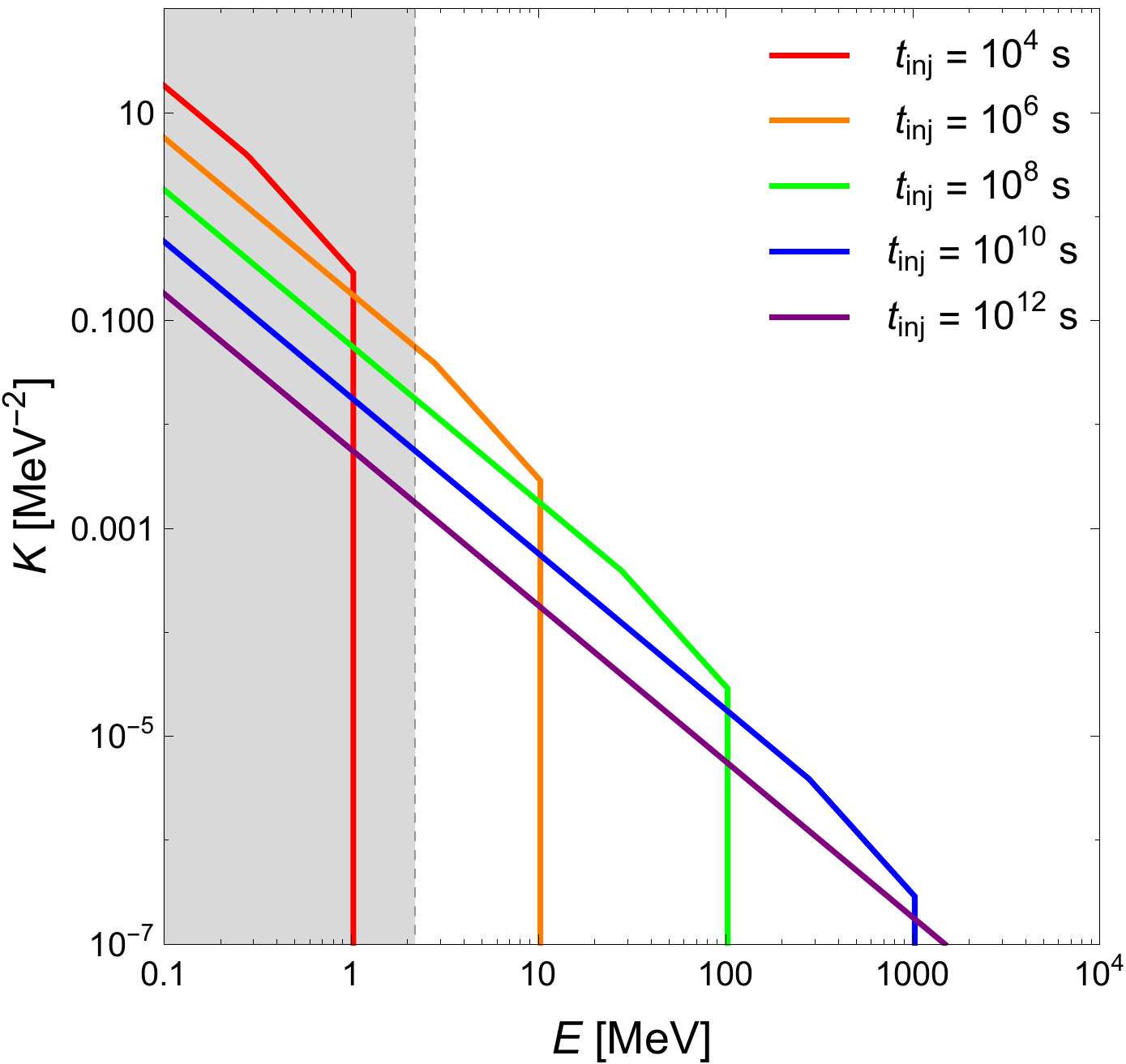}~~~
\caption{The function $K(E,\tinj)$,
    plotted as a function of photon energy $E$ for several different values of injection time $\tinj$.  
   As discussed in the text,
  this function determines the shape of
  the reprocessed photon spectrum for an instantaneous injection of photons of energy $E$ 
    at time $\tinj$.
   The dashed vertical
  line at $E\approx 2.2$~MeV indicates the lowest threshold energy 
  associated with any photoproduction or photodisintegration reaction which can 
  significantly alter the abundance of any light nucleus whose primordial 
  abundance is tightly constrained by observation.
  \label{fig:ReprocessedSpectrum}}
\end{figure}
\end{center}

The reprocessed photon spectrum in Eq.~(\ref{eq:spectrum}) is the spectrum which
effectively contributes to the photoproduction and/or photodisintegration of light 
elements after the BBN epoch.  In order to illustrate how the shape of
this spectrum depends on the injection time $\tinj$, 
we plot in Fig.~\ref{fig:ReprocessedSpectrum} the function 
$K(E,\tinj)$ which determines the shape of this spectrum as a function 
of $E$ for several different values of the injection time $\tinj$.  
Since the ultraviolet cutoff $E_C$ in the photon spectrum 
increases with $\tinj$, injection at later times can initiate photoproduction and 
photodisintegration reactions with higher energy thresholds.   
The dashed vertical line, which we include
for reference, represents the lowest threshold energy associated with any such reaction 
which can have a significant impact on the primordial abundance of any light nucleus 
which is tightly constrained by observation.  As discussed in 
Sect.~\ref{sec:LightElProdDest}, this reaction turns out to be the 
deuterium-photodisintegration reaction $\ce{D} + \gamma \rightarrow n + p$, which has a 
threshold energy of roughly 2.2~MeV.~  Thus, the portion of the photon
spectrum which lies within the gray shaded region in Fig.~\ref{fig:ReprocessedSpectrum} 
has no effect on the abundance of any relevant nucleus.  Since $E_C$ lies below this 
threshold for $\tinj \approx 10^4$~s, electromagnetic injection between the end of BBN and 
this timescale has essentially no effect on the abundances of light nuclei.     
  
One additional complication that we must take into account in assessing the effect of 
injection on the abundances of light nuclei is that the reprocessed photon spectrum 
established by Class-I processes immediately after injection at $\tinj$ is subsequently
degraded by Class-III processes, which slowly act to bring this reprocessed spectrum into 
thermal equilibrium with the background photons in the radiation bath.  
The timescale $\delta t_{\mathrm{th}}(E,\tinj)$ on which these processes act on a 
photon of energy $E$ is roughly
\begin{equation}
  \delta t_{\mathrm{th}} (E,\tinj) ~\approx~ \left[n_B(\tinj) 
    \sigma_{\mathrm{th}}(E)\right]^{-1}~,
\end{equation} 
where $n_B(\tinj)$ is the number density of baryons at the time of injection 
and where $\sigma_{\mathrm{th}}(E)$ is the characteristic cross-section for the 
relevant scattering processes, which include Compton scattering and $e^+ e^-$ pair 
production off nuclei.  The cross-sections for all relevant individual contributing 
processes can be found in Ref.~\cite{KawasakiMoroi}.  Note that while at low energies 
$\sigma_{\mathrm{th}}(E)$ is well approximated by the Thomson cross-section, this 
approximation breaks down at higher energies as other processes become relevant.  

Given these observations, the spectrum of the resulting non-thermal population of photons at time $t$ 
not only represents the sum of all contributions from injection at all times $\tinj < t$
but also reflects the subsequent degradation of these contributions by the Class-III
processes which serve to thermalize this population of photons with the 
radiation bath.  This overall non-thermal photon spectrum takes the form   
\begin{eqnarray}
  \frac{dn_{\gamma}(E,t)}{dE} &=& \int_0^t 
  \frac{dn_{\gamma}(E,\tinj)}{dE d\tinj} G(t,\tinj) d\tinj~,
  \label{eq:OverallReprocSpect}
\end{eqnarray}
where
\begin{equation}
   G(t,\tinj) ~\equiv~ e^{- (t - \tinj) / \delta t_{\mathrm{th}}(E,\tinj)} 
     \Theta (t - \tinj)
\end{equation}
is the Green's function which solves the differential equation
\begin{equation}
  \frac{dG(t,\tinj)}{dt} + \frac{G(t,\tinj)}{\delta t_{\mathrm{th}}(E,\tinj)} 
    ~=~ \delta(t-\tinj)~.
\end{equation}
The population of non-thermal photons described by Eq.~(\ref{eq:OverallReprocSpect}) 
serves as the source for the initial photoproduction and photodisintegration reactions
ultimately responsible for the modification of light-element abundances after BBN.~   
We shall therefore henceforth refer to photons in this population as ``primary'' photons.

In what follows, we will find it useful 
to employ what we shall call the ``uniform-decay approximation.''
Specifically, we shall approximate the full exponential decay of each dark-sector species $\chi_i$
as if the entire population of such particles throughout the universe were to decay precisely 
at the same time $\tau_i$.
As we shall see, this will prove critical in allowing us to formulate our ultimate analytic
approximations.
We shall nevertheless find that the results of our approximations 
are generally in excellent agreement with the results of a full numerical analysis. 

Within the uniform-decay approximation, the contribution to $d\rho/dt(\tinj)$  
from the decay of a single unstable particle species $\chi$ takes the form of a Dirac 
$\delta$-function:
\begin{equation}
  \frac{d\rho(\tinj)}{d\tinj} ~=~ \rho_\chi(\tinj) \epsilon_\chi 
    \delta(\tinj - \tau_\chi)~,
  \label{eq:drhodtinjUniformDec}
\end{equation}
where $\rho_\chi(\tinj)$ is the energy density of $\chi$ at time $\tinj$ and where 
$\epsilon_\chi$ is the fraction of the energy density released by $\chi$ decays which 
is transferred to photons.  It therefore follows that in this approximation, the primary-photon 
spectrum in Eq.~(\ref{eq:OverallReprocSpect}) reduces to
\begin{eqnarray}
  \frac{dn_{\gamma}(E,t)}{dE} &=&
    \rho_\chi(\tau_\chi) \epsilon_\chi K(E,\tau_\chi) \nonumber \\
    & & \times 
    \, e^{- (t - \tau_\chi) / \delta t_{\mathrm{th}}(E,\tinj)} \Theta(t - \tau_\chi)~.~~~~~~~~
  \label{eq:NonThermalPhotSpecApprox}
\end{eqnarray}

%%%%%%%%%%%%%%%%%%%%%%%%%%%%%%%%%%%%%%%%%%%%%%%%%%%%%%%%%%%%%%%%%%%%%%%%%%%%%%%%%%%%%%
\FloatBarrier
\subsection{Light-Element Production/Destruction\label{sec:LightElProdDest}}
%%%%%%%%%%%%%%%%%%%%%%%%%%%%%%%%%%%%%%%%%%%%%%%%%%%%%%%%%%%%%%%%%%%%%%%%%%%%%%%%%%%%%%

Generally speaking, the overall rate of change of the number density $n_a$ of a 
nuclear species $N_a$ due to the injection of electromagnetic energy 
at late times is governed by a Boltzmann equation of the form 
\begin{equation}
  \frac{d n_a}{d t} + 3H n_a ~=~ \mathcal{C}^{(p)}_a + \mathcal{C}^{(s)}_a~,
  \label{DensityInstForn}
\end{equation}
where $H$ is the Hubble parameter and where $\mathcal{C}_a^{(p)}$ and 
$\mathcal{C}_a^{(s)}$ are the collision terms associated with two different classes 
of scattering processes which contribute to this overall rate of change.  
We shall describe these individual collision terms in detail below.  Since $n_a$ is
affected by Hubble expansion, it is more convenient to work with
the corresponding comoving number density $Y_a \equiv n_a / n_B$,
where $n_B$ denotes the total number density of baryons.  The Boltzmann 
equation for $Y_a$ then takes the form
\begin{equation}
  \frac{d Y_a}{d t} ~=~ \frac{1}{n_B}\left[\mathcal{C}^{(p)}_a 
    + \mathcal{C}^{(s)}_a\right]~.~~~
\label{eq:DensityInstforY}
\end{equation} 
 
The collision term $\mathcal{C}_a^{(p)}$ represents the collective contribution from
Class-II processes directly involving the population of primary photons 
described by Eq.~(\ref{eq:NonThermalPhotSpecApprox}).
The principal processes which contribute to $\mathcal{C}_a^{(p)}$ are photoproduction 
processes of the form $N_b + \gamma \rightarrow N_a + N_c$ and photodisintegration processes 
of the form $N_a + \gamma \rightarrow N_b + N_c$, where $N_b$ and $N_c$ are other nuclei 
in the thermal plasma.  The collision term associated with these processes takes the
form 
\begin{eqnarray}
  \mathcal{C}_a^{(p)} &=& \sum_{b,c} \left[
     Y_b n_B \! \int_{E_{b}^{(ac)}}^{E_C} \!\!\!
    \frac{dn_\gamma (E,t)}{dE}  
    \sigma_{b}^{(ac)}\!(E) dE \right. \nonumber \\ & &
  - \left. Y_a n_B\! \int_{E_{a}^{(bc)}}^{E_C}  
    \!\!\!\frac{dn_\gamma (E,t)}{dE}  
    \sigma_{a}^{(bc)}\!(E)dE \right]~,~~~~~~~
\label{eq:CollisionsPrimary}
\end{eqnarray}
where the indices $b$ and $c$ run over the nuclei present in the plasma, where
$\sigma_{b}^{(ac)}(E)$ and $\sigma_{a}^{(bc)}(E)$ respectively 
denote the cross-sections for the corresponding photoproduction and 
photodisintegration processes discussed above, and where 
$E_{b}^{(ac)}$ and $E_{a}^{(bc)}$ are the respective energy thresholds 
for these processes.  Expressions for these cross-sections and 
values for the corresponding energy thresholds can be found, \eg, in 
Ref.~\cite{CyburtEllisUpdated}.  

By contrast, $\mathcal{C}_a^{(s)}$ represents the collective contribution from 
additional, secondary processes which involve not the primary photons themselves 
but rather a non-thermal population of energetic nuclei produced by interactions 
involving those primary photons.  In principle, these secondary processes include
both reactions that produce nuclei of species $N_a$ and reactions which destroy them. 
In practice, however, because the non-thermal population of any given species $N_b$ 
generated by processes involving primary photons is comparatively small, the effect of 
secondary processes on the populations of most nuclear species is likewise small.  As we 
shall discuss in more detail in Sect.~\ref{sec:ConstraintsPrimord}, the only exception is 
$\ce{^{6}Li}$, which is not produced in any significant amount during the BBN epoch 
but which can potentially be produced by secondary processes initiated by photon injection 
at subsequent times.  Since these processes involve the production rather than 
the destruction of $\ce{^{6}Li}$, we focus on the effect of secondary processes on 
nuclei which appear in the final state rather than the initial state in what follows.

The energetic nuclei which participate in secondary processes are the products of the 
same kinds of reactions which lead to the collision term in Eq.~(\ref{eq:CollisionsPrimary}).
Thus, the kinetic-energy spectrum $d \widetilde{n}_b(E_b,t)/dE_b$ of the non-thermal       
population of a nuclear species $N_b$ produced in this manner is in large part determined  
by the energy spectrum of the primary photons.
In calculating this spectrum, one must in principle account for the fact that a photon of 
energy $E_\gamma$ can give rise to a range of possible $E_b$ values due to the range of
possible scattering angles between the three-momentum vectors of the incoming photon and 
the excited nucleus in the center-of-mass frame.  However, it can be 
shown~\cite{SecProdKinematicsToAppear} that a reasonable 
approximation for the collision term $\mathcal{C}^{(s)}_a$ for $\ce{^{6}Li}$ is 
nevertheless obtained by taking $E_b$ to be a one-to-one function of $E_\gamma$ of the 
form~\cite{JedamzikLi6} 
\begin{equation}
  E_b ~=~ \mathcal{E}(E_\gamma) ~\equiv~ \frac{1}{4} 
    \Big[E_\gamma - E_c^{(bd)}\Big]~,
  \label{eq:MonochMap}
\end{equation}   
where $E_c^{(bd)}$ is the energy threshold for the primary process 
$N_c + \gamma \rightarrow N_b + N_d$.  In this approximation, 
$d \widetilde{n}_b(E_b,t)/dE_b$ takes the form~\cite{CyburtEllisUpdated}
\begin{eqnarray}
  \frac{d \widetilde{n}_b(E_b,t)}{dE_b} &=& \sum_{c,d} \frac{n_c}{b (E_b,t)} 
    \int_{E_{\mathrm{min}} (E_b)}^{E_C} \Bigg[  
    \frac{dn_\gamma (E_\gamma',t)}{dE_\gamma'} ~~~\nonumber \\ & & ~\times   
      \sigma^{(bd)}_c(E_\gamma') e^{-\beta_b(E_\gamma',t)} \Bigg] dE_\gamma'~,~~~~~
  \label{eq:ExcitedNucleusSpectrum}
\end{eqnarray}
where $b(E_b,t)$ is the energy-loss rate for $N_b$ due to Coulomb scattering  
with particles in the thermal background plasma.  The exponential factor 
$\beta_b(E_\gamma',t)$ accounts for
the collective effect of additional processes which act to reduce the 
number of nuclei of species $N_b$.  The lower limit of integration in
Eq.~(\ref{eq:ExcitedNucleusSpectrum}) is given by 
$E_{\mathrm{min}}(E_b) = \max [\mathcal{E}^{-1}(E_b),E_{c}^{(bd)}]$, where 
$\mathcal{E}^{-1}(E_b)$ is the inverse of the function defined in
Eq.~(\ref{eq:MonochMap}).  In other words, $\mathcal{E}^{-1}(E_b)$ is
the photon energy which corresponds to a kinetic energy $E_b$ for the 
excited nucleus. 

In principle, the processes which contribute to $\beta_a(E_\gamma',t)$ include both 
decay processes (in the case in which $N_b$ is unstable) and photodisintegration 
processes of the form $N_b + \gamma \rightarrow N_c + N_d$ involving a primary
photon.  In practice, the photodisintegration rate due to these processes is 
much slower that the energy-loss rate due to Coulomb scattering for any species
of interest.  Moreover, as we shall see in Sect.~\ref{sec:ConstraintsPrimord}, the 
only nuclear species whose non-thermal population has a significant effect on the 
$\ce{^{6}Li}$ abundance are tritium ($\ce{T}$) and the helium isotope $\ce{^{3}He}$. 
Because these two species are mirror nuclei, the secondary processes
in which they participate affect the $\ce{^{6}Li}$ abundance in the same way and 
have almost identical cross-sections and energy thresholds.  Thus, in terms of
their effect on the production of $\ce{^{6}Li}$, the populations of $\ce{T}$ 
and $\ce{^{3}He}$ may effectively be treated together as if they were the population 
of a single nuclear species.  Although tritium is unstable and decays via beta 
decay to $\ce{^{3}He}$ with a lifetime of $\tau_{\ce{T}} \approx 5.6 \times 10^8$~s, 
these decays have no impact on the combined population of $\ce{T}$ and 
$\ce{^{3}He}$.  We may therefore safely approximate $\beta_b(E_\gamma',t) \approx 0$ 
for this combined population of excited nuclei in what follows.     

The most relevant processes through which an energetic nucleus $N_b$ in this 
non-thermal population can alter the abundance of another nuclear species $N_b$
are scattering processes of the form $N_b + N_f \rightarrow N_a + X$, in which an 
energetic nucleus $N_b$ from the non-thermal population generated by primary processes 
scatters with a background nucleus $N_f$, resulting in the production of a nucleus of 
species $N_a$ and some other particle $X$ (which could be either an additional nucleus 
or a photon).  In principle, processes of the form $N_b + N_a \rightarrow N_f + X$ 
can also act to reduce the abundance of $N_a$.  However, as discussed above, this 
reduction has a negligible impact on $Y_a$ for any nuclear species 
$N_a$ which already has a sizable comoving number density at the end of BBN.~
Thus, we focus here on production rather than destruction
when assessing the impact of secondary processes on the primordial abundances of 
light nuclei. 

With this simplification, the collision term associated with secondary production 
processes takes the form
\begin{eqnarray}
  \mathcal{C}_a^{(s)} &=& \sum_{b,f} Y_f n_B \! 
    \int_{E_{bf}^{(aX)}}^{\mathcal{E}(E_C)} \Bigg[ 
    \frac{d \widetilde{n}_b (E_b,t)}{dE_b} \nonumber \\ & & ~~~~~~~~ \times  
    \sigma_{bf}^{(aX)}(E_b) |v(E_b)| \Bigg] dE_b ~,~~~
  \label{eq:CollisionsSecondary}
\end{eqnarray}
where $v(E_b)$ is the (non-relativistic) relative velocity of nuclei $N_b$ and $N_f$ in the
background frame, where $d \widetilde{n}_b(E_b,t) / dE_b$ is the differential energy spectrum of 
the non-thermal population of $N_b$, where $\sigma_{bf}^{(aX)}$ is the cross-section for 
the scattering process $N_b + N_f \rightarrow N_a + X$ with corresponding threshold energy 
$E_{bf}^{(aX)}$, and where $\mathcal{E}(E_C)$ is the cutoff in 
$d \widetilde{n}_b(E_b,t)/dE_b$ produced by primary processes.

%%%%%%%%%%%%%%%%%%%%%%%%%%%%%%%%%%%%%%%%%%%%%%%%%%%%%%%%%%%%%%%%%%%%%%%%%%%%%%%%%%%%%%
\FloatBarrier
\subsection{Constraints on Primordial Light-Element Abundances\label{sec:ConstraintsPrimord}}
%%%%%%%%%%%%%%%%%%%%%%%%%%%%%%%%%%%%%%%%%%%%%%%%%%%%%%%%%%%%%%%%%%%%%%%%%%%%%%%%%%%%%%

The nuclear species whose primordial abundances are the most tightly constrained 
by observation ---  and which are therefore relevant for constraining the late decays of 
unstable particles ---  are $\ce{D}$, $\ce{^{4}He}$, $\ce{^{6}Li}$, and $\ce{^{7}Li}$. 
The abundance of $\ce{^{3}He}$ during the present cosmological epoch has also been 
constrained by observation~\cite{BaniaHe3,GeissHe3}.  However, uncertainties in the 
contribution to this abundance from stellar sources make it difficult to translate 
the results of these measurements into bounds on the primordial $\ce{^{3}He}$ 
abundance~\cite{ChiappiniHe3,VangioniFlamHe3}.  The effect of these uncertainties can
be mitigated in part if we consider the ratio $(\ce{D} + \ce{^{3}He})/\ce{H}$ 
rather than $\ce{^{3}He}/\ce{H}$, as the former is expected to be largely unaffected
by stellar processing~\cite{Ellis:1984er,IoccoBBNReview,CocBBNReview}. 
In this paper, we focus our 
attention on $\ce{D}$, $\ce{^{4}He}$, $\ce{^{6}Li}$, and $\ce{^{7}Li}$, as the 
relationship between the measured abundances of these nuclei and their corresponding
primordial abundances is more transparent.   

The observational constraints on the primordial abundances of these nuclei can be 
summarized as follows.  Bounds on the primordial $\ce{^{4}He}$ abundance are typically
phrased in terms of the primordial helium mass fraction 
$Y_p\equiv (\rho_{\ce{^{4}He}}/{\rho_B})_p$, where 
$\rho_{\ce{^{4}He}}$ is the mass density of $\ce{^{4}He}$, where $\rho_B$ is
the total mass density of baryonic matter, and where the subscript $p$ signifies
that it is only the primordial contribution to $\rho_{\ce{^{4}He}}$ which is used in 
calculating $Y_p$, with subsequent modifications to this quantity due to stellar synthesis, 
\etc, ignored.  The $2\sigma$ limits on $Y_p$ are~\cite{AverHe4} 
\begin{equation}
  0.2369 ~<~ Y_p ~<~ 0.2529~.
  \label{eq:ObsBoundHe4}
\end{equation}
The observational $2\sigma$ limits on the $\ce{^{7}Li}$ abundance 
are~\cite{SbordoneLi7}
\begin{equation}
  1.0 \times 10^{-10} ~<~ \left( \frac{\ce{^{7}Li}}{\ce{H}} \right)_p 
    ~<~ 2.2 \times 10^{-10}~,
  \label{eq:ObsBoundLi7}
\end{equation}
where the symbols $\ce{^{7}Li}$ and $\ce{H}$ denote the primordial number 
densities of the corresponding nuclear species. 
In this connection, we note that significant tension exists between these 
observational bounds and the predictions of theoretical calculations 
of the $\ce{^{7}Li}$ abundance, which are roughly a factor of three larger.  
While it is not our aim in this paper to address this discrepancy, 
electromagnetic injection from the late decays of 
unstable particles may play a
role~\cite{KusakabeAxion,EllisParticleDecaysAndLi7,KusakabeLi6andLi7} 
in reconciling these predictions with observational data.

Constraining the primordial abundance of $\ce{D}$ is complicated by a mild
tension which currently exists between the observational results for $\ce{D}/\ce{H}$
derived from measurements of the line spectra of low-metallicity
gas clouds~\cite{CookeDUpper} and the results obtained from numerical analysis of the
Boltzmann equations for BBN~\cite{MarcucciDLower} with input from Planck
data~\cite{Planck2015}, which predict a slightly lower value for this ratio. 
We account for these tensions by choosing our central value and lower limit on
$\ce{D}/\ce{H}$ in accord with the central value and $2\sigma$ lower limit from
numerical calculations, while at the same time adopting the $2\sigma$ observational
upper limit as our own upper limit on this ratio.  Thus, we take our bounds on the
$\ce{D}$ abundance to be
\begin{equation}
  2.317 \times 10^{-5} ~<~ \left( \frac{\ce{D}}{\ce{H}} \right)_p 
    ~<~ 2.587 \times 10^{-5}~.
  \label{eq:ObsBoundD}
\end{equation} 
   
An upper bound on the ratio $\ce{^{6}Li}/\ce{H}$ can likewise be derived from
observation by combining observational upper bounds on the more directly 
constrained quantities $\ce{^{6}Li}/\ce{^{7}Li}$ and $\ce{^{7}Li}/\ce{H}$.
By combining the upper bound on $\ce{^{6}Li}/\ce{^{7}Li}$ from 
Ref.~\cite{AsplundLi6toLi7} with 
the upper bound from Eq.~(\ref{eq:ObsBoundLi7}), we obtain 
\begin{equation}
  \left( \frac{\ce{^{6}Li}}{\ce{H}} \right)_p ~<~ 2 \times 10^{-11}~.
  \label{eq:ObsBoundLi6}
\end{equation} 
While this upper bound is identical to the corresponding 
constraint quoted in Ref.~\cite{CyburtEllisUpdated}, this is a numerical
accident resulting from a higher estimate of the $\ce{^{6}Li}/\ce{^{7}Li}$ 
ratio (due to the recent detection of additional $\ce{^{6}Li}$ in 
low-metallicity stars) and a reduction in the upper bound on the 
$\ce{^{7}Li}/{\ce{H}}$ ratio.  

Having assessed the observational constraints on $\ce{D}$, $\ce{^{4}He}$, 
$\ce{^{6}Li}$, and $\ce{^{7}Li}$, we now turn to consider the effect that 
the late-time injection of electromagnetic 
radiation has on the abundance of each of these nuclear species relative 
to its initial abundance at the conclusion of the BBN epoch. 

Of all these species, $\ce{^{4}He}$ is by far the most abundant.  
For this reason, reactions involving $\ce{^{4}He}$ nuclei
in the initial state play an outsize role in the production of other nuclear 
species.  Moreover, since the abundances of all other such species in the thermal
bath are far smaller than that of $\ce{^{4}He}$, reactions involving these other 
nuclei in the initial state have a negligible impact on the $\ce{^{4}He}$ abundance.  
Photodisintegration processes initiated directly by primary photons 
are therefore the only processes which have an appreciable effect on the primordial 
abundance of $\ce{^{4}He}$.  A number of individual such processes contribute to the 
overall photodisintegration rate of $\ce{^{4}He}$, all of which have threshold energies 
$E_{\mathrm{thresh}}\gtrsim 20~\mev$.

The primordial abundance of $\ce{^{7}Li}$, like that of $\ce{^{4}He}$, evolves
in response to photon injection primarily as a result of photodisintegration 
processes initiated directly by primary photons.  At early times, when the energy 
ceiling $E_C$ in Eq.~(\ref{eq:ECandEX}) for the spectrum of these photons is relatively 
low, the process $\ce{^{7}Li} + \gamma \rightarrow \ce{^{4}He} + \ce{T}$, which has 
a threshold energy of only $\sim 2.5~\mev$, dominates the photodisintegration rate.
By contrast, at later times, additional processes with higher threshold energies,
such as $\ce{^{7}Li} + \gamma \rightarrow \ce{^{6}Li} + n$ and 
$\ce{^{7}Li} + \gamma \rightarrow \ce{^{4}He} + 2n + p$, become relevant. 

While the reactions which have a significant impact on the $\ce{^{4}He}$ and 
$\ce{^{7}Li}$ abundances all serve to reduce these abundances, the reactions 
which have an impact on the $\ce{D}$ abundance include both processes 
which create deuterium nuclei and processes which destroy them.       
At early times, photodisintegration processes initiated by primary photons --- and 
in particular the process $\ce{D} + \gamma \rightarrow n + p$, which has a 
threshold energy of only 
$E_{\mathrm{thresh}} \approx 2.2~\mev$ --- dominate and serve to deplete the initial 
$\ce{D}$ abundance.  At later times, however, additional processes with
higher energy thresholds turn on and serve to counteract this initial depletion.  
The dominant such process is the photoproduction process
$\ce{^{4}He} + \gamma \rightarrow \ce{D} + n + p$, which has a threshold energy of 
$E_{\mathrm{thresh}} \approx 25~\mev$.

Unlike $\ce{^{4}He}$, $\ce{^{7}Li}$, and $\ce{D}$, the nucleus $\ce{^{6}Li}$ 
is not generated to any significant degree by BBN.~ 
However, a population of $\ce{^{6}Li}$ nuclei can be generated after 
BBN as a result of photon injection at subsequent times.  The most relevant 
processes 
are $\ce{^{7}Li} + \gamma \rightarrow \ce{^{6}Li} + n$ 
and the secondary 
production processes $\ce{^{4}He} + \ce{^{3}He} \rightarrow \ce{^{6}Li} + p$ and 
$\ce{^{4}He} + \ce{T} \rightarrow \ce{^{6}Li} + n$, where $\ce{T}$ denotes 
a tritium nucleus. 
All of these processes have energy thresholds $E_{\mathrm{thresh}} \approx 7~\mev$.  
The abundances of $\ce{^{3}He}$ and $\ce{T}$, which serve as reactants in these 
secondary processes, are smaller at the end of BBN than the abundance of $\ce{^{4}He}$ 
by factors of $\mathcal{O}(10^4)$ and $\mathcal{O}(10^6)$, 
respectively (for reviews, see, \eg, Ref.~\cite{PospelovReview}).  At the same time, 
the non-thermal populations of $\ce{^{3}He}$ and $\ce{T}$ generated via the 
photodisintegration of $\ce{^{4}He}$ are much larger than the non-thermal population 
of $\ce{^{4}He}$, which is generated via the photodisintegration of other, far less 
abundant nuclei.  Thus, to a very good approximation, the reactions which contribute 
to the secondary production of $\ce{^{6}Li}$ involve an excited $\ce{^{3}He}$ or 
$\ce{T}$ nucleus and a ``background'' $\ce{^{4}He}$ nucleus in thermal equilibrium with 
the radiation bath.     

In Table~\ref{tab:BBNprocs}, we provide a list of the relevant reactions which 
can serve to alter the abundances of light nuclei as a consequence of photon
injection at late times, along with their corresponding energy thresholds.
Expressions for the cross-sections for these processes are given 
in Ref.~\cite{CyburtEllisUpdated}.  
We note that alternative parametrizations for some of the relevant cross-section formulae have been 
proposed~\cite{PramHe4PhotodestructXSec} 
on the basis of recent nuclear experimental results,
though tensions still exist among data from different sources.
While there exist additional nuclear processes beyond those listed in 
Table~\ref{tab:BBNprocs}
that in principle contribute to the collision terms in Eq.~(\ref{eq:DensityInstforY}),
these processes do not have a significant impact on the $Y_a$ of 
any relevant nucleus when the injected energy density is small and can therefore be neglected.
In should be noted that 
the population of excited $\ce{T}$ and $\ce{^{3}He}$ nuclei which participate in 
the secondary production of $\ce{^{6}Li}$ are generated primarily by 
the same processes which contribute to the destruction of $\ce{^{4}He}$.
We note that we have not included processes which contribute to the destruction of 
$\ce{^{6}Li}$.  The reason is that
the collision terms for these processes are proportional to $Y_{\ce{^{6}Li}}$ itself 
and thus only become important in the regime in which the rate of electromagnetic 
injection from unstable-particle decays is large.  By contrast, for reasons that shall 
be discussed in greater detail below, we focus in what follows primarily on the regime 
in which injection is small and $\ce{^{6}Li}$-destruction processes are unimportant.    
However, we note that these processes can have an important effect on $Y_{\ce{^{6}Li}}$ 
in the opposite regime, rendering the bound in Eq.~(\ref{eq:ObsBoundLi6}) essentially 
unconstraining for sufficiently large injection rates~\cite{CyburtEllisUpdated}. 

\begin{table}[t]
\centering
\smaller
\begin{tabular}{||c|c|r||}
\hline
\hline
  Process & Associated Reactions  & $E_{\mathrm{thresh}}$~ \\
\hline
\hline
  \ce{D} Destruction  
   & $\ce{D} + \gamma \rightarrow n + p$  & $ 2.2 \mev$  \\
\hline
  \ce{D}  Production  
   & $\ce{^{4}He} + \gamma \rightarrow \ce{D} + n + p $ & $26.1 \mev$  \\
\hline
 \ce{^{4}He} Destruction 
   & $\ce{^{4}He} + \gamma \rightarrow \ce{T} + p$  & $19.8 \mev$ \\
   & $\ce{^{4}He} + \gamma \rightarrow \ce{^{3}He} + n$  & $ 20.6 \mev$ \\
\hline
 \ce{^{7}Li} Destruction 
   & $\ce{^{7}Li} + \gamma \rightarrow \ce{^{4}He} + \ce{T}$ & $ 2.5 \mev$ \\
   & $\ce{^{7}Li} + \gamma \rightarrow \ce{^{6}Li} + n$ & $ 7.2 \mev$  \\
   & $\ce{^{7}Li} + \gamma \rightarrow  \ce{^{4}He} + 2n + p$ & $ 10.9 \mev$  \\
\hline
 Primary \ce{^{6}Li} Production
   & $\ce{^{7}Li} + \gamma \rightarrow \ce{^{6}Li} + n$ & $ 7.2 \mev$  \\
\hline
 ~Secondary \ce{^{6}Li} Production~  
    & $\ce{^{4}He} + \ce{^{3}He} \rightarrow \ce{^{6}Li} + p$  & $ 7.0 \mev$ \\
    & $\ce{^{4}He} + \ce{T} \rightarrow \ce{^{6}Li} + n$  & $ 8.4 \mev$ \\
\hline
\hline
\end{tabular}
\caption{~Reactions which alter the abundances of
  $\ce{D}$, $\ce{^{4}He}$, $\ce{^{6}Li}$, and $\ce{^{7}Li}$ in scenarios
  involving photon injection at late times, along with the corresponding 
  energy threshold $E_{\mathrm{thresh}}$ for each process.  We note that 
  the excited $\ce{T}$ and 
  $\ce{^{3}He}$ nuclei which participate in the secondary production
  of $\ce{^{6}Li}$ are generated primarily by the processes listed above 
  which contribute to the destruction of $\ce{^{4}He}$.
\label{tab:BBNprocs}}
\end{table}

Finally, we note that
the rates and energy thresholds for 
$\ce{^{4}He} + \ce{^{3}He} \rightarrow \ce{^{6}Li} + p$ and 
$\ce{^{4}He} + \ce{T} \rightarrow \ce{^{6}Li} + n$ are very similar, as are the 
rates and energy thresholds for the $\ce{^{4}He}$-destruction processes which 
produce the non-thermal 
populations of $\ce{^{3}He}$ and $\ce{T}$~\cite{CyburtEllisUpdated}.  In what follows, 
we shall make the simplifying approximation that $\ce{^{3}He}$ and $\ce{T}$ are 
``interchangeable'' in the sense that we treat these rates --- and hence also the 
non-thermal spectra $d\widetilde{n}_a(E,t)/dE$ of $\ce{^{3}He}$ and $\ce{T}$ --- as 
identical.  Thus, although $\ce{T}$ decays via beta decay to $\ce{^{3}He}$ on a timescale 
$\tau_{\ce{T}} \approx 10^{8}$~s, we neglect the effect of the decay kinematics on
the resulting non-thermal $\ce{^{3}He}$ spectrum.  As we shall see, these simplifying
approximations do not significantly impact our results.

%%%%%%%%%%%%%%%%%%%%%%%%%%%%%%%%%%%%%%%%%%%%%%%%%%%%%%%%%%%%%%%%%%%%%%%%%%%%%%%%%%%%%%
\FloatBarrier
\subsection{Towards an Analytic Approximation:  Linearization and Decoupling\label{sec:Linearization}}
%%%%%%%%%%%%%%%%%%%%%%%%%%%%%%%%%%%%%%%%%%%%%%%%%%%%%%%%%%%%%%%%%%%%%%%%%%%%%%%%%%%%%%

In order to assess whether a particular injection history is consistent with 
the constraints discussed in the previous section, we must evaluate the overall 
change $\delta Y_a(t) \equiv Y_a(t) - Y_a^{\mathrm{init}}$ in the comoving number density 
$Y_a$ of a given nucleus at time $t = \tnow$, where $Y_a^{\mathrm{init}}$ denotes the
initial value of $Y_a$ at the end of BBN.~  In principle, this involves
solving a system of coupled differential equations,
one for each nuclear species present in the thermal bath, each of the form given in Eq.~(\ref{eq:DensityInstforY}).  

In practice, however, we can obtain reasonably 
reliable estimates for the $\delta Y_a$ without having to resort to a full 
numerical analysis.  This is possible ultimately because observational constraints 
require $|\delta Y_a|$ to be quite small for all relevant nuclei, as 
we saw in Sect.~\ref{sec:ConstraintsPrimord}.  The equations governing the evolution
of the $Y_a$ are coupled due to feedback effects in which a change in the comoving 
number density of one nuclear species $N_a$ alters the reaction rates 
associated with the production of other nuclear species.  However, if the change
in $Y_a$ is sufficiently small for all relevant species, these feedback effects can 
be neglected and the evolution equations effectively decouple.

In order for the evolution equations for a particular nuclear species $N_a$ to 
decouple, the {\it linearity criterion}\/ $|\delta Y_b| \ll Y_b$ must be satisfied for 
any other nuclear species $N_b \neq N_a$ which serves as a source for reactions that 
significantly affect the abundance of $N_a$ at all times after the 
conclusion of the BBN epoch.  In principle, there are two ways in 
which this criterion could be enforced by the observational constraints and consistency
conditions discussed in Sect.~\ref{sec:ConstraintsPrimord}.  The first is simply 
that the applicable bound on each $Y_b$ which serves as a source for $N_a$ is sufficiently 
stringent that this bound is always violated before the linearity criterion 
$|\delta Y_b| \ll Y_b$ fails.  The second possibility is that while the direct
bound on $Y_b$ may not in and of itself require that $|\delta Y_b|$ be small, the comoving 
number densities $Y_a$ and $Y_b$ are nevertheless directly related in such a way that the 
applicable bound on $Y_a$ is always violated before the linearity criterion 
$|\delta Y_b| \ll Y_b$ fails.  If one of these two conditions is satisfied for every 
species $N_b$ which serves as a source for $N_a$, we may treat the evolution equation for
$N_a$ as effectively decoupled from the equations which govern the evolution of all other 
nuclear species.  We emphasize that $N_a$ itself need not satisfy the linearity criterion
in order for its evolution equation to decouple in this way.

%============ figure
\begin{center}
\begin{figure*}[t]
\includegraphics[width=0.8\textwidth]{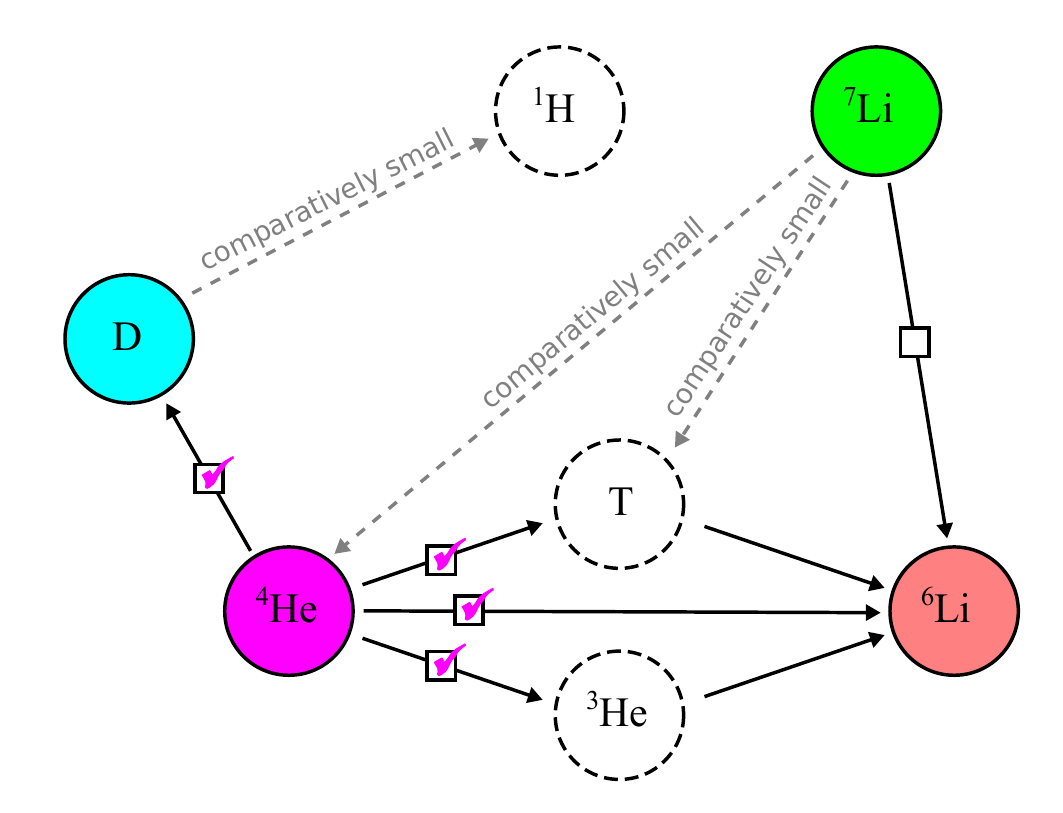}~~~
\caption{Schematic illustrating the network of reactions precipitated by electromagnetic injection
after the end of the BBN epoch.  The nodes in the network represent different nuclear species:
nodes represented by solid circles correspond to nuclei whose primordial comoving number
densities $Y_a$ are reliably constrained by data, while nodes represented by dashed circles correspond
to other nuclear species involved in these reactions.  An arrow pointing from one node to
another indicates that the nucleus associated with the node from which the arrow originates serves
as a source for the nucleus associated with the node to which the arrow points.  A solid arrow
indicates that the corresponding reaction can potentially have a non-negligible impact on the
abundance of the product nucleus, while a dashed arrow indicates that the effect of the corresponding
reaction is always negligible.  
Note that $\ce{^{1}H}$ nuclei --- \ie, protons --- are 
generated as a byproduct of many of the other reactions shown.  
However, the impact of these contributions to the $\ce{^{1}H}$ abundance is 
negligible and the corresponding arrows have been omitted for clarity.
A checked box
superimposed on the arrows emerging from a node indicates that observational constraints enforce
the linearity criterion $|\delta Y_a| \ll Y_a$ for the corresponding nucleus.  By contrast, an open
box indicates that this criterion is {\it not}\/ satisfied.
The Boltzmann equation for a given nucleus $N_a$ effectively decouples,
in the sense that feedback effects can be neglected in calculating $Y_a$, whenever the linearity criterion is
satisfied for all $N_b \neq N_a$ which serve as a source for $N_a$ (though not necessarily for $N_a$ itself). 
Since no other nucleus serves as a source for $\ce{^{4}He}$ or $\ce{^{7}Li}$, the Boltzmann equations for both
of these species trivially decouple.  The Boltzmann equation for $\ce{D}$ also decouples because $\ce{^{4}He}$,
the one nucleus which serves as a significant source for $\ce{D}$, is required to satisfy the linearity
criterion.  However, the Boltzmann equation for $\ce{^{6}Li}$ does not decouple, as one of its source
nuclei --- in particular, $\ce{^{7}Li}$ --- does not satisfy the linearity criterion.  Further details
are described in the text.
\label{fig:ReactionNetwork}}
\end{figure*}
\end{center}
%============ end of figure

We now turn to examine whether and under what circumstances our criterion for
the decoupling of the evolution equations is satisfied in practice for all relevant 
nuclear species.  
In Fig.~\ref{fig:ReactionNetwork} we illustrate the 
network of reactions which can have a significant effect on the values of $Y_a$ 
for these species.
The nuclei which appear in the initial state of one of the primary production 
processes listed in Table~\ref{tab:BBNprocs} are $\ce{^{4}He}$, which serves as a source 
for $\ce{D}$ and $\ce{^{6}Li}$, and $\ce{^{7}Li}$, which serves as a source for 
$\ce{^{6}Li}$.  We note that while $\ce{^{3}He}$ and $\ce{T}$ each appear in the 
initial state of one of the secondary production processes for $\ce{^{6}Li}$, it is 
the non-thermal population of each nucleus which plays a significant role in these 
reactions.  Since the non-thermal populations of both $\ce{^{3}He}$ and $\ce{T}$ are 
generated primarily as a byproduct of $\ce{^{4}He}$ destruction, requiring that the 
linearity criterion be satisfied for $\ce{^{4}He}$ and $\ce{^{7}Li}$ is sufficient to 
ensure that the evolution equations for all relevant nuclear species decouple.  

We begin by assessing whether the direct constraints on $Y_{\ce{^{4}He}}$ and 
$Y_{\ce{^{7}Li}}$ themselves are sufficient to enforce the linearity criterion.
We take the initial values of these comoving number densities at the end of the 
BBN epoch to be those which correspond to the central observational values for 
$Y_p$ and $\ce{^{7}Li}/\ce{H}$ quoted in Ref.~\cite{CyburtBBN2015}, namely 
$Y_p = 0.2449$ and $\ce{^{7}Li}/\ce{H} = 1.6 \times 10^{-10}$.
Since neither $\ce{^{4}He}$ nor $\ce{^{7}Li}$ is produced at a significant rate by 
interactions involving other nuclear species, the evolution equation in 
Eq.~(\ref{eq:DensityInstforY}) for each of these nuclei takes
the form 
\begin{equation}
  \frac{dY_b}{dt} ~=~ -Y_b(t) \Gamma_b(t)~,
  \label{eq:DepletionRateYa}
\end{equation}
where the quantity $\Gamma_b(t) \equiv -\mathcal{C}_b^{(p)}/Y_b n_B \geq 0$ represents 
the rate at which $Y_b$ is depleted as a result of photodisintegration processes.  
This depletion rate varies in time, but depends neither on $Y_b$ nor on the comoving number 
density of any other nucleus.  Since $Y_b$ decreases monotonically in this case, it 
follows that if this comoving quantity lies within the observationally-allowed range 
today, it must also lie within this range at all times since the end of the BBN epoch.

The bound on $Y_{\ce{^{4}He}}$ which follows from Eq.~(\ref{eq:ObsBoundHe4}) is 
sufficiently stringent that $|\delta Y_{\ce{^{4}He}}| \ll Y_{\ce{^{4}He}}$ is indeed
required at all times since the end of BBN for consistency with observation.  Thus, our 
linearity criterion is always satisfied for $\ce{^{4}He}$.  By contrast, the bound on 
$Y_{\ce{^{7}Li}}$ in Eq.~(\ref{eq:ObsBoundLi7}) is far weaker in the sense that 
$|\delta Y_{\ce{^{7}Li}}|$ need not necessarily be small in relation to $Y_{\ce{^{7}Li}}$ 
itself.  Moreover, while the contribution to $\delta Y_{\ce{^{6}Li}}$ from primary 
production is indeed directly related to $\delta Y_{\ce{^{7}Li}}$, we find that the 
observational bound on $Y_{\ce{^{6}Li}}$ is not always violated before the linearity 
criterion $|\delta Y_{\ce{^{7}Li}}| \ll Y_{\ce{^{7}Li}}$ fails --- even if we assume
$Y_{\ce{^{6}Li}} \approx 0$ at the end of BBN.~  The reason is that not 
every $\ce{^{7}Li}$ nucleus destroyed by primary photodisintegration processes produces 
a $\ce{^{6}Li}$ nucleus.  Indeed, $\ce{^{7}Li} + \gamma \rightarrow \ce{^{4}He} + \ce{T}$
and $\ce{^{7}Li} + \gamma \rightarrow \ce{^{4}He} + 2n + p$ contribute to the 
depletion of $Y_{\ce{^{7}Li}}$ as well.
    
Since the energy threshold for $\ce{^{7}Li} + \gamma \rightarrow \ce{^{4}He} + \ce{T}$
is lower than the threshold for $\ce{^{7}Li} + \gamma \rightarrow \ce{^{6}Li} + n$,
there will be a range of $\tinj$ within which injection contributes to the 
destruction of $\ce{^{7}Li}$ without producing $\ce{^{6}Li}$ at all.   
Furthermore, even at later injection times $\tinj \gtrsim 4.9 \times 10^5$~s, when 
the primary-photon spectrum from injection includes photons with energies above the 
threshold for $\ce{^{7}Li} + \gamma \rightarrow \ce{^{6}Li} + n$, the 
$\ce{^{7}Li}$-photodisintegration rates associated with this process and the
rates associated with $\ce{^{7}Li} + \gamma \rightarrow \ce{^{4}He} + \ce{T}$ and 
$\ce{^{7}Li} + \gamma \rightarrow \ce{^{4}He} + 2n + p$ are comparable.  
Consequently, only around 30\% of $\ce{^{7}Li}$ nuclei destroyed by primary 
photodisintegration for $\tinj \gtrsim 4.9 \times 10^5$~s produce a 
$\ce{^{6}Li}$ nucleus in the process.  Thus, a large $|\delta Y_{\ce{^{7}Li}}|$ 
invariably results in a much smaller contribution to $\delta Y_{\ce{^{6}Li}}$.
Thus, $\ce{^{7}Li}$ does not satisfy the linearity criterion when serving as a source
for $\ce{^{6}Li}$.

That said, while our linearity criterion is not truly satisfied for $\ce{^{7}Li}$,
we can nevertheless derive meaningful constraints on decaying particle ensembles 
from observational bounds on $\ce{^{6}Li}$ by neglecting feedback effects on 
$Y_{\ce{^{7}Li}}$ in calculating $Y_{\ce{^{6}Li}}$. 
Since the Boltzmann equation for $Y_{\ce{^{7}Li}}$ takes the form given in 
Eq.~(\ref{eq:DepletionRateYa}), $Y_{\ce{^{7}Li}}$ is always less than or equal 
to its initial value $Y_{\ce{^{7}Li}}^{\mathrm{init}}$ at the end of BBN.~  
This in turn implies that the collision term $\mathcal{C}_{\ce{^{6}Li}}^{(p)}$ 
in the Boltzmann 
equation for $\ce{^{6}Li}$ is always less than or equal to the value that it would 
have had if the linearity criterion for $\ce{^{7}Li}$ {\it had}\/ been satisfied.  
It therefore follows that the contribution to $\delta Y_{\ce{^{6}Li}}$ from 
primary production which we would obtain if we were to approximate 
$Y_{\ce{^{7}Li}}$ by $Y_{\ce{^{7}Li}}^{\mathrm{init}}$ at all times subsequent to
the end of BBN is always an overestimate.  In this sense, then, the bound on 
electromagnetic injection which we would obtain by invoking this linear
approximation for $Y_{\ce{^{7}Li}}$ represents a conservative bound.      
Moreover, it turns out that because of the relationship between $Y_{\ce{^{7}Li}}$ 
and $Y_{\ce{^{6}Li}}$, the bound on decaying ensembles from the destruction of 
$\ce{^{7}Li}$ is always more stringent than the bound from the primary production 
of $\ce{^{6}Li}$.  Thus, adopting the linear approximation for $\ce{^{7}Li}$ in 
calculating $\mathcal{C}_{\ce{^{6}Li}}^{(p)}$ does not artificially exclude any region of
parameter space for such ensembles once the combined constraints from all relevant 
nuclear species are taken into account.   

Motivated by these considerations, in what follows we shall therefore adopt the linear 
approximation in which $Y_b \approx Y_b^{\mathrm{init}}$ in calculating the collision
terms $\mathcal{C}_a^{(p)}$ and $\mathcal{C}_a^{(s)}$ for any nuclear species 
$N_a \neq N_b$ for which $N_b$ serves as a source.  As we have seen, this approximation
is valid for all species except for $\ce{^{7}Li}$, which serves as a source for primary  
$\ce{^{6}Li}$ production.  Moreover, adopting this approximation for $\ce{^{7}Li}$
in calculating $\mathcal{C}_{\ce{^{6}Li}}^{(p)}$ yields a conservative bound on
electromagnetic injection from decaying particle ensembles.  

As discussed above, the advantage of working within the linear approximation is that 
the Boltzmann equations for all relevant $N_a$ effectively decouple and may be solved 
individually in order to yield analytic approximations for $\delta Y_a$. 
In the simplest case, in which the collision terms $C_a^{(p)}$ and $C_a^{(s)}$ in
the Boltzmann equation for $N_a$ include only source terms and not sinks,
the right side of Eq.~(\ref{eq:DensityInstforY}) is independent of $Y_a$ 
itself.  Thus, within the linear approximation, this equation may be integrated
directly, yielding
\begin{equation}
  \delta Y_a(t) ~\approx~ \int_{t_0}^t \frac{dY_a}{dt'} dt'~,
  \label{eq:IntegratedYdtGeneral}
\end{equation}  
where $t_0$ represents the time at the conclusion of the BBN epoch beyond which 
the initial abundance $Y_a(t_0) = Y_a^{\mathrm{init}}$ generated by standard
primordial nucleosynthesis remains essentially fixed in the absence of any
subsequent injection. 
Moreover, even in cases in which $C_a^{(p)}$ and $C_a^{(s)}$ include both source and 
sink terms, we may still evaluate $\delta Y_a$ in this way, provided that observational 
constraints restrict $\delta Y_a$ to the region $|\delta Y_a| \ll Y_a$ and therefore allow us to ignore feedback
effects and approximate $Y_a \approx Y_a^{\mathrm{init}}$ as a constant on  
the right side of Eq.~(\ref{eq:DensityInstforY}).

Since the Boltzmann equation for $\ce{^{6}Li}$ contains no non-negligible sink terms,
and since observational constraints require that $|\delta Y_a| \ll Y_a$ for both 
$\ce{^{4}He}$ and $\ce{D}$, it follows that $\delta Y_a$ is well approximated by 
Eq.~(\ref{eq:IntegratedYdtGeneral}) for these species.  Indeed, the only relevant nucleus 
which does not satisfy these criteria for direct integration is $\ce{^{7}Li}$.
Nevertheless, since $\ce{^{7}Li}$ is destroyed by a number of primary photodisintegration 
processes but not produced in any significant amount, the Boltzmann equation for 
this nucleus takes the particularly simple form specified in Eq.~(\ref{eq:DepletionRateYa}).  
This first-order differential equation can easily be solved for $Y_a$, 
yielding an expression for the comoving number density
at any time $t \geq t_0$:
\begin{equation}
  Y_a(t) ~=~ Y_a^{\mathrm{init}}
    \exp \left[-\int_{t_0}^{t}\!\! \Gamma_a(t') dt' \right]~.
\end{equation}
When this relation is expressed in terms of $\delta Y_a$ rather than $Y_a$, we find that
it may be recast in the more revealing form
\begin{eqnarray}
  Y_a^{\mathrm{init}} \ln \left[1+ \frac{\delta Y_a(t)}{Y_a^{\mathrm{init}}}\right] &=& 
    - Y_a^{\mathrm{init}}\! \int_{t_0}^{t}\!\! \Gamma_a(t') dt' ~.~~~
  \label{eq:deltaYaDecouplingRegime}
\end{eqnarray}
In situations in which the linearity criterion $|\delta Y_a(t)| \ll Y_a^{\mathrm{init}}$ 
is satisfied for the nucleus $N_a$ {\it itself}\/ at all times $t \geq t_0$,  
Taylor expansion of the left side of this equation yields
\begin{eqnarray}
  \delta Y_a(t) & \approx &  -Y_a^{\mathrm{init}} \int_{t_0}^{t}\!\! \Gamma_a(t') dt' ~,
  \label{eq:deltaYaDecouplingRegimeTaylorExp}
\end{eqnarray}
which is also the result obtained by direct integration of the Boltzmann equation for
$N_a$ in the approximation that $Y_a \approx Y_a^{\mathrm{init}}$.  

Comparing Eqs.~(\ref{eq:deltaYaDecouplingRegime}) 
and~(\ref{eq:deltaYaDecouplingRegimeTaylorExp}), we see that if we were to 
neglect feedback and take $Y_a \approx Y_a^{\mathrm{init}}$ when evaluating 
$\delta Y_a$ for a species for which this approximation is not particularly good,
the na\"\i ve result that we would obtain for $\delta Y_a$ would in fact correspond to the
value of the quantity $Y_a^{\mathrm{init}} \ln [1+ \delta Y_a(t)/Y_a^{\mathrm{init}}]$.
Thus, given that a dictionary exists between the value 
of $\delta Y_a$ obtained from Eq.~(\ref{eq:deltaYaDecouplingRegime}) and the value 
obtained from Eq.~(\ref{eq:deltaYaDecouplingRegimeTaylorExp}), 
for simplicity in what follows we shall 
derive our analytic approximation for $\delta Y_a$ using
Eq.~(\ref{eq:deltaYaDecouplingRegimeTaylorExp}) and simply note that 
the appropriate substitution should be made for the case of $\ce{^{7}Li}$.
That said, we also find that the constraint on $\Omega_\chi$ that we would derive 
from Eq.~(\ref{eq:deltaYaDecouplingRegimeTaylorExp}) 
in single-particle injection scenarios from the observational bound on 
$Y_{\ce{^{7}Li}}$ differs from the 
constraint that we would derive from the more accurate approximation in 
Eq.~(\ref{eq:deltaYaDecouplingRegime}) by only $\mathcal{O}(10\%)$.  Thus, results
obtained by approximating $\delta Y_{\ce{^{7}Li}}$ by the expression in
Eq.~(\ref{eq:deltaYaDecouplingRegimeTaylorExp}) are nevertheless fairly reliable 
in such scenarios --- and indeed can be expected to be reasonably reliable in 
scenarios involving decaying ensembles as well.

%%%%%%%%%%%%%%%%%%%%%%%%%%%%%%%%%%%%%%%%%%%%%%%%%%%%%%%%%%%%%%%%%%%%%%%%%%%%%%%%%%%%%%
\FloatBarrier
\subsection{Analytic Approximation:  Contribution from Primary Processes\label{sec:LinearizationForPrimProd}}
%%%%%%%%%%%%%%%%%%%%%%%%%%%%%%%%%%%%%%%%%%%%%%%%%%%%%%%%%%%%%%%%%%%%%%%%%%%%%%%%%%%%%%

Having discussed how the Boltzmann equations for the relevant $N_a$ effectively 
decouple in the linear regime, we now proceed to derive a set of approximate 
analytic expressions for $\delta Y_a$ from these decoupled equations. 
We begin by considering the contribution to $\delta Y_a$ that arises from primary 
photoproduction or photodisintegration processes.  The contribution from secondary 
processes, which is relevant only for $\ce{^{6}Li}$, will be discussed in 
Sect.~\ref{sec:LinearizationForSecProd}. 
 
Our ultimate goal is to derive an approximate analytic expression for the
total contribution to $\delta Y_a$ due the injection of photons from an entire 
ensemble of decaying states.  However, our first step in this derivation shall be 
to consider the simpler case in which the injection is due to the decay of a single 
unstable particle species $\chi$ with a lifetime $\tau_\chi$.  We shall work within the 
uniform-decay approximation, in which the non-thermal photon spectrum takes the 
particularly simple form in Eq.~(\ref{eq:NonThermalPhotSpecApprox}).  In this
approximation, the lower limit of integration in Eq.~(\ref{eq:IntegratedYdtGeneral}) 
may be replaced by $\tau_\chi$, while the upper limit can be taken to be any time well 
after photons at energies above the thresholds $E_{b}^{(ac)}$ and $E_{a}^{(bc)}$ for all 
relevant photoproduction and photodisintegration processes have thermalized.  Thus,
we may approximate the change in the comoving number density of each relevant nuclear
species as
\begin{equation}
  \delta Y_a ~\approx~ \int_{\tau_\chi}^\infty \frac{dY_a}{dt} dt~.
  \label{eq:IntegratedYdt}
\end{equation}

In evaluating each $\delta Y_a$, we may also take advantage 
of the fact that the rates for the relevant reactions discussed in 
Sect.~\ref{sec:ConstraintsPrimord} turn out to be such that the first (source) and 
second (sink) terms in Eq.~(\ref{eq:CollisionsPrimary}) are never simultaneously 
large for any relevant nuclear species.  Indeed, the closest thing to an exception 
occurs during a very small time interval within which the source and sink terms for  
$\ce{D}$ are both non-negligible.  Thus, depending on the value of $\tau_\chi$ and its
relationship to the timescales associated with these reactions, we may
to a very good approximation treat the effect of injection from a single decaying 
particle as either producing or destroying $N_a$.

With these approximations, the integral over $t$ in 
Eq.~(\ref{eq:IntegratedYdt}) may be evaluated in closed form.  In particular,
when the source term in Eq.~(\ref{eq:CollisionsPrimary}) dominates, we find that 
$\delta Y_a$ is given by
\begin{eqnarray}
  \delta Y_a &\approx&  \sum_{b,c}
    \frac{Y_b \epsilon_\chi \rho_\chi(\tau_\chi)}{n_B(\tau_\chi)} \nonumber \\ & &~~\times 
    \int_{E_{b}^{(ac)}}^{E_C} dE K(E,\tau_\chi) S_b^{(ac)}(E)~,~~~
\label{eq:AbundanceChangeSource}
\end{eqnarray}
where we have defined 
\begin{eqnarray}
  S_b^{(ac)}(E) ~\equiv~ \frac{\sigma_{b}^{(ac)}(E)}{\sigma_{\mathrm{th}}(E)}~.
  \label{eq:DefofScbd}
\end{eqnarray}
By contrast, when the sink term dominates, we find that
$\delta Y_a$  is given by
\begin{eqnarray}
  \delta Y_a &\approx& -\sum_{b,c}   
    \frac{Y_a \epsilon_\chi \rho_\chi(\tau_\chi)}{n_B(\tau_\chi)} \nonumber \\ & &~~\times
    \int_{E_{a}^{(bc)}}^{E_C} dE K(E,\tau_\chi) S_a^{(bc)}(E)~.
\label{eq:AbundanceChangeSink}
\end{eqnarray}

While the expressions for $\delta Y_a$ in Eqs.~(\ref{eq:AbundanceChangeSink}) 
and~(\ref{eq:AbundanceChangeSource}) pertain to the case of a single unstable 
particle within the uniform-decay approximation, it is straightforward to generalize 
these results to more complicated scenarios.  Indeed, within the linear approximation, 
the total change $\delta Y_a$ in $Y_a$ which results from 
multiple instantaneous injections over an extended time interval is well approximated 
by the sum of the individual contributions from these injections.
In the limit in which this set of discrete injections becomes a continuous 
spectrum, this sum becomes an integral over the injection time $\tinj$.  Thus, in the
continuum limit, $\delta Y_a$ is well approximated by 
\begin{equation}
  \delta Y_a ~\approx~ \int_0^\infty \frac{dY_a}{d\tinj} d\tinj~,
  \label{eq:DeltaYaContinuumInjectionLinear}
\end{equation}
where $dY_a/d\tinj$ is the differential change in $Y_a$ due to an infinitesimal 
injection of energy in the form of photons at time $\tinj$.  

The approximation in Eq.~(\ref{eq:DeltaYaContinuumInjectionLinear}) 
allows us to account for the full exponential time-dependence of the 
electromagnetic injection due to particle decay in calculating $\delta Y_a$ for any 
given nucleus.  By extension, Eq.~(\ref{eq:DeltaYaContinuumInjectionLinear}) gives
us the ability to compare the results for $\delta Y_a$ obtained both with and 
without invoking the uniform-decay approximation, thereby providing us insight 
into how reliably $\delta Y_a$ can be computed with this approximation.

As an example, let us consider the effect of a single decaying particle $\chi$ with 
lifetime $\tau_\chi$ on the comoving number density of $\ce{{^4}He}$.  Within the
uniform-decay approximation, $\delta Y_{\ce{^{4}He}}$ is given by 
Eq.~(\ref{eq:AbundanceChangeSink}) because the sink term in 
Eq.~(\ref{eq:CollisionsPrimary}) dominates.  By contrast, when the full 
exponential nature of $\chi$ decay is taken into account, the corresponding 
result is 
\begin{eqnarray}
  \delta Y_a
    &\approx& - \sum_{b,c}\int_0^\infty d\tinj 
    \Bigg[ \frac{Y_a \epsilon_\chi}{n_B(\tinj)} \frac{d\rho_\chi(\tinj)}{d\tinj}
     \nonumber \\ 
    & &    ~~~~~ \times \int_{E_{a}^{(bc)}}^{E_C}dE K(E,\tinj)
    S_c^{(bd)}(E) \Bigg]~,~~~~~~
\label{eq:AbundanceChangeContinuum}
\end{eqnarray}
where $d\rho_\chi(\tinj)/d\tinj$ denotes the rate of change in the energy density of 
$\chi$ per unit time $\tinj$.  Prior to $\tMRE$, the energy density of an unstable particle with an 
extrapolated abundance $\Omega_\chi$ may be written
\begin{equation}
  \rho_\chi(t) ~=~ \rhocrit(\tnow)
    \left(\frac{\tnow}{\tMRE}\right)^2\left(\frac{\tMRE}{t}\right)^{3/2}
    \Omega_\chi e^{-t/\tau_\chi}~,
  \label{eq:rhoForDecay}
\end{equation}
where $\rhocrit(\tnow)$ is the critical density of the universe at present time and
where $\tMRE$ is the time of matter-radiation equality.  The corresponding
rate of change in the energy density, properly evaluated in the comoving frame and
then transformed to the physical frame, is
\begin{eqnarray}
  \frac{d\rho_\chi(\tinj)}{d\tinj} & = & -\rhocrit(\tnow)
    \left(\frac{\tnow}{\tMRE}\right)^2\left(\frac{\tMRE}{\tinj}\right)^{3/2}
    \nonumber \\ & & \times\,
     \frac{\Omega_\chi}{\tau_\chi} 
    e^{-\tinj/\tau_\chi}~.
  \label{eq:drhodtinjForDecay}
\end{eqnarray}
The corresponding expressions for continuum injection in cases in which the 
source term in Eq.~(\ref{eq:CollisionsPrimary}) dominates are completely 
analogous and can be derived in a straightforward way.

\begin{figure}[t!]
\includegraphics*[width=0.49 \textwidth]{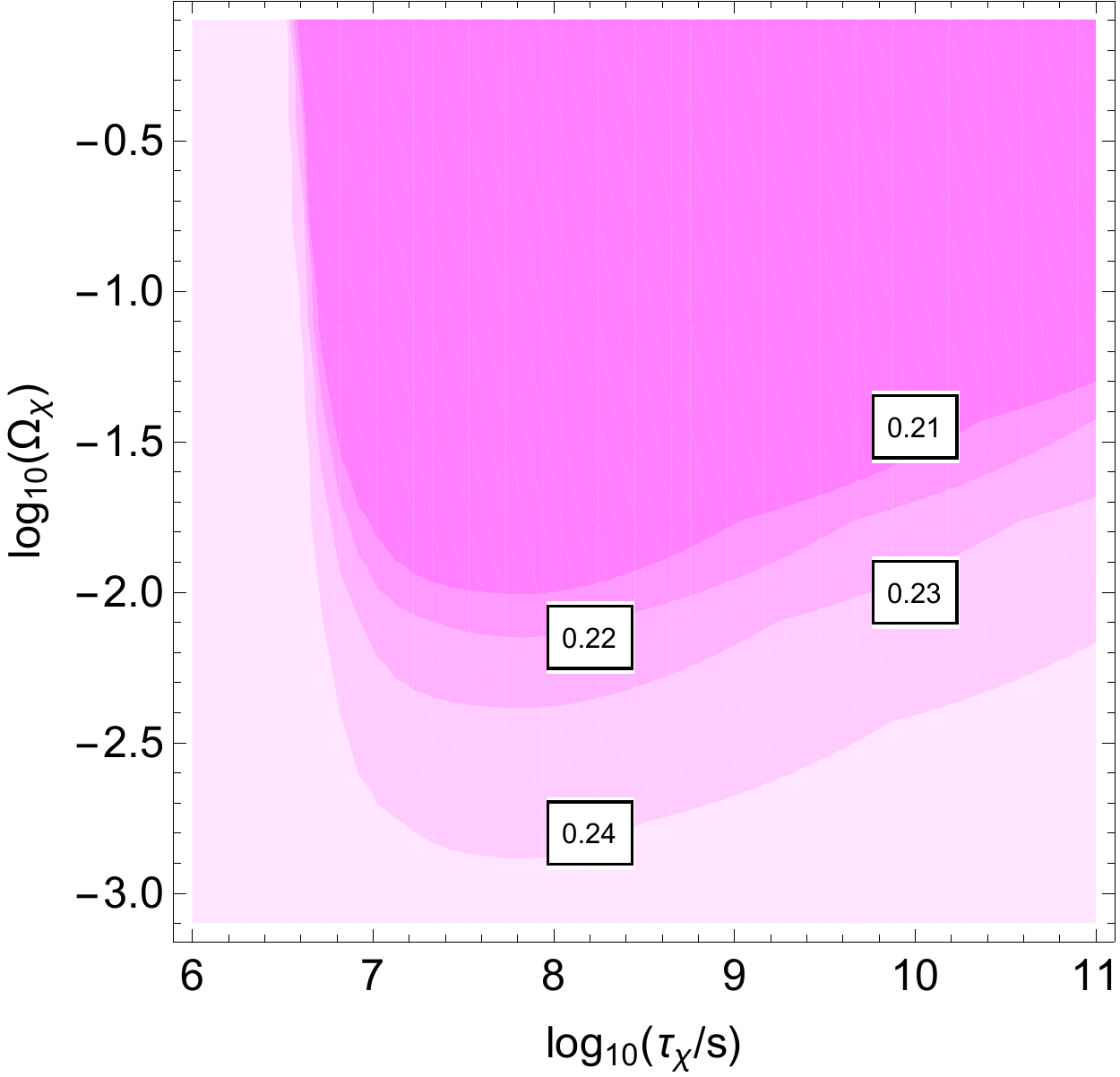}
\includegraphics*[width=0.49\textwidth]{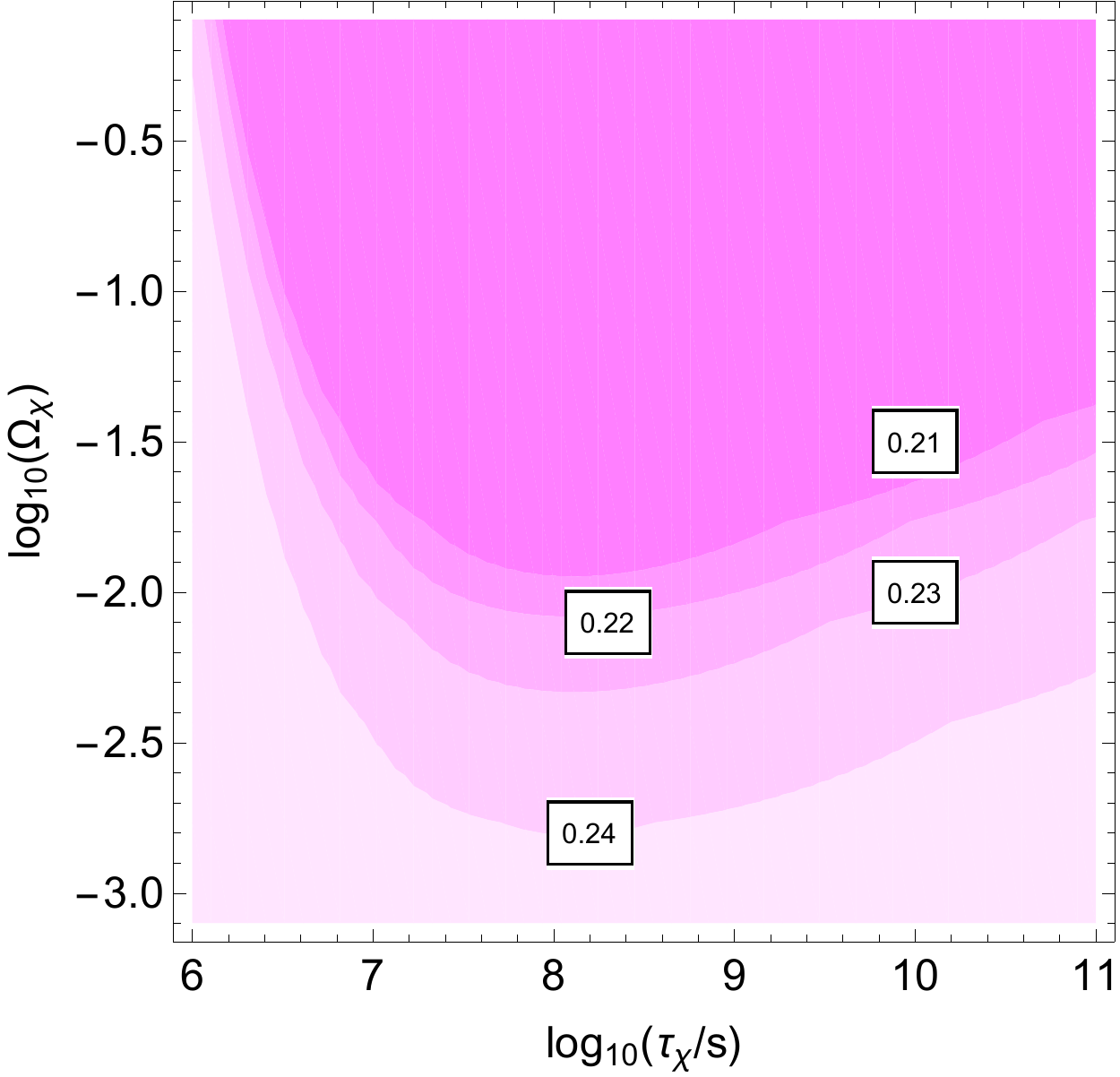}
\caption{Contours of the primordial $\ce{^{4}He}$ mass fraction $Y_p$ obtained in 
  the presence of a single decaying particle species $\chi$ with lifetime $\tau_\chi$
  and extrapolated abundance $\Omega_{\chi}$, plotted within the $(\Omega_{\chi},\tau_\chi)$ plane.  
  Note that $\Omega_\chi$ represents the abundance that $\chi$ would have had at 
  present time if it had not decayed.  The upper panel shows the results 
  obtained within the uniform-decay approximation, while the lower panel shows the corresponding 
  results obtained with the full exponential time-dependence of $\chi$ decay taken into 
  account.  Comparing the results shown in the two panels, we see that
  the uniform-decay approximation indeed provides a reasonable approximation 
  for $Y_p$.
\label{fig:HeFrac}}
\end{figure}

In Fig.~\ref{fig:HeFrac}, we compare the results obtained for $\delta Y_{\ce{^{4}He}}$
within the uniform-decay approximation to the results obtained with the full 
exponential time-dependence of $\chi$ decay taken into account.  In particular, 
within the $(\Omega_{\chi},\tau_\chi)$ plane,  
we display contours of the corresponding $\ce{^{4}He}$ mass fraction $Y_p$ obtained within the 
uniform-decay approximation (upper panel) and through the full exponential 
calculation (lower panel). 
For concreteness, 
in calculating these contours we have assumed an initial value $Y_p = 0.2449$ for the 
$\ce{^{4}He}$ mass fraction at the end of BBN, following Ref.~\cite{CyburtBBN2015}.

Comparing the two panels of Fig.~\ref{fig:HeFrac}, we see that the results for $Y_p$ 
obtained within the uniform-decay approximation are indeed very similar 
throughout most of the parameter space shown  
to the results obtained through the full  calculation. 
Nevertheless, we observe that discrepancies do arise.
For example, we note that the constraints obtained within the uniform-decay approximation 
are slightly stronger for particles with lifetimes 
within the range $10^7\mbox{~s} \lesssim \tau_\chi \lesssim 10^8\mbox{~s}$ and slightly weaker for 
$\tau_\chi$ above this range than the constraints obtained through the full calculation.
These discrepancies are ultimately due to the fact that the reprocessed photon 
spectrum in Eq.~(\ref{eq:spectrum}) depends on the temperature of the radiation bath.
Thus, there is a timescale $t_{\sigma,\mathrm{\max}}$ for which the reaction rate for 
a particular photoproduction or photodisintegration process is maximized for fixed 
$d\rho(\tinj)/d\tinj$.  Since all of the energy density $\rho_\chi$ initially associated 
with the decaying particle is injected precisely at $\tau_\chi$ within the uniform-decay 
approximation, this approximation yields slightly more stringent constraints than
those obtained through the full calculation when 
$\tau_\chi \sim t_{\sigma,\mathrm{\max}}$.  By the same token, the constraints 
obtained within the uniform-decay approximation are slightly less stringent than
those obtained through the full calculation when $\tau_\chi$ differs significantly from 
$t_{\sigma,\mathrm{\max}}$.   

We also observe that for lifetimes $\tau_\chi \lesssim 10^7$~s, the uniform-decay 
approximation likewise yields constraints that are weaker than those obtained through the full
exponential calculation.  The reason for this is that the upper energy cutoff $E_C$ in 
the reprocessed photon spectrum is proportional to $\tinj^{1/2}$.  For sufficiently 
early injection times, $E_C$ lies below the threshold energy $E_a^{(bc)}$ for the 
photodisintegration reactions which contribute to $\delta Y_{\ce{^{4}He}}$.  Injection 
at such early times therefore has essentially no impact on $Y_p$.  This implies that
within the uniform-decay approximation, a particle with a lifetime in this regime likewise has
no effect on $Y_p$.  By contrast, within the full exponential calculation,
injection occurs at a significant rate well after $\tau_\chi$, leading to a 
non-negligible change in $Y_p$ even when the lifetime of the decaying particle is 
short. 

In summary, the results shown in Fig.~\ref{fig:HeFrac} indicate that other than
in the regime where $\tau_\chi$ is sufficiently short that $E_C$ lies below the 
threshold energy for $\ce{^{4}He}$ destruction, the result for 
$\delta Y_{\ce{^{4}He}}$ obtained within the uniform-decay approximation is very
similar to the result obtained through the full exponential calculation for the same
$\tau_\chi$ and $\Omega_\chi$.  We find this to be the case for the other relevant 
nuclei as well.
Thus, having shown that the results for $\delta Y_a$ obtained within the uniform-decay 
approximation accord well with those obtained through the full exponential 
calculation, at least for sufficiently long $\tau_\chi$, we shall adopt this 
approximation in deriving our constraints on ensembles of electromagnetically-decaying 
particles.  As we shall see, the advantage of working within the uniform-decay approximation 
is that within this approximation it is possible to write down a simple analytic expression 
for $\delta Y_a$.  However, as we shall discuss in more detail below, we shall adopt an 
alternative strategy for approximating $\delta Y_a$ within the regime in which 
$\tau_\chi$ is short and the results obtained within the uniform-decay approximation do 
not agree with those obtained through the full exponential calculation.

In order to write down our analytic expressions for $\delta Y_a$, we shall make
one additional approximation: 
we shall treat the ratio of cross-sections $S_c^{(bd)}(E)$ as not varying significantly
as a function of energy between the threshold energy $E_a^{(bc)}$ and $E_C$.  
When this approximation holds, we may treat this ratio as a constant and pull it 
outside the integral over photon energies.  In order to justify this approximation, 
we begin by noting that the cross-sections $\sigma_a^{(bc)}(E)$ for primary 
processes typically peak at a value slightly above $E_a^{(bc)}$ but fall 
precipitously with $E$ beyond that point (see, \eg, Ref.~\cite{CyburtEllisUpdated}).  
In the vicinity of the peak, however, the variation of $\sigma_a^{(bc)}(E)$ is 
relatively gentle.  Since the thermalization cross-section 
$\sigma_{\mathrm{th}}(E)$ for primary photons varies much less rapidly 
across the relevant region of $E$, the functional dependence of 
$S_a^{(bc)}(E)$ on $E$ is principally determined by the behavior
of $\sigma_a^{(bc)}(E)$.  Moreover, because $K(E,\tau_\chi)$ falls rapidly 
with $E$, the dominant contribution to the energy integral in 
Eq.~(\ref{eq:AbundanceChangeSink}) arises from photons just above threshold 
even within the approximation that $S_a^{(bc)}(E)$ is constant.  Thus, to a good 
approximation, we may replace $S_c^{(bd)}(E)$ by a constant on the order
of its peak value and take this quantity outside the energy integral.

Within this approximation, it is now possible to analytically evaluate the integral in 
Eq.~(\ref{eq:AbundanceChangeSource}).
The form of the result depends on the relationship between the threshold energy 
$E_a^{(bc)}$ for the scattering process and the energy scales $E_C(\tinj)$ and
$E_X(\tinj)$ which determine the shape of the photon spectrum at time $\tinj=\tau_\chi$.
There are three cases of interest:  the case in which $E_a^{(bc)} > E_C (\tau_\chi)$,
the case in which $E_C (\tau_\chi) > E_a^{(bc)} > E_X (\tau_\chi)$, and the 
case in which $E_X (\tau_\chi) > E_a^{(bc)}$.  Moreover, the relations in Eq.~(\ref{eq:ECandEX}) imply
that each of these cases corresponds to a specific range of $\tau_\chi$.   
In particular, the respective lifetime regimes are $\tau_\chi < t_{Ca}^{(bc)}$, 
$t_{Ca}^{(bc)} < \tau_\chi < t_{Xa}^{(bc)}$, and $t_{Xa}^{(bc)} < \tau_\chi$, 
where we have defined   
\begin{eqnarray}
  t_{Ca}^{(bc)} &\approx & (10^8\, {\s})\times 
    \left[\frac{E_a^{(bc)}}{103\, \mev} \right]^2 \nonumber\\
  t_{Xa}^{(bc)}  &\approx & (10^8\, {\s})\times 
    \left[\frac{E_a^{(bc)}}{28\,\mev} \right]^2~.
  \label{eqn:uniformTimes}
\end{eqnarray}

Within the $\tau_\chi < t_{Ca}^{(bc)}$ regime, none of the photons 
produced by $\chi$ decay exceed the threshold for the photodisintegration process.  
The contribution to $\delta Y_a$ within the uniform-decay approximation is therefore 
formally zero.  As discussed above, this is the regime in which the uniform-decay
approximation fails to reproduce the results obtained through the full exponential 
calculation.  Thus, in order to derive a meaningful bound on decaying particles 
with lifetimes in this regime, we instead model the injection of photons
using the full continuum expression in Eq.~(\ref{eq:AbundanceChangeContinuum}) 
with $d\rho(\tinj)/d\tinj$ given by Eq.~(\ref{eq:drhodtinjForDecay}).  However, 
in order to arrive at a simple analytic expression for $\delta Y_a$, we 
include only the contribution from injection times in the range 
$t_{Ca}^{(bc)} < \tinj < t_{Ca}^{(bc)} + \tau_\chi$ and drop terms in 
$\tau_\chi/t_{Ca}^{(bc)}$ beyond leading order in the resulting expression.  With 
these approximations, in this regime we find 
\begin{equation}
  \delta Y_a ~=~ \sum_{b,c} A_a^{(bc)}\Omega_\chi 
    \left(\frac{\tau_\chi}{t_{Ca}^{(bc)}} \right) 
    e^{-t_{Ca}^{(bc)} / \tau_\chi}~,
  \label{eq:UniformBBNCaseI}
\end{equation}
where the proportionality constant $A_a^{(bc)}$ for each contributing
reaction is independent of the properties of the decaying particle.  This treatment
ensures that we obtain a more reliable estimate for the contribution to 
$\delta Y_a$ from particles with $\tau_\chi$ in this regime. 

Within the remaining two lifetime regimes, the contribution to 
$\delta Y_a$ within the uniform-decay approximation is non-vanishing.  Thus, within these
regimes, we obtain our approximation for $\delta Y_a$ by integrating 
Eq.~(\ref{eq:AbundanceChangeSource}), as discussed above.  For 
$t_{Ca}^{(bc)} < \tau_\chi < t_{Xa}^{(bc)}$ 
we find 
\begin{equation}
  \delta Y_a ~=~ \sum_{bc}B_a^{(bc)}\Omega_\chi 
    \left[1 - \beta \left(\frac{\tau_\chi}{t_{Xa}^{(bc)}}\right)^{\! -1/2} \right]~,
  \label{eq:UniformBBNCaseII}
\end{equation}
where $B_a^{(bc)}$ is a proportionality constant and where 
$\beta \equiv E_X / E_C \approx 0.27$ is the $\tau_\chi$-independent ratio of 
the energy scales in Eq.~(\ref{eq:ECandEX}).  Likewise, for $t_{Xa}^{(bc)} < \tau_\chi$ we
find 
\begin{eqnarray}
  \delta Y_a &=& \sum_{bc}B_a^{(bc)} \Omega_\chi 
    \left( \frac{\tau_\chi}{t_{Xa}^{(bc)}}\right)^{\! -1/4} \nonumber \\
    & & \times
    \left[2 - (1 + \beta) \left(\frac{\tau_\chi}{t_{Xa}^{(bc)}}\right)^{\! -1/4}\right]~.
  \label{eq:UniformBBNCaseIII}
\end{eqnarray}
We emphasize that the proportionality constant $B_a^{(bc)}$ for a given process in
Eq.~(\ref{eq:UniformBBNCaseIII}) is the same as the corresponding proportionality 
constant in Eq.~(\ref{eq:UniformBBNCaseII}).  However, in general $B_a^{(bc)}$ 
differs from the corresponding proportionality constant $A_a^{(bc)}$ appearing 
in Eq.~(\ref{eq:UniformBBNCaseI}).  

Strictly speaking, Eq.~(\ref{eq:UniformBBNCaseIII}) does not hold for
arbitrarily large $\tau_\chi$, since photons produced by extremely late decays 
are not efficiently reprocessed by Class-I processes into the spectrum in 
Eq.~(\ref{eq:spectrum}).  
In order to account for this in what follows, we shall 
consider each term in the sum in Eq.~(\ref{eq:UniformBBNCaseIII}) to be valid only 
for injection times $\tinj < t_{fa}^{(bc)}$, where $t_{fa}^{(bc)}$ is a characteristic 
cutoff timescale associated with the reaction.  Photons injected after this cutoff 
timescale are assumed to have no effect on $Y_a$.  For most reactions, it is 
appropriate to take $t_{fa}^{(bc)} \approx 10^{12}$~s, as this is the timescale
beyond which certain crucial Class-I processes effectively begin to shut off and 
the reprocessed photon spectrum is no longer reliably described by 
Eq.~(\ref{eq:spectrum}).
  
\smallskip
%%%%%%%%%%%%%%%%%%%%%%%%%%%%%%%%%%%%%%%%%%%%%%%%%%%%%%%%%%%%%%%%%%%%%%%%%%%%%%%%%%%%%%
\FloatBarrier
\subsection{Analytic Approximation:  Contribution from Secondary Processes\label{sec:LinearizationForSecProd}}
%%%%%%%%%%%%%%%%%%%%%%%%%%%%%%%%%%%%%%%%%%%%%%%%%%%%%%%%%%%%%%%%%%%%%%%%%%%%%%%%%%%%%%  
  
The approximate analytic expressions for $\delta Y_a$ which we have derived in
Sect.~\ref{sec:LinearizationForPrimProd} are applicable to all of the primary
photoproduction or photodisintegration processes relevant for constraining
electromagnetic injection from unstable-particle decays after BBN.~  
However, since secondary production can contribute non-negligibly to the production
of $\ce{^{6}Li}$, we must derive analogous expressions for $\delta Y_a$ in 
the case of secondary production as well.  Moreover, since secondary production is 
fundamentally different from primary production in terms of particle kinematics,
there is no reason to expect that these expressions should have the same functional 
dependence on $\tau_\chi$ as that exhibited by the expressions 
in Eqs.~(\ref{eq:UniformBBNCaseI})--(\ref{eq:UniformBBNCaseIII}).  Indeed, as we shall see, they do not.

We begin this undertaking by observing that within the uniform-decay approximation, the 
contribution to $\delta Y_a$ from secondary production is given by 
\begin{widetext}
\begin{equation}
  \delta Y_a
    ~\approx~\!\! \sum_{b,c,d,f}\int_{\tau_\chi}^\infty\!\!\!\!\! dt 
    \Bigg\{Y_f \epsilon_\chi \rho_\chi(\tau_\chi) 
    \!\int_{E_{bf}^{(aX)}}^{\mathcal{E}(E_C)}\!\!\! dE_b \Bigg[
    \frac{n_c(t) \sigma_{bf}^{(aX)}(E)|v(E_b)|}{b(E_b,t)} 
    \!\int_{E_{\mathrm{min}}(E_b)}^{E_C} 
    \!\!\!\!\!\!\!\!\!\! dE_\gamma K(E_\gamma,\tau_\chi)
    \sigma_c^{(bd)}(E_\gamma) 
    e^{-\frac{(t-\tau_\chi)}{\delta t_{\mathrm{th}}(E_\gamma,\tau_\chi)}}
     \Bigg]\Bigg\}~.
   \label{eq:dYFullFromSecondaryProd}
\end{equation}
\end{widetext}
We may simplify this expression by noting that the time-dependence of the 
energy-loss rate $b(E_b,t)$ of excited nuclei due to Coulomb scattering is primarily 
due to the dilution of the number density of electrons $n_e(t)$.  Since $n_e(t)$ 
scales with $t$ in the same manner as $n_c(t)$, the quantity
\begin{equation}
  \Upsilon_{bcf}^{(aX)}(E_b) ~\equiv~ 
    \frac{n_c(t)\sigma_{bf}^{(aX)}(E)|v(E_b)|}{b(E_b,t)}
\end{equation} 
is, to a good approximation, a comoving quantity, and hence independent of
$t$.  Thus, $\Upsilon_{bcf}^{(aX)}(E)$ may be pulled outside the time integral
in Eq.~(\ref{eq:dYFullFromSecondaryProd}).  Within this approximation,
the contribution to $\delta Y_a$ reduces to
\begin{eqnarray}
  \delta Y_a
    &\approx&\!\!\! \sum_{b,c,d,f}
    \!\! \frac{Y_f \epsilon_\chi \rho_\chi(\tau_\chi)}{n_B(\tau_\chi)} 
    \!\int_{E_{bc}^{(aX)}}^{\mathcal{E}(E_C)}\!\!\! dE_b \Bigg[
    \Upsilon_{bcf}^{(aX)}(E_b) 
    \nonumber \\ & & \times
    \int_{E_{\mathrm{min}}(E_b)}^{E_C} 
    \!\!\!\!\!\!\!\!\!\! dE_\gamma K(E_\gamma,\tau_\chi)
    S_c^{(bd)}(E_\gamma)
     \!\Bigg]~.
   \label{eq:dYSimpFromSecondaryProd}
\end{eqnarray}

In order to proceed further,
we must first assess 
the dependence of the quantities $S_c^{(ab)}(E_\gamma)$ and 
$\Upsilon_{bcf}^{(aX)}(E_b)$ on the respective energy scales $E_\gamma$ and 
$E_b$.  However, we find that $\Upsilon_{bcf}^{(aX)}(E_b)$ varies reasonably
slowly over the relevant range of $E_b$~\cite{JedamzikLi6,CyburtEllisUpdated}.  
Thus, to a good approximation, this quantity may also be pulled outside the integral 
over $E_b$.  

By contrast, we find that $S_c^{(ab)}(E_\gamma)$ cannot reliably be approximated as a 
constant over the range of $E_\gamma$ relevant for secondary production.  This 
deserves further comment --- especially because we have approximated 
$S_c^{(ab)}(E_\gamma)$ as a constant in deriving the expressions in 
Eqs.~(\ref{eq:UniformBBNCaseI})--(\ref{eq:UniformBBNCaseIII}) for primary
production.  As we shall now make clear, 
there are important differences between the kinematics of primary 
and secondary production which enable us to approximate $S_c^{(ab)}(E_\gamma)$ as
independent of $E_\gamma$ in the former case but not in the latter.  

For primary production, as discussed in Sect.~\ref{sec:LinearizationForPrimProd}, 
the rapid decrease of $K(E_\gamma,\tau_\chi)$ with $E_\gamma$ suppresses the contribution
to $\delta Y_a$ from photons with $E_\gamma \gg E_b^{(ac)}$.  The dominant
contribution to $\delta Y_a$ therefore comes from a narrow region of the spectrum just
above threshold within which $S_b^{(ac)}(E_\gamma)$ varies reasonably slowly,
while photons with energies well above $E_b^{(ac)}$ have little collective impact on 
$\delta Y_a$.  Thus, in approximating the overall contribution to $\delta Y_a$ from
primary production, it is reasonable to treat $S_b^{(ac)}(E_\gamma)$ as a constant. 

By contrast, for secondary production --- or at least for the secondary production of 
$\ce{^{6}Li}$, the one nuclear species in our analysis for which secondary production 
can have a significant impact on $\delta Y_a$ --- photons with $E_\gamma$ well above 
the threshold energy $E_c^{(bd)}$ for any relevant primary process play a more 
important role.  One reason for this is that the energy threshold $E_{bc}^{(aX)}$ for each 
secondary processes which contributes meaningfully to $\ce{^{6}Li}$ production corresponds 
to a primary-photon energy $\mathcal{E}^{-1}[E_{bf}^{(aX)}]$ well above the
associated primary-process threshold $E_c^{(bd)}$.  Indeed, the kinetic-energy
thresholds for $\ce{^{4}He} + \ce{^{3}He} \rightarrow \ce{^{6}Li} + p$ and
$\ce{^{4}He} + \ce{T} \rightarrow \ce{^{6}Li} + n$ given in Table~\ref{tab:BBNprocs}
are quite similar and both correspond to photon energies of roughly 
$\mathcal{E}^{-1}[E_{bf}^{(aX)}] \sim {\cal O}(50~{\rm MeV})$  --- energies well above the peak in
$S_c^{(bd)}(E_\gamma)$.  Thus, it is really the energy threshold for the secondary process 
which sets the minimum value of $E_\gamma$ relevant for the secondary production 
of $\ce{^{6}Li}$.  Since $S_c^{(bd)}(E_\gamma)$ varies more rapidly with $E_\gamma$ 
at these energies than it does around its peak value, it follows that this variation 
cannot be neglected in determining the overall dependence of $\delta Y_a$ on $\tau_\chi$ 
in this case. 
  
There is, however, another reason why the variation of $S_c^{(bd)}(E_\gamma)$ with 
$E_\gamma$ cannot be neglected in the case of secondary production --- a deeper reason which is
rooted more in fundamental differences between primary and secondary production than
in the values of the particular energy thresholds associated with processes 
pertaining to $\ce{^{6}Li}$.  The overall contribution to $\delta Y_a$ from primary production 
in Eq.~(\ref{eq:AbundanceChangeSource}) involves a single integral over $E_\gamma$.
Thus, for primary production, the fall-off in $K(E_\gamma,\tau_\chi)$ itself with 
$E_\gamma$ is sufficient to suppress the partial contribution to $\delta Y_a$ from 
photons with $E_\gamma \gg E_c^{(bd)}$.  By contrast, the overall contribution to 
$\delta Y_a$ from secondary production involves integration not only over $E_\gamma$ 
but also over $E_b$.  In this case, the fall-off in $K(E_\gamma,\tau_\chi)$ with 
$E_\gamma$ is not sufficient to suppress the partial contribution to $\delta Y_a$ from 
photons with energies well above threshold.  Thus, for any secondary production
process, an accurate estimate for $\delta Y_a$ can only be obtained when the
variation of $S_c^{(bd)}(E_\gamma)$ with $E_\gamma$ --- a variation which can be quite significant
at such energies --- is taken into account.

We must therefore explicitly incorporate the functional dependence of 
$S_c^{(bd)}(E_\gamma)$ on $E_\gamma$ into our calculation
of $\delta Y_a$.  We recall that 
$S_c^{(bd)}(E_\gamma)$, as we have defined it in Eq.~(\ref{eq:DefofScbd}), 
represents the ratio of the cross-section $\sigma_c^{(bd)}(E_\gamma)$ for the 
primary process $N_c + \gamma \rightarrow N_b + N_d$
which produces the population of excited nuclei to the cross-section 
$\sigma_{\mathrm{th}}(E_\gamma)$ for the Class-III processes which serve to thermalize the 
primary-photon spectrum.  The cross-sections for the two primary processes
$\ce{^{4}He} + \gamma  \rightarrow \ce{^{3}He} + n$ and
$\ce{^{4}He} + \gamma  \rightarrow \ce{T} + p$ relevant for 
secondary $\ce{^{6}Li}$ production, expressed as a functions
of the photon energy $E_\gamma$, both take the form~\cite{CyburtEllisUpdated}
\begin{equation}
  \sigma_c^{(bd)}(E_\gamma) ~\approx ~ 
      \sigma_{c,0}^{(bd)}     
      \left[\frac{E_\gamma}{E_c^{(bd)}} - 1\right] 
      \left[\frac{E_\gamma}{E_c^{(bd)}}\right]^{-9/2}~,
\end{equation}    
where $\sigma_{c,0}^{(bd)} \approx 20.6$~mb and $\sigma_{c,0}^{(bd)} \approx 19.8$~mb
for these two processes, respectively.  By contrast, $\sigma_{\mathrm{th}}(E_\gamma)$
includes two individual contributions.  The first contribution is due to $e^+ e^-$ 
pair-production off nuclei via Bethe-Heitler processes of the form 
$\gamma + N_a \rightarrow X + e^+ + e^-$, where $N_a$ denotes a 
background nucleus and $X$ denotes some hadronic final state.  The cross-section for 
this process is
\begin{equation}
  \sigma_{\mathrm{BH}}(E_\gamma) ~=~ \frac{3\alpha}{8\pi} \sigma_T 
     \left[\frac{28}{9}\ln\left(\frac{2E_\gamma}{m_e}\right) 
       - \frac{218}{27}\right]~,
\end{equation}  
where $\alpha$ is the fine-structure constant, where $m_e$ is the electron mass, and 
where $\sigma_T \approx 661$~mb is the Thomson cross-section. 
The second contribution to $\sigma_{\mathrm{th}}(E_\gamma)$ is due to Compton
scattering, for which the cross-section is 
\begin{eqnarray}
  \sigma_{\mathrm{CS}}(E_\gamma) &=& \frac{3 m_e}{8E_\gamma} \sigma_T 
    \Bigg\{\left[1 - \frac{2m_e}{E_\gamma} - \frac{4m_e^2}{E_\gamma^2}
      \ln\left(1 + \frac{2E_\gamma}{m_e}\right)\right] \nonumber \\ & & 
      + \frac{4m_e}{E_\gamma} 
      + \frac{1}{2}\left[1 - \left(1 + \frac{2E_\gamma}{m_e}\right)^{-2}\right]
      \Bigg\}~.
\end{eqnarray}

\begin{figure}[t!]
\includegraphics[width=0.49 \textwidth]{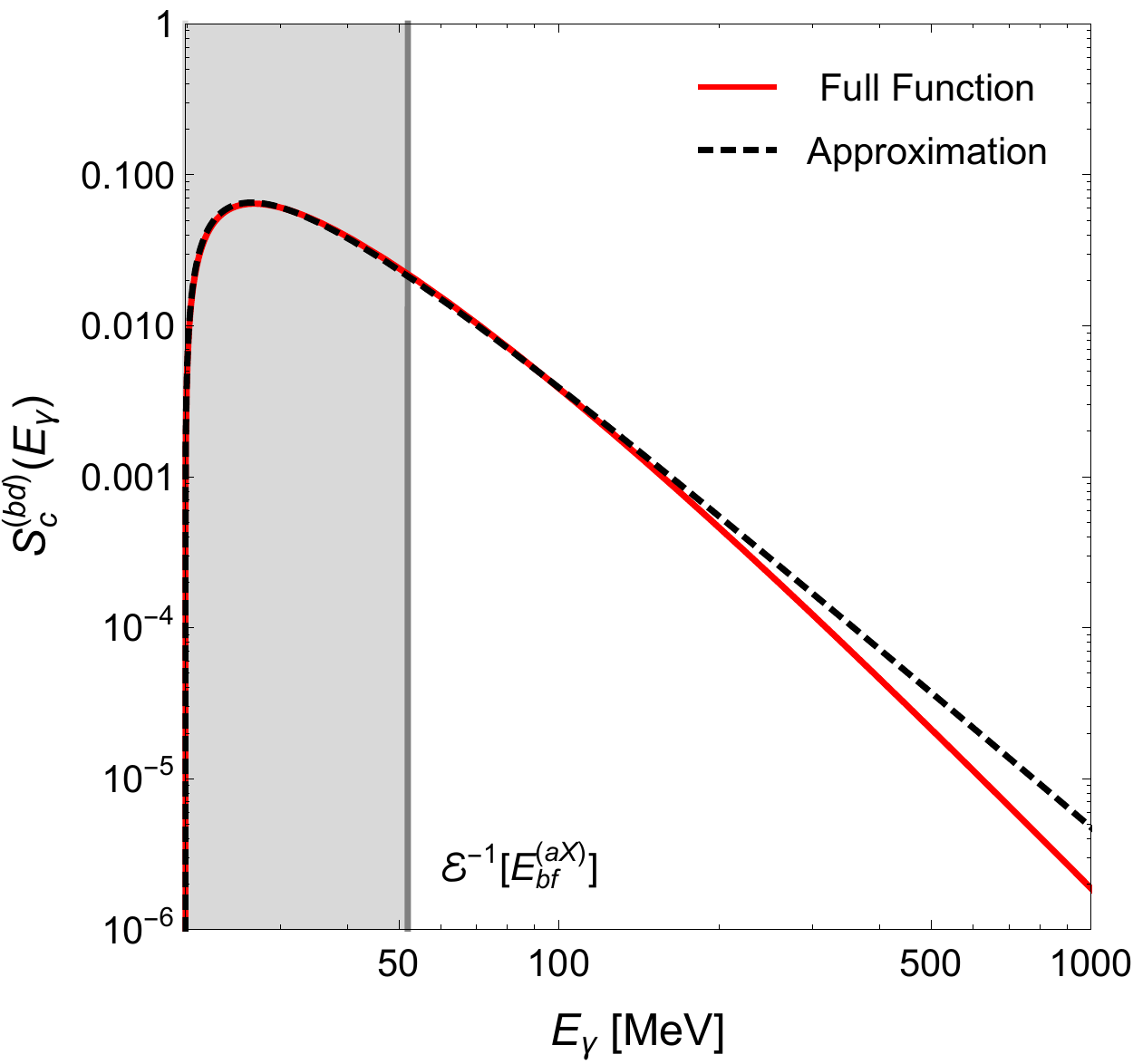}
\caption{The exact value of the function $S_c^{(bd)}(E_\gamma)$
  for the process $\ce{^{4}He} + \gamma \rightarrow \ce{T} + p$ (red curve)
  and our approximation to this function in Eq.~(\protect\ref{eq:ScbdApprox}) 
  (black dashed curve), plotted over the range $E_c^{(bd)} < E_\gamma \lsim 1$~GeV.~  The 
  gray vertical line corresponds to the minimum photon energy 
  $\mathcal{E}^{-1}[E_{bf}^{(aX)}]$ for the production of an excited $\ce{T}$ 
  nucleus above the kinetic-energy threshold for the secondary process
  $\ce{^{4}He} + \ce{T} \rightarrow \ce{^{6}Li} + n$.  Thus, photons in the 
  gray shaded region play no role in the secondary production of $\ce{^{6}Li}$.  
  We note that the corresponding $S_c^{(bd)}(E_\gamma)$ curves and thresholds for 
  $\ce{^{4}He} + \gamma \rightarrow \ce{^{3}He} + n$ (the other primary process 
  relevant for secondary $\ce{^{6}Li}$ production) are nearly identical to those
  shown here.   
\label{fig:ScbdCompare}}
\end{figure}

Given the dependence of these cross-sections on $E_\gamma$, we find that 
over the photon-energy range $E_c^{(bd)} <  E_\gamma \lesssim 1$~GeV,
the ratio $S_c^{(bd)}(E_\gamma)$ for each relevant process is well approximated by 
a simple function of the form
\begin{eqnarray}
  S_c^{(bd)}(E_\gamma) &=& \frac{\sigma_c^{(bd)}(E_\gamma)}
    {\sigma_{\mathrm{BH}}(E_\gamma) + \sigma_{\mathrm{CS}}(E_\gamma)} \nonumber \\     
  & \approx & S_{c,0}^{(bd)} 
    \left(\frac{E_\gamma}{E_c^{(bd)}} - 1\right) 
    \left(\frac{E_\gamma}{E_c^{(bd)}}\right)^{-4}~,~~~~~~~~
  \label{eq:ScbdApprox} 
\end{eqnarray} 
where $S_{c,0}^{(bd)}$ is a constant.  
In Fig.~\ref{fig:ScbdCompare} we show a comparison between the exact value of 
$S_c^{(bd)}(E_\gamma)$ for the individual process 
$\ce{^{4}He} + \gamma \rightarrow \ce{T} + p$ and our approximation 
in Eq.~(\ref{eq:ScbdApprox}) over this same range of $E_\gamma$. 
We see from this figure that 
our approximation indeed provides a good fit to $S_c^{(bd)}(E_\gamma)$ over most of 
this range.  Moreover, while the discrepancy between these 
two functions becomes more pronounced as $E_\gamma \sim \mathcal{O}(1\mbox{~GeV})$,
we emphasize that $S_c^{(bd)}(E_\gamma)$ is comparatively negligible at these energies.
Thus, although the primary-photon spectrum can include photons with energies
$\mathcal{O}(1\mbox{~GeV})$ --- indeed for $\tau_\chi \approx 10^{12}$~s, the timescale 
beyond which Eq.~(\ref{eq:Kfndef}) ceases to provide a reliable description of this 
spectrum, we find that $E_C \sim {\cal O}(10~{\rm GeV}\/)$ --- these photons have little impact on $\delta Y_a$.
Thus, for the purposes of approximating $\delta Y_a$, any discrepancy 
between Eq.~(\ref{eq:ScbdApprox}) and the exact expression for $S_c^{(bd)}$ at such 
energies may safely be ignored.

Armed with our approximation for $S_c^{(bd)}(E_\gamma)$ in Eq.~(\ref{eq:ScbdApprox}),
it is now straightforward to evaluate the integral in Eq.~(\ref{eq:dYSimpFromSecondaryProd})
and obtain an approximate analytic expression for $\delta Y_a$.
Just as it does for primary photoproduction and photodisintegration,
the functional dependence of $\delta Y_a$ on $\tau_\chi$ for secondary production 
depends on the relationship between $\tau_\chi$ and a pair of characteristic timescales
determined by the energy thresholds for the relevant processes.  
As we have discussed above, it is $\mathcal{E}^{-1}[E_{bf}^{(aX)}]$ rather than 
$E_c^{(bd)}$ which sets the minimum $E_\gamma$ required of a photon in order for it to 
contribute to the secondary production of $\ce{^{6}Li}$.  It therefore follows that 
the characteristic timescales for the secondary production of this nucleus are determined
by $\mathcal{E}^{-1}[E_{bf}^{(aX)}]$ rather than $E_c^{(bd)}$.  Thus, for 
secondary production, we define   
\begin{eqnarray}
  t_{Ca}^{(bcf)} &\approx & (10^8\, {\s})\times 
    \left( \frac     {\mathcal{E}^{-1}[E_{bf}^{(aX)}]    }{103\,\mev} \right)^2 \nonumber\\
  t_{Xa}^{(bcf)}  &\approx & (10^8\, {\s})\times 
    \left(\frac{     \mathcal{E}^{-1}[E_{bf}^{(aX)}]    }{28\,\mev} \right)^2~.
  \label{eqn:uniformTimesSecondary}
\end{eqnarray}

For $\tau_\chi < t_{Ca}^{(bcf)}$ within the uniform-decay 
approximation, $\delta Y_a$ is formally zero, as it is in the corresponding 
lifetime regime for primary production.  Thus, in order to
derive a meaningful constraint on unstable particles with lifetimes within this
regime, we follow the same procedure as we employed in order to calculate 
the contribution to $\delta Y_a$ from primary production for a particle with 
$\tau_\chi < t_{Ca}^{(bc)}$.  We model the injection of photons using the 
expression for secondary production appropriate for continuum injection,
with $d\rho(\tinj)/d\tinj$ given by Eq.~(\ref{eq:drhodtinjForDecay}).  In
analogy to our treatment of the primary process in this regime,
we include only the contribution from injection times in the range 
$t_{Ca}^{(bcf)} < \tinj < t_{Ca}^{(bcf)} + \tau_\chi$.  Dropping terms 
beyond leading order in the dimensionless variable 
$w_{C\chi} \equiv \tau_\chi/t_{Ca}^{(bcf)}$, we find 
\begin{equation}
  \delta Y_a ~=~ \sum_{b,c,f} A_a^{(bcf)}\Omega_\chi w_{C\chi} 
    e^{-w_{C\chi}^{-1}}~,
  \label{eq:UniformBBNSecondaryCaseI}
\end{equation}
where the proportionality constant $A_a^{(bcf)}$ for each combination of primary 
and secondary processes is independent of $\tau_\chi$.  We note 
that the dependence of $\delta Y_a$ on $\tau_\chi$ in this expression is exactly the 
same as in Eq.~(\ref{eq:UniformBBNCaseI}).  By contrast, for 
$t_{Ca}^{(bcf)} < \tau_\chi < t_{Xa}^{(bcf)}$, we find
\begin{eqnarray}
  \delta Y_a  &=& \sum_{b,c,f}
    B_a^{(bcf)}\Omega_\chi \Bigg[
      (5 - 3 \theta) 
      - 20 \beta^3 w_{X\chi}^{-3/2}  
        \nonumber \\ & &
      + 15 \beta^4 (1+ \theta) w_{X\chi}^{-2} 
        \!\!
      - 12 \beta^5 \theta w_{X\chi}^{-5/2} \Bigg]~,~~~
  \label{eq:UniformBBNSecondaryCaseII}
\end{eqnarray}
where we have defined $w_{X\chi} \equiv \tau_\chi / t_{Xa}^{(bcf)}$ in analogy
with $w_{C\chi}$ and where
\begin{equation}
  \theta ~\equiv~ \frac{E_{c}^{(bd)} }{\mathcal{E}^{-1}\big[E_{bf}^{(aX)}\big]}
\end{equation}
represents the ratio of the photon-energy threshold for primary production to the minimum 
photon energy needed to produce an excited nucleus of species $N_b$ with a kinetic energy 
above the energy threshold for the secondary process.  Numerically, we find 
$\theta \approx 0.46$ for 
$\ce{^{4}He} + \gamma  \rightarrow \ce{T} + p$ followed by
$\ce{^{4}He} + \ce{T}  \rightarrow \ce{^{6}Li} + n$, while $\theta \approx 0.52$ for
$\ce{^{4}He} + \gamma  \rightarrow \ce{^{3}He} + n$ followed by
$\ce{^{4}He} + \ce{^{3}He}  \rightarrow \ce{^{6}Li} + p$.  Finally, for 
$t_{Xa}^{(bcf)} < \tau_\chi$, we find
\begin{eqnarray}
  \delta Y_a  &=& \sum_{b,c,f}
      B_a^{(bcf)}\Omega_\chi \Bigg[
      \frac{16}{21} (9 - 5\theta) w_{X\chi}^{-1/4} \!\!
      \nonumber \\ & &
      - 4 (1 + 5\beta^3) w_{X\chi}^{-3/2} 
      + \frac{15}{7}(1 + \theta)(1 + 7\beta^4)w_{X\chi}^{-2} 
     \nonumber \\ & & 
      - \frac{4}{3} \theta (1 + 9\beta^5) w_{X\chi}^{-5/2}
   \Bigg]~. 
  \label{eq:UniformBBNSecondaryCaseIII} 
\end{eqnarray} 

Thus, to summarize, 
we have derived a set of simple, analytic 
approximations for the change $\delta Y_a$ in the abundance of a given nuclear 
species $N_a$ due to the late injection of photons by a decaying particle $\chi$.  For 
primary photoproduction or photodisintegration processes, we find that $\delta Y_a$ is 
given by Eq.~(\ref{eq:UniformBBNCaseI}), Eq.~(\ref{eq:UniformBBNCaseII}), or 
Eq.~(\ref{eq:UniformBBNCaseIII}), depending on the lifetime $\tau_\chi$ of the particle.
Likewise, for secondary production, we find that $\delta Y_a$ is 
given by Eq.~(\ref{eq:UniformBBNSecondaryCaseI}), Eq.~(\ref{eq:UniformBBNSecondaryCaseII}), 
or Eq.~(\ref{eq:UniformBBNSecondaryCaseIII}) within the corresponding lifetime
regimes.

%%%%%%%%%%%%%%%%%%%%%%%%%%%%%%%%%%%%%%%%%%%%%%%%%%%%%%%%%%%%%%%%%%%%%%%%%%%%%%%%%%%%%%
\FloatBarrier
\subsection{The Fruits of Linearization:~
Light-Element Constraints on Ensembles of Unstable Particles \label{sec:ConstMultUnstable}}
%%%%%%%%%%%%%%%%%%%%%%%%%%%%%%%%%%%%%%%%%%%%%%%%%%%%%%%%%%%%%%%%%%%%%%%%%%%%%%%%%%%%%%

We now turn to the task of extending these results 
to the case of an ensemble of decaying particles $\chi_i$ with lifetimes $\tau_i$ and extrapolated 
abundances $\Omega_i$.  Indeed, we have seen that if the linearity criterion is satisfied 
both for $N_a$ itself and for all of its source nuclei $N_b \neq N_a$, all feedback
effects on $Y_a$ can be neglected.  Thus, in this regime, the overall change $\delta Y_a$ 
in the abundance of a light nucleus is well approximated by the sum of the individual 
contributions associated with the individual $\chi_i$ from each pertinent process.  
Indeed, it is only because we have entered a linear regime that such a direct sum
is now appropriate.
These, then, are the fruits of linearization.

While certain nuclei in our analysis --- namely $\ce{^{6}Li}$ and $\ce{^{7}Li}$ --- do 
not have the property that the linearity criterion is always satisfied for both the 
nucleus itself and its source nuclei, we emphasize that we are nevertheless able to 
derive meaningful bounds on the comoving number densities of these nuclei.
As noted in Sect.~\ref{sec:Linearization}, artificially adopting the linearity 
criterion for $Y_{\ce{^{7}Li}}$ in the Boltzmann equation for $\ce{^{6}Li}$ always 
yields a conservative bound on $\delta Y_{\ce{^{6}Li}}$, regardless of the injection 
history.  For $\ce{^{7}Li}$, the issue is that feedback effects on the photodisintegration
rate due to changes in the comoving number density of $\ce{^{7}Li}$ itself are not 
necessarily small.  Thus, strictly speaking, $\delta Y_{\ce{^{7}Li}}$ is not  
well approximated by a direct sum of the contributions
from the individual $\chi_i$.  However, since this direct sum is simply an approximation
of the integral in Eq.~(\ref{eq:deltaYaDecouplingRegime}), it is 
equivalent to the quantity $Y_{\ce{^{7}Li}}^{\mathrm{init}} 
\ln [1+ \delta Y_{\ce{^{7}Li}}(t)/Y_{\ce{^{7}Li}}^{\mathrm{init}}]$. 
Thus, for all relevant $N_a$, we find that 
an ensemble of decaying particles 
makes an overall contribution to 
$\delta Y_a$ --- or, in the case of $\ce{^{7}Li}$, to the quantity $Y_a^{\mathrm{init}} 
\ln [1+ \delta Y_a(t)/Y_a^{\mathrm{init}}]$ --- 
which can be approximated 
as a direct sum of the individual contributions from the individual ensemble constituents.
Indeed, this yields  either a reliable estimate for the true value of $\delta Y_a$ or a reliably 
conservative bound on $\delta Y_a$.

In principle, each of these individual contributions to a given $\delta Y_a$
can involve a large number of reactions with different energy thresholds and 
scattering kinematics, each with its own distinct fit
parameters $A_a^{(bc)}$, $B_a^{(bc)}$, $t_{Xa}^{(bc)}$, $t_{Ca}^{(bc)}$,
\etc\,  In practice, however, the number of reactions which contribute significantly
to $\delta Y_a$ for any one of the four relevant nuclear species is quite small, as
can be seen from Table~\ref{tab:BBNprocs}.~  Moreover, it is often the case that
many if not all of the reactions which have a non-negligible impact on a given 
$\delta Y_a$ have very similar energy thresholds and scattering kinematics.  When this is the 
case, the contribution to $\delta Y_a$ from these processes can, to a very good approximation, 
collectively be modeled using a single set of fit parameters.         
For example, the only reactions which have a significant impact on 
$\delta Y_{\ce{^{4}He}}$ are the primary photodisintegration processes 
$\ce{^{4}He} + \gamma \rightarrow \ce{T} + p$ and 
$\ce{^{4}He} + \gamma \rightarrow \ce{^{3}He} + n$, which have very similar 
energy thresholds and scattering kinematics.  Thus, a fit involving a single set 
of parameters yields an accurate 
approximation for $\delta Y_{\ce{^{4}He}}$.  The reactions which contribute to 
$\delta Y_{\ce{^{7}Li}}$ differ more significantly in terms of their energy thresholds 
and scattering kinematics.  Nevertheless, as we shall see, the collective effect of these 
processes is also well modeled by a single set of fit parameters.  By contrast,
for $\ce{D}$ there are two processes with qualitatively different 
energy thresholds and scattering kinematics which must be modeled using
separate sets of fit parameters: primary photodisintegration via 
$\ce{D} + \gamma \rightarrow  n + p$ 
and primary photoproduction via $\ce{^{4}He} + \gamma \rightarrow  \ce{D} + n + p$.  
Likewise, for $\ce{^{6}Li}$, two sets of fit parameters are required:  one for 
primary photoproduction via $\ce{^{7}Li} + \gamma \rightarrow \ce{^{6}Li} + n$ and 
one for secondary production via both 
$\ce{^{4}He} + \ce{^{3}He}\rightarrow  \ce{^{6}Li} + p$ and 
$\ce{^{4}He} + \ce{T} \rightarrow \ce{^{6}Li} + n$.

\begin{table*}[t!]
\centering
\smaller
\begin{tabular}{||c|c|c|c|c|c|c|c|c||}
\hline
\hline
  ~Nucleus~ & Process & $t_{Ba} (s) $ & $t_{Ca} (s) $ &
    $t_{Xa} (s) $  & $A_a $ & $B_a $ &
    $\delta Y_a^{\mathrm{min}}$ & $\delta Y_a^{\mathrm{max}}$ \\
\hline
\hline
  \ce{^{4}He} & Destruction  & $ ~8.9 \times 10^{6}$~ &
    $ ~3.2 \times 10^{6}$~ & $ ~4.2 \times 10^{7}~$ &
    $-3.3 \times 10^{-1} $ & $-1.5 \times 10^{0}$ & $-3.0 \times 10^{-3}$ & --- \\
\hline
  \ce{^{7}Li} & Destruction & $ 4.5 \times 10^{5}$ &
    $ 5.7 \times 10^{4}$ & $ 1.6 \times 10^{6}$ &
    ~$-2.7 \times 10^{-10}~ $ & ~$-3.8 \times 10^{-9}$~ & ~$-5.5 \times 10^{-11}$~ & --- \\
\hline
  \multirow{2}{*}{\ce{D}} & Destruction  & $ 2.2 \times 10^{5}$ &
    $ 4.6 \times 10^{4}$ &  $ 1.1  \times 10^{6} $  &
    $-1.2 \times 10^{-4}$ & $-1.2 \times 10^{-3}$  &
    \multirow{2}{*}{$-8.9 \times 10^{-7}$} &
    \multirow{2}{*}{$ 1.6 \times 10^{-6}$} \\
\cline{2-7}
  & Production* & $ 1.9 \times 10^{7}$ &
    $ 8.0 \times 10^{6}$ & $ 1.1  \times 10^{8} $ &
    $1.8 \times 10^{-2} $ & $8.2 \times 10^{-2} $  & & \\
\hline
  \multirow{2}{*}{\ce{^{6}Li}} & Primary Production  &
    $1.8 \times 10^{6}$ & $ 6.0 \times 10^{5}$ & $1.0 \times 10^{7} $&
    $1.6 \times 10^{-10} $ & $9.8 \times 10^{-10}$ & \multirow{2}{*}{---} &
    \multirow{2}{*}{$1.8 \times 10^{-11}$} \\
\cline{2-7}
  & ~~Secondary Production* ~~& $ 8.9 \times 10^{7}$ &
    $ 1.9 \times 10^{7}$ & $ 4.6 \times 10^{8}$ &
    $8.0 \times 10^{-8} $ & $1.8 \times 10^{-7}$ & & \\
\hline
\hline
\end{tabular}
\caption{~Values for the fit parameters appearing in Eqs.~(\ref{eq:BBNFitsTerm1})--(\ref{eq:BBNFitsLi6}) 
which characterize the contribution to the change $\delta Y_a$ 
  in the comoving number density of nucleus $N_a$ due to the electromagnetic decays 
  of unstable particles after BBN.~   
  An asterisk appearing after the name of the process indicates that the values 
  listed in the table correspond to the parameters $A_a^\ast$, $B_a^\ast$, 
  $t_{Ca}^\ast$, $t_{Xa}^\ast$, \etc, associated with $\delta Y_a^{(2)}$ for the
  nucleus in question rather than the parameters $A_a$, $B_a$, $t_{Ca}$, $t_{Xa}$, 
  \etc, associated with $\delta Y_a^{(1)}$.       
  The observational upper and lower limits $\delta Y_a^{\mathrm{min}}$ and 
  $\delta Y_a^{\mathrm{max}}$ on the value of $\delta Y_a$ are also shown. 
\label{tab:BBN}}
\end{table*}

In general, then, we require at most two distinct sets of parameters in order to model the     
overall contribution to $\delta Y_a$ for each relevant nucleus due to electromagnetic 
injection from an ensemble of decaying particles.  Thus, in general, we may express this 
overall contribution as the sum of two terms
\begin{equation}
  \delta Y_a ~=~ \delta Y_a^{(1)} + \delta Y_a^{(2)}~.
  \label{eq:BBNFits}
\end{equation}
For each relevant nuclear species, one of these terms is associated with a
primary process: primary photoproduction in the case of $\ce{^{6}Li}$ and 
primary photodisintegration in the case of $\ce{^{4}He}$, $\ce{^{7}Li}$, and $\ce{D}$.
Thus, $\delta Y_a^{(1)}$ takes a universal form for all relevant nuclear species.
In particular, Eqs.~(\ref{eq:UniformBBNSecondaryCaseI})--(\ref{eq:UniformBBNSecondaryCaseIII}) suggest that we may model
$\delta Y_a^{(1)}$ with a function of the form
\begin{eqnarray}
  \delta Y_a^{(1)} &=& \!\!\!\!\!\!\!
    \sum_{t_{Aa} < \tau_i  < t_{Ba} } \!\!\!\!\!\!\!\!
        A_a \Omega_i \left(\frac{\tau_i}{t_{Ca}}\right) e^{-t_{Ca}/\tau_i} \nonumber\\
      + &&\!\!\!\!\!\!\!\sum_{t_{Ba} < \tau_i < t_{Xa}} \!\!\!\!\!
        B_a \Omega_i \left[1 - \beta \left(\frac{t_{Xa}}{\tau_i}\right)^{\! 1/2} 
        \right] \nonumber\\
      + &&\!\!\!\!\!\!\!\sum_{t_{Xa} < \tau_i < t_{fa}} \!\!\!\!\!\!\!\!
        B_a \Omega_i \left(\frac{t_{Xa}}{\tau_i}\right)^{\! 1/4}\!
        \left[2 - (1 + \beta)\left(\frac{t_{Xa}}{\tau_i}\right)^{\! 1/4}  
        \right]~, \nonumber \\
  \label{eq:BBNFitsTerm1}
\end{eqnarray}
where $A_a$, $B_a$, $t_{Ca}$, $t_{Xa}$, \etc, are model parameters whose 
assignments we shall discuss below.  By contrast, the form of the second 
term in Eq.~(\ref{eq:BBNFits}) differs depending on the nucleus in question.  
For $\ce{^{4}He}$ and $\ce{^{7}Li}$, we simply have $\delta Y_a^{(2)} = 0$.  
For $\ce{D}$, this term is associated with primary production and thus takes
exactly the same functional form as $\delta Y_a^{(1)}$.  In other words,
we have
\begin{eqnarray}
  \delta Y_a^{(2)} &=& \!\!\!\!\!\!\!\!
     \sum_{t_{Aa}^* < \tau_i  < t_{Ba}^* } \!\!\!\!\!\!\!\!\!
        A_a^* \Omega_i \left(\frac{\tau_i}{t_{Ca}^*}\right) e^{-t_{Ca}^*/\tau_i}
        \nonumber\\
      &\,+&\!\!\!\!\! \sum_{t_{Ba}^* < \tau_i < t_{Xa}^*} \!\!\!\!\!\!\!\!\!
        B_a^* \Omega_i \left[1 - \beta \left(\frac{t_{Xa}^*}{\tau_i}\right)^{\! 1/2} 
        \right] \nonumber\\
      &\,+&\!\!\!\!\! \sum_{t_{Xa}^* < \tau_i < t_{fa}^*} \!\!\!\!\!\!\!\!\!
        B_a^* \Omega_i \left(\frac{t_{Xa}^*}{\tau_i}\right)^{\! 1/4}\!
        \left[2 - (1 + \beta) \left(\frac{t_{Xa}^*}{\tau_i} \right)^{\! 1/4}  
        \right]~, \nonumber \\
  \label{eq:BBNFitsTerm2}
\end{eqnarray}
where $A_a^\ast$, $B_a^\ast$, $t_{Ca}^\ast$, $t_{Xa}^\ast$, \etc, represent an
additional set of model parameters distinct from the parameters 
$A_a$, $B_a$, $t_{Ca}$, $t_{Xa}$, \etc, in Eq.~(\ref{eq:BBNFitsTerm1}).
For $\ce{^{6}Li}$, the term $\delta Y_a^{(2)}$ is associated
with secondary production and therefore takes the form
\begin{widetext} 
\begin{eqnarray}
  \delta Y_a^{(2)} &=& \!\!\!\!\!\!\!\!
    \sum_{t_{Aa}^* < \tau_i  < t_{Ba}^* } \!\!\!\!\!\!\!\!\!
        A_a^* \Omega_i \left(\!\frac{\tau_i}{t_{Ca}^*}\!\right) e^{-t_{Ca}^*/\tau_i}
      \nonumber \\
      &\,+&\!\!\!\!\!\!\!\! \sum_{t_{Ba}^* < \tau_i < t_{Xa}^*} \!\!\!\!\!\!\!\!\!
        B_a^* \Omega_i  \Bigg[     (5 - 3 \theta) 
      - 20 \beta^3 \left(\!\frac{t_{Xa}^*}{\tau_i}\!\right)^{\! 3/2} \!\!\!\!
      + 15 \beta^4 (1+ \theta) \left(\!\frac{t_{Xa}^*}{\tau_i}\!\right)^{\! 2} \!\!
      - 12 \beta^5 \theta \left(\!\frac{t_{Xa}^*}{\tau_i}\!\right)^{\! 5/2} \Bigg] 
        \nonumber \\
      &\,+&\!\!\!\!\!\!\!\! \sum_{t_{Xa}^* < \tau_i < t_{fa}^*} \!\!\!\!\!\!\!\!\!
        B_a^* \Omega_i 
      \Bigg[
      \frac{16}{21} (9 - 5\theta) \left(\!\frac{t_{Xa}^*}{\tau_i}\!\right)^{\! 1/4} \!\!\!\!
      - 4 (1 + 5\beta^3) \left(\!\frac{t_{Xa}^*}{\tau_i}\!\right)^{\! 3/2} \!\!\!\!
      + \frac{15}{7}(1 + \theta)(1 + 7\beta^4)
        \left(\!\frac{t_{Xa}^*}{\tau_i}\!\right)^{\! 2} \!\! 
      - \frac{4}{3} \theta (1 + 9\beta^5) 
        \left(\!\frac{t_{Xa}^*}{\tau_i}\!\right)^{\! 5/2}
   \Bigg]~. \nonumber \\
\label{eq:BBNFitsLi6}
\end{eqnarray}
\end{widetext}
We note that for simplicity and compactness of notation, we have implicitly taken 
$\epsilon_i = 1$ for each $\chi_i$ in formulating the expressions appearing in 
Eqs.~(\ref{eq:BBNFitsTerm1})--(\ref{eq:BBNFitsLi6}).  However, it is straightforward 
to generalize these results to the case in which $\epsilon_i$ differs from 
unity for one or more of the $\chi_i$.  In particular, the corresponding expressions 
for $\delta Y_a^{(1)}$ and $\delta Y_a^{(2)}$ in this case may be obtained by replacing 
$\Omega_i$ for each species with the product $\Omega_i \epsilon_i$.  Moreover,
we remind the reader that for the case of $\ce{^{7}Li}$, a more accurate estimate of 
$\delta Y_a$ can be obtained by replacing $\delta Y_a^{(1)}$ with 
$Y_a^{\mathrm{init}}\ln [1+ \delta Y_a^{(1)}(t)/Y_a^{\mathrm{init}}]$ in 
Eq.~(\ref{eq:BBNFitsTerm1}).

We now determine the values of the parameters in Eq.~(\ref{eq:BBNFitsTerm1}).  
We begin by noting that the parameter $t_{Aa}$ for each relevant nucleus $N_a$ may be 
assigned essentially any value between the end of the BBN epoch and the 
timescale at which the first reaction which contributes to $\delta Y_a$
becomes efficient.  Thus, for simplicity, we choose a universal value 
$t_{Aa} = 10^4$~s for all $N_a$, which corresponds to the timescale 
at which the first reaction in Table~\ref{tab:BBNprocs} --- namely the 
photodisintegration process $\ce{D} + \gamma \rightarrow n + p$ --- effectively turns 
on.  For $t_{fa}$, we likewise choose a universal value $t_{fa} = 10^{12}$~s for all
$N_a$.  
As discussed in Sect.~\ref{sec:DiffusePhoton},
the reason for this choice is that many of the Class-I processes which serve to establish the 
reprocessed photon spectrum in Eq.~(\ref{eq:spectrum}) start to become inefficient
around this timescale.  As a result, the photon spectrum 
that results from electromagnetic injection at 
times $\tinj \gtrsim 10^{12}$~s differs from that in Eq.~(\ref{eq:spectrum}),
and the analytic approximation for $\delta Y_a^{(1)}$ in Eq.~(\ref{eq:BBNFitsTerm1}) 
is no longer reliable.  The values of $t_{Aa}^\ast$ and $t_{fa}^\ast$ appearing in 
Eqs.~(\ref{eq:BBNFitsTerm2}) and (\ref{eq:BBNFitsLi6}) are determined in 
the same way.

Given these results, 
we now determine the values of the remaining parameters 
in Eqs.~(\ref{eq:BBNFitsTerm1})--(\ref{eq:BBNFitsLi6}) by demanding that the constraint contour we obtain from each
of these equations be consistent with the contour obtained from a more complete numerical 
analysis of the Boltzmann system in the limiting case in which our decaying ``ensemble'' 
comprises only a single particle species $\chi$.  In this single-particle case, our analytic 
expression for $\delta Y_a$ becomes a function of only two variables: the abundance 
$\Omega_\chi$ and lifetime $\tau_\chi$ of $\chi$.  We determine the constraint contours 
corresponding to the observational limits quoted in Sect.~\ref{sec:ConstraintsPrimord} 
by surveying ($\Omega_\chi$,$\tau_\chi$) space and 
numerically solving the full, coupled system of Boltzmann equations for $\delta Y_a$ at 
each point.  The undetermined parameters in 
Eqs.~(\ref{eq:BBNFitsTerm1})--(\ref{eq:BBNFitsLi6}) are then chosen such that our analytic expression provides a 
good fit to the corresponding constraint contour for each nucleus.  We list the values 
for all parameters obtained in this manner in Table~\ref{tab:BBN}. 

\begin{center}
\begin{figure*}[h]
\includegraphics[width=0.4\textwidth]{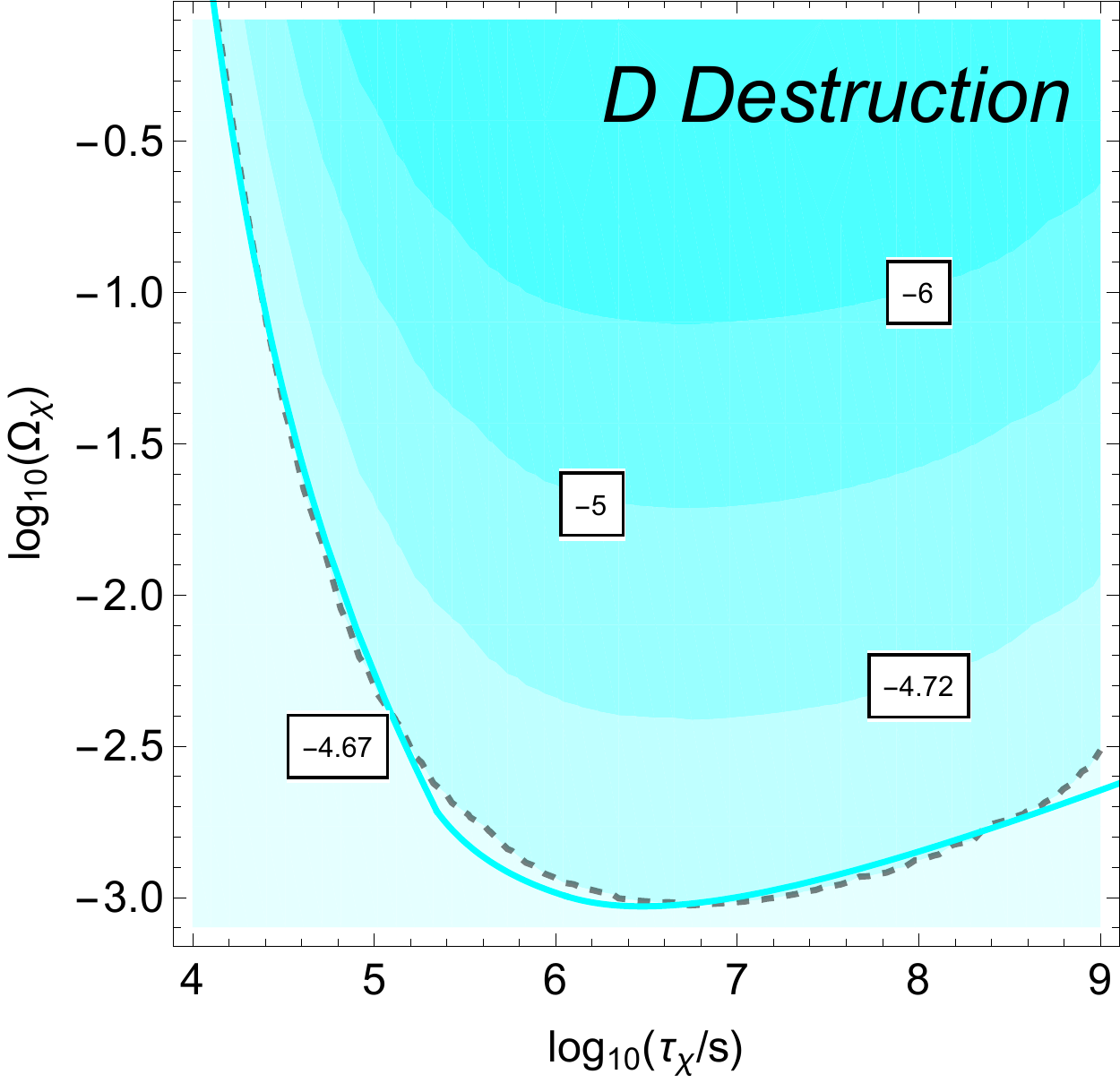}~~~
\includegraphics[width=0.4\textwidth]{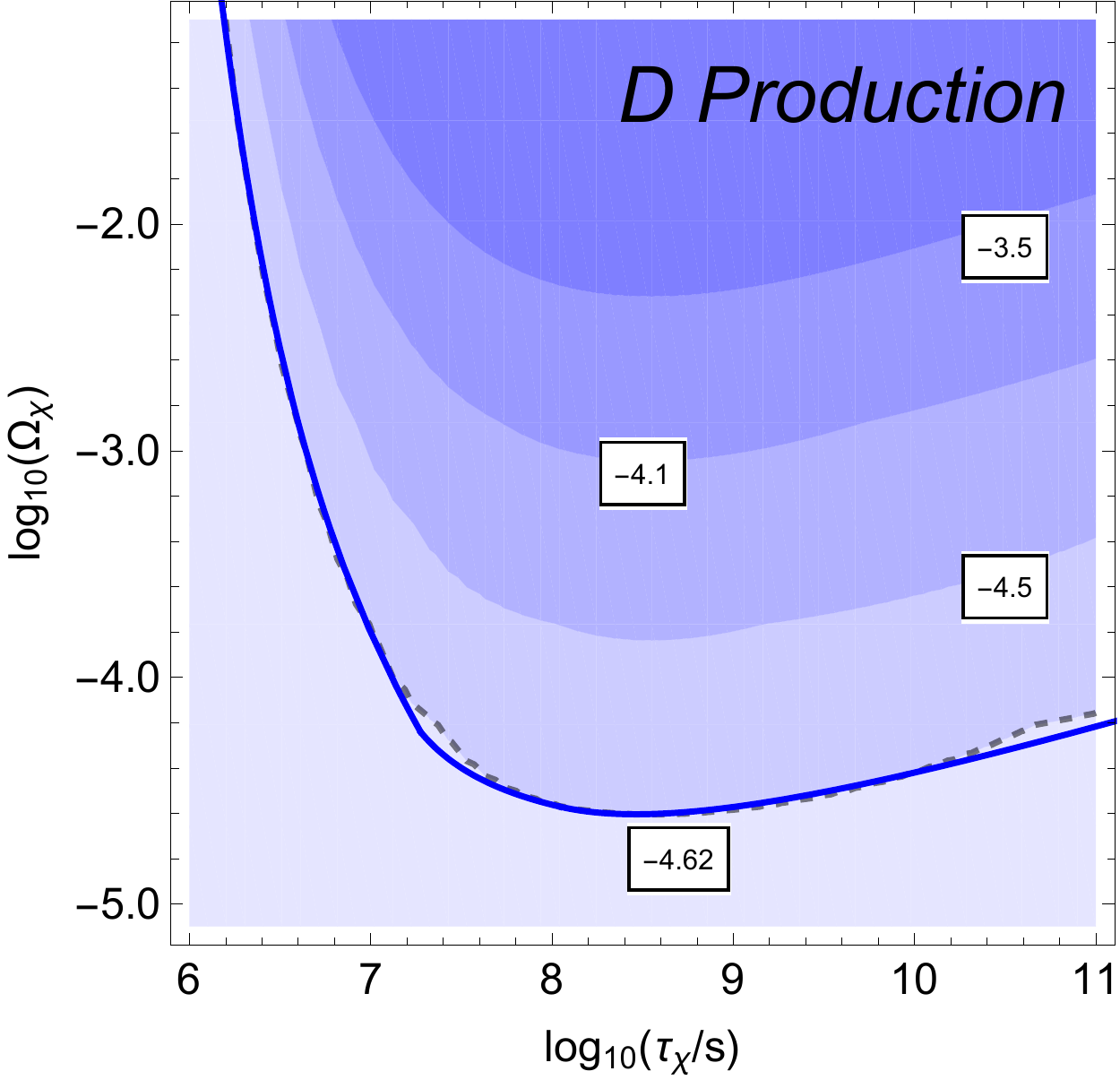} \\
\includegraphics[width=0.4\textwidth]{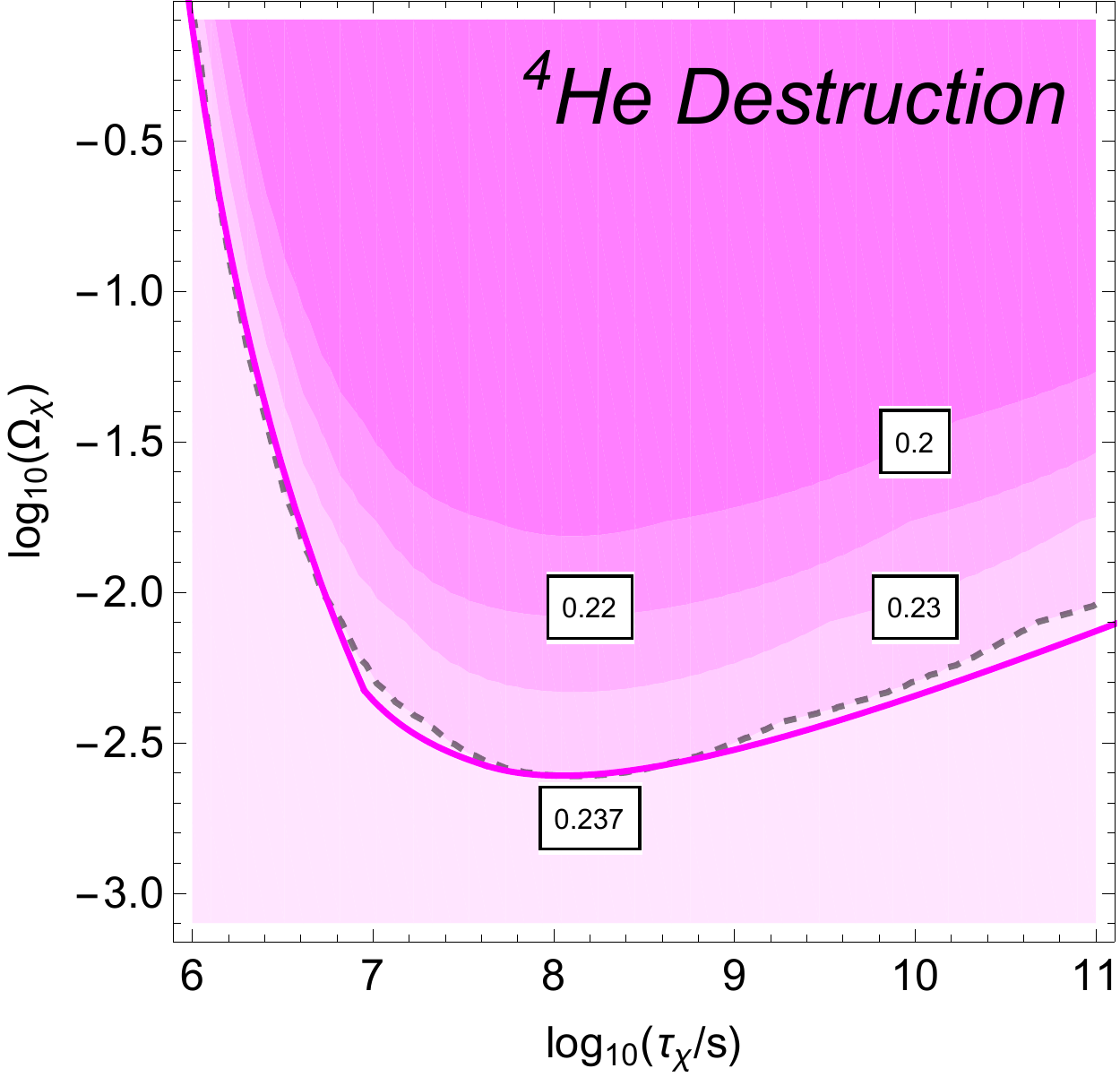}~~~ 
\includegraphics[width=0.4\textwidth]{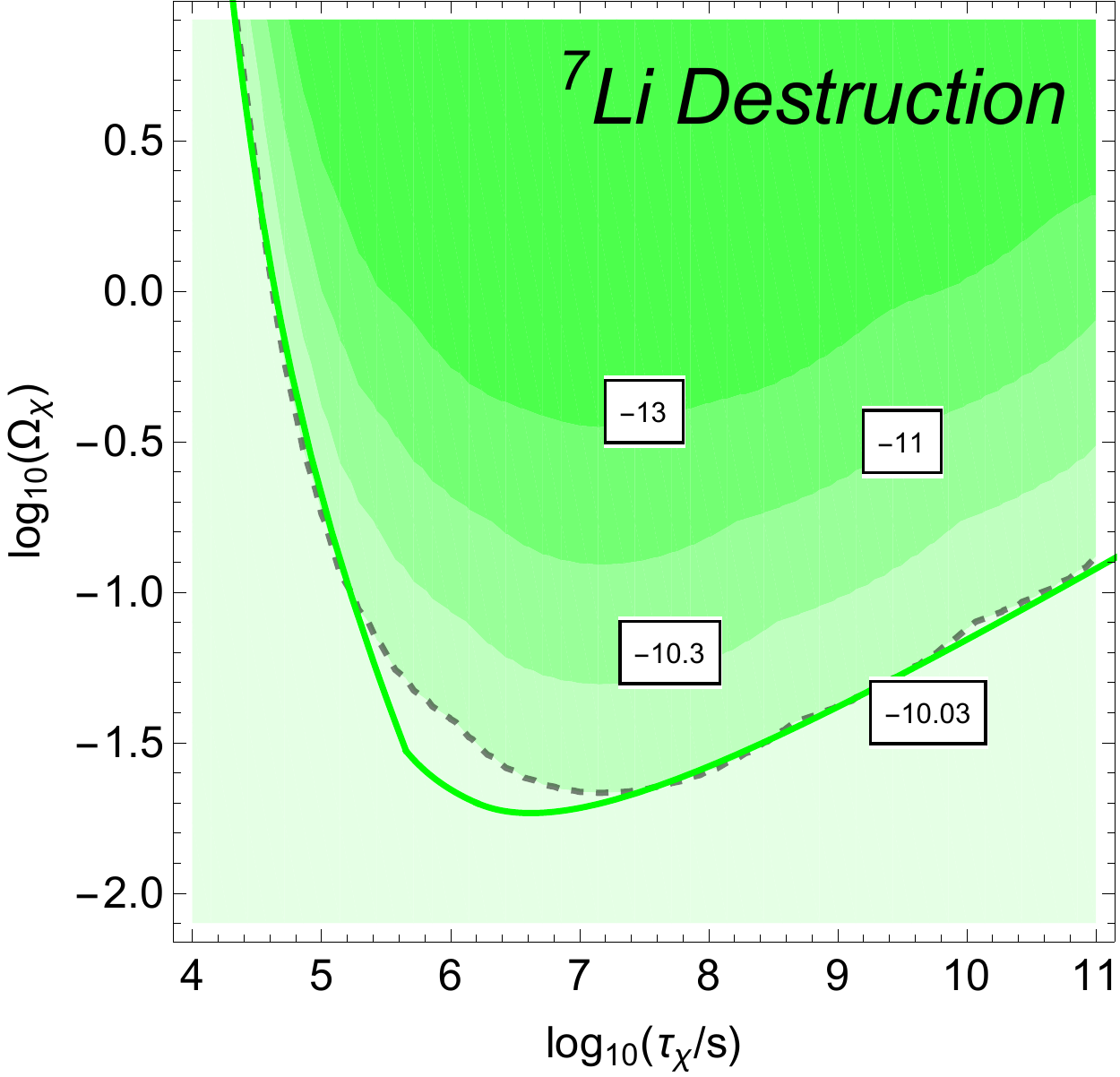} \\
\includegraphics[width=0.4\textwidth]{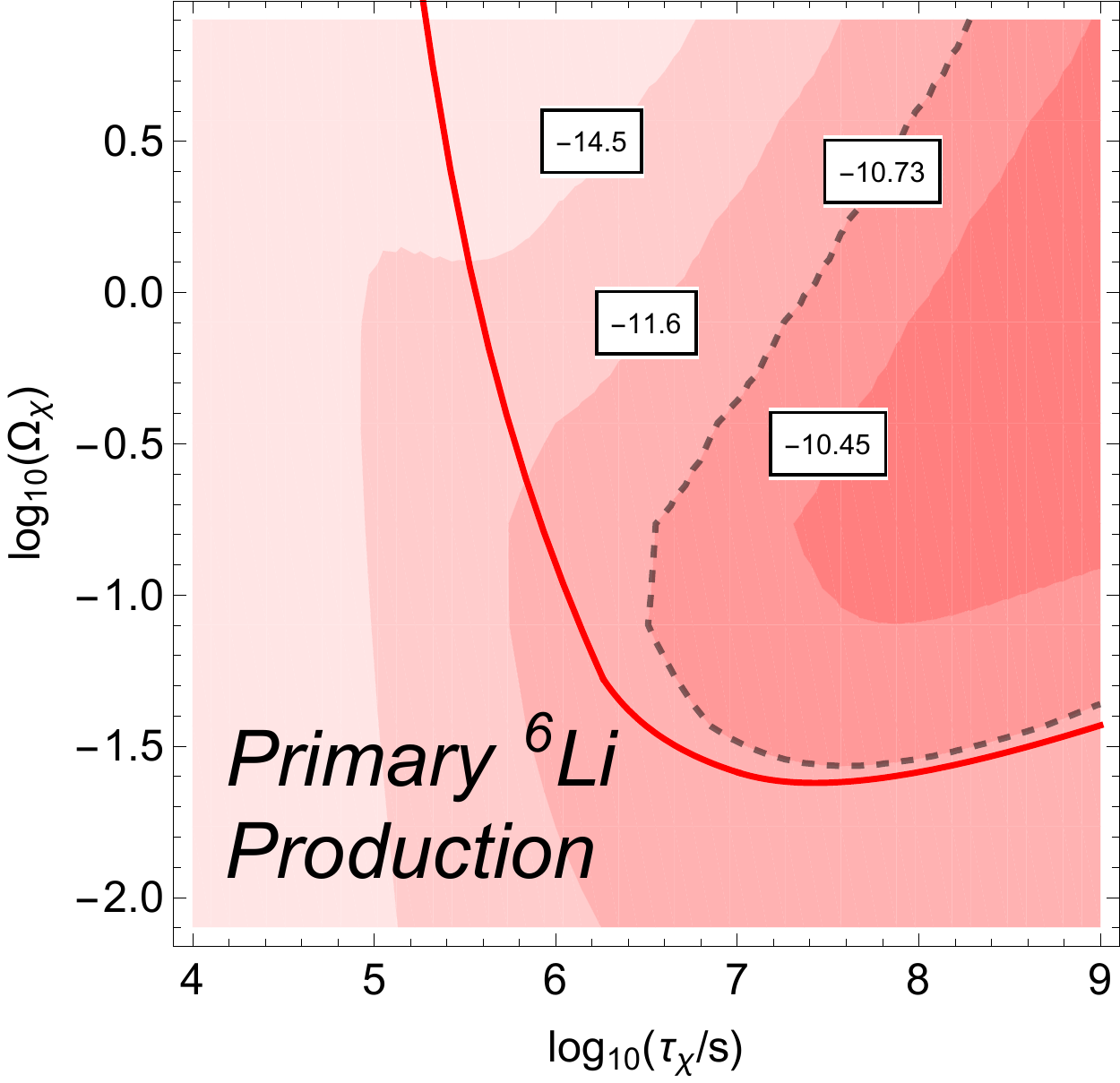}~~~ 
\includegraphics[width=0.4\textwidth]{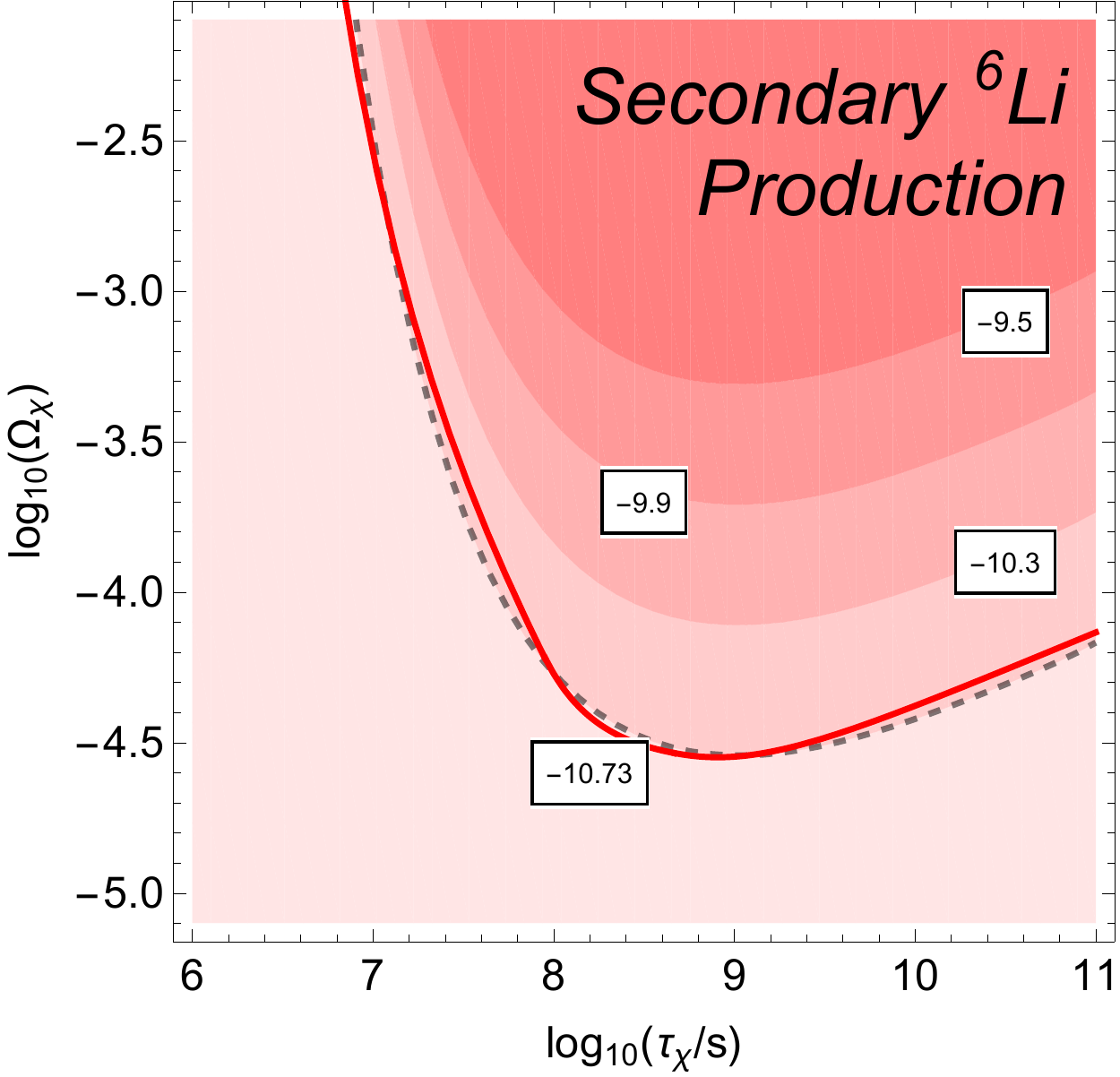}
\caption{Numerical results for $\delta Y_a$ for each relevant nucleus in the presence of a 
  single electromagnetically-decaying particle $\chi$ with extrapolated abundance 
  $\Omega_\chi$ and lifetime $\tau_\chi$.  
   In all panels except that pertaining to $\ce{^{4}He}$ destruction, we plot
      contours of $\log_{10}Y_a$, while for 
   $\ce{^{4}He}$ destruction we plot contours of $Y_p$.
   Note that for later reference, we have plotted separate contours for $\ce{D}$ production and
  destruction; likewise, we show separate contours for primary and secondary
  $\ce{^{6}Li}$ production. 
   Regions of the $(\tau_\chi,\Omega_\chi)$ plane which are found by numerical analysis to be excluded
are those above and/or to the right of the gray dashed line in each panel, while the solid colored curve
   in each panel demarcates the analogous region
            found to be excluded by fitting 
    the free parameters in our analytic approximations for $\delta Y_a$ in  
  Eqs.~(\protect\ref{eq:BBNFitsTerm1})--(\protect\ref{eq:BBNFitsLi6})
  to our numerical results.   We see that 
   the full numerical results and our analytic approximations
    agree very well in all cases except that of primary $\ce{^{6}Li}$ production, where
        our analytic approximation diverges from the numerical result at large $\Omega_\chi$ and
       thus provides an even more conservative bound on the allowed parameter space.
\label{fig:combfits}}
\end{figure*}
\end{center}
 
In deriving these constraint contours, it is necessary to translate 
the observational constraints in Eqs.~(\ref{eq:ObsBoundHe4})--(\ref{eq:ObsBoundLi6})
on the relevant $Y_a$ at the end of the BBN epoch into bounds of the form
\begin{equation}
  \delta Y_a^{\mathrm{min}} ~<~ \delta Y_a ~<~ \delta Y_a^{\mathrm{max}}
  \label{eq:DeltaaConstraint}
\end{equation}
on the corresponding subsequent change $\delta Y_a$ in each $Y_a$.  
In so doing, we must specify a set of initial values of $Y_a$ at the end of BBN.~
For $Y_{\ce{^{4}He}}$ and $Y_{\ce{^{7}Li}}$, we take $Y_p = 0.2449$ and 
$\ce{^{7}Li}/\ce{H} = 1.6 \times 10^{-10}$, as discussed in Sect.~\ref{sec:Linearization}.  
For $Y_{\ce{D}}$, we take the value which corresponds to the central theoretical 
prediction $\ce{D}/\ce{H} = 2.413 \times 10^{-5}$ for the primordial $\ce{D}/\ce{H}$ 
ratio derived in Ref.~\cite{CookeDUpper}.  Finally, since $\ce{^{6}Li}$ is not produced 
in any significant amount during BBN, we take $Y_{\ce{^{6}Li}} = 0$.
The values for the observational upper and lower limits $\delta Y_a^{\mathrm{min}}$ 
and $\delta Y_a^{\mathrm{max}}$ on $\delta Y_a$ which correspond to these choices
for the initial $Y_a$ are listed in Table~\ref{tab:BBN} alongside the values for
our fit parameters.

In Fig.~\ref{fig:combfits}, we display contours showing the results of our 
numerical calculation of $\delta Y_a$ in $(\Omega_\chi,\tau_\chi)$ space for each 
relevant nucleus.  More specifically, we display contours of $Y_p$ for $\ce{^{4}He}$, 
whereas we display contours of $\log_{10}Y_a$ for $\ce{D}$, $\ce{^{7}Li}$, 
and $\ce{^{6}Li}$.  Moreover, for purposes of illustration, we display separate contours 
for $\ce{D}$ production and destruction --- \ie, contours evaluated considering 
either production or destruction processes in $\mathcal{C}_{\ce{D}}^{(p)}$ only, 
with the cross-sections for the opposite set of processes artificially set to zero.
Likewise, we also display separate contours for primary and secondary $\ce{^{6}Li}$
production.  We emphasize, however, that in assessing our overall constraints on
decaying ensembles, we consider these processes together, allowing for the possibility 
of a cancellation between the two individual contributions to $\delta Y_a$ (in the case 
of $\ce{D}$) or a reinforcement (in the case of $\ce{^{6}Li}$).  The gray dashed line 
in each panel of Fig.~\ref{fig:combfits} indicates the upper bound on
$\Omega_\chi$ obtained through a numerical calculation of $\delta Y_a$ which 
follows from the observational limits in 
Eqs.~(\ref{eq:ObsBoundHe4})--(\ref{eq:ObsBoundLi6}).  By contrast, the solid
curve in each panel represents the corresponding constraint contour    
obtained through our analytic approximation for $\delta Y_a$ in Eq.~(\ref{eq:BBNFits}).       

We see from Fig.~\ref{fig:combfits}
that in all cases except for that of primary $\ce{^{6}Li}$ production,
the constraint contours we obtain using the linear and uniform-decay approximations
are quite similar to those obtained 
in the more realistic case in which we account
for the fact that $\chi$ has a constant decay rate and 
in which we account for feedback effects
in the collision terms $\mathcal{C}_a^{(p)}$ and $\mathcal{C}_a^{(s)}$
for each relevant nucleus.  Moreover, as discussed in Sect.~\ref{sec:Linearization}, 
we see that for primary $\ce{^{6}Li}$ production the constraint contour obtained
using these approximations indeed represents a conservative bound on $\Omega_\chi$. 
We also note that with this sole exception, the results of our analytic approximations 
are in good agreement with the corresponding results in Ref.~\cite{CyburtEllisUpdated} 
within the regime in which $\Omega_\chi$ is small and the linear approximation holds.
Once again, we note that the effects of additional processes which we do not incorporate 
into this analysis can have important effects on the $Y_a$ in the regime in which 
$\Omega_\chi$ --- and therefore the overall magnitude of electromagnetic injection --- is 
large.  These include, for example, photodisintegration processes which contribute to the 
depletion of a previously generated population of $\ce{^{6}Li}$ as well as additional processes 
which contribute to the production or destruction of $\ce{D}$~\cite{CyburtEllisUpdated}.
However, since the regions of $(\Omega_\chi,\tau_\chi)$ space in which these 
effects on $\ce{D}$ and $\ce{^{6}Li}$ are relevant are excluded by constraints on the 
primordial abundances of other nuclei, our neglecting these processes will have essentially
no impact on the overall constraints we derive on ensembles of decaying particles.

\begin{center}
\begin{figure}[t]
\includegraphics[width=0.49\textwidth]{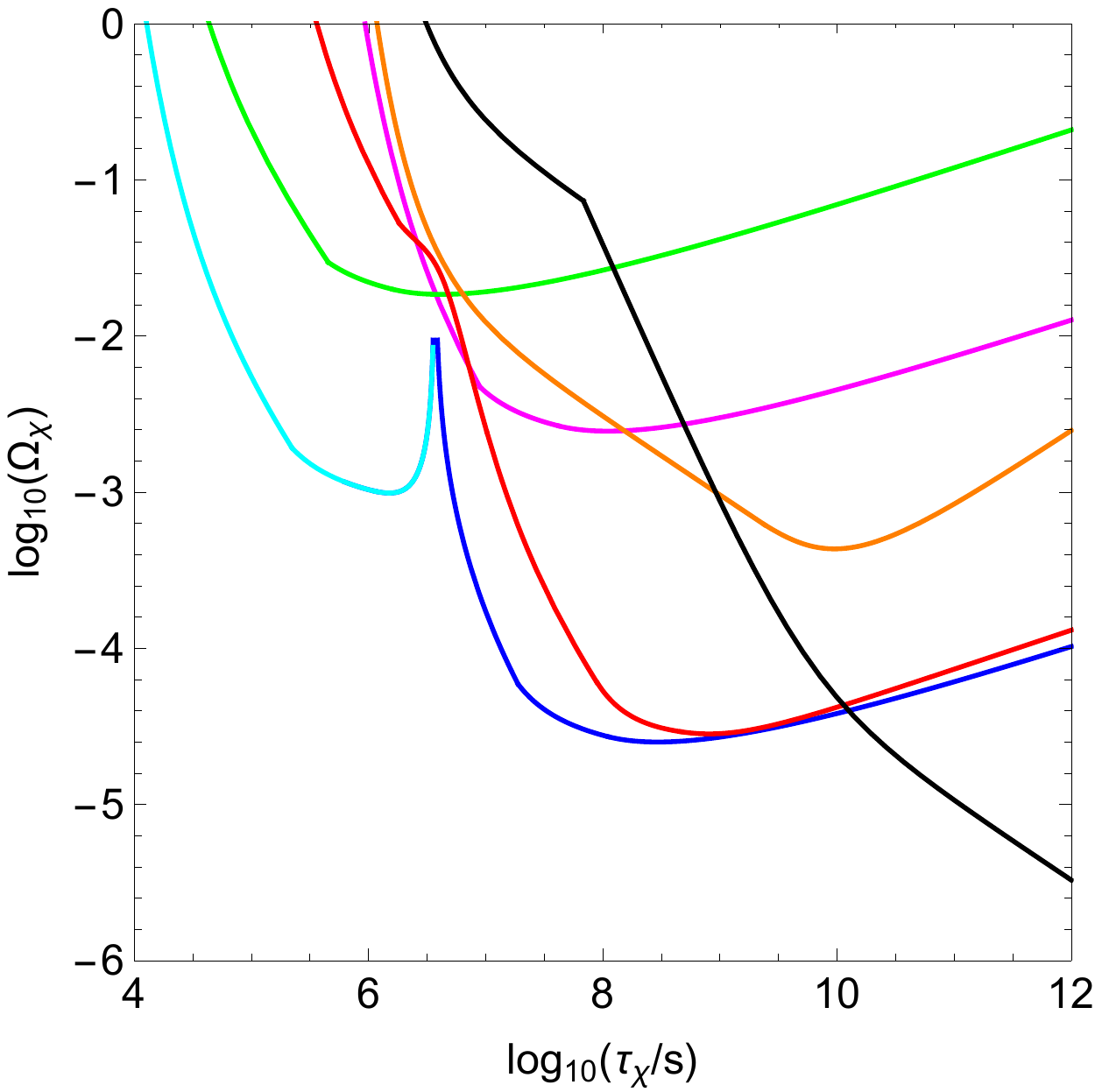}
\caption{Compilation of constraints on electromagnetic injection
      from bounds on the abundances of light nuclei.
   The individual contours shown 
   indicate the constraints stemming from considerations of 
    $\ce{D}$ production (dark blue) and destruction (light blue), 
  $\ce{^{7}Li}$ destruction (green), $\ce{^{4}He}$ destruction (magenta), and 
  $\ce{^{6}Li}$ production (red), the latter including both primary and secondary contributions.
  In addition, we also plot constraints stemming from 
  limits on potential distortions of the CMB-photon spectrum, both 
   $\mu$-type (orange) and $y$-type (black),
as will be discussed in Sect.~\protect\ref{sec:CMBDistortions}.~
      In each case, the regions of parameter space above each contour are excluded.
\label{fig:comb}}
\end{figure}
\end{center}

In Fig.~\ref{fig:comb} we plot our analytic approximations to the constraint 
contours obtained for each relevant nucleus together in 
$(\Omega_\chi,\tau_\chi)$ space.  Note that the $\ce{^{6}Li}$ contour 
includes both primary and secondary contributions to $\delta Y_{\ce{^{6}Li}}$.
In determining the $\ce{D}$ contour, we have included contributions from both photoproduction
and photodisintegration processes, the interference between which is responsible for the
``funnel'' region at $\tau_\chi \approx 10^{6.5}$~s within which the bound on $\Omega_\chi$ is 
considerably weakened.  However, since the upper and lower bounds on $\delta Y_{\ce{D}}$
derive from different considerations, different colors are used to distinguish the part of 
the contour which corresponds to the upper bound (dark blue) from the part which
corresponds to the lower bound (light blue).  We note that in all cases, these 
contours are in good agreement with the corresponding results in 
Ref.~\cite{CyburtEllisUpdated} within the regime in which $\Omega_\chi$ is small and 
the linear approximation holds.

%%%%%%%%%%%%%%%%%%%%%%%%%%%%%%%%%%%%%%%%%%%%%%%%%%%%%%%%%%%%%%%%%%%%%%%%%%%%%%%%%%%%%%

\FloatBarrier
\section{Spectral Distortions in the CMB\label{sec:CMBDistortions}}

%%%%%%%%%%%%%%%%%%%%%%%%%%%%%%%%%%%%%%%%%%%%%%%%%%%%%%%%%%%%%%%%%%%%%%%%%%%%%%%%%%%%%%

In addition to altering the abundances of light nuclei, electromagnetic injection 
from the decays of unstable particles in the early universe can also have observable 
effects on the spectrum of CMB photons, distorting that spectrum away from a pure 
blackbody profile.  Observational limits on these distortions therefore also constrain
such decays.  Once again, while the constraints on a single decaying particle species are well known, 
the corresponding constraints on an ensemble of decaying particles are less well
established.  Our aim in this section is to derive approximate analytic expressions for 
the CMB distortions which can arise due to entire ensembles of decaying particles ---
expressions analogous to those we derived in Sect.~\ref{sec:LightElemAbundances} 
for the change $\delta Y_a$ in the comoving number density in each relevant
nucleus after BBN.

One of the subtleties which arises in constraining distortions in the CMB-photon 
spectrum from ensembles of decaying particles with a broad range of lifetimes is that 
both the manner in which that spectrum is distorted and the 
magnitude of that distortion depend on the timescale over which the injection takes
place.  At early times, energetic photons produced by these decays are rapidly brought 
into both kinetic and thermal equilibrium by the Class-III processes discussed in 
Sect.~\ref{sec:InjectionConstraints}.  As a result, the spectrum of photons retains
a blackbody shape.  However, at later times, photon-number-changing processes such as 
double-Compton scattering and bremsstrahlung ``freeze out,'' in the sense that
the rates for these processes drop below the expansion rate $H$ of the universe.  Once
this occurs, injected photons are no longer able to attain thermal equilibrium with 
photons in the radiation bath.  Nevertheless, they are able to attain kinetic equilibrium
with these photons through elastic Compton scattering. 
As a result, the photon spectrum no longer retains its original blackbody shape; rather, 
it is distorted into a Bose-Einstein distribution characterized by a pseudo-degeneracy 
parameter $\mu$.  

Eventually, at an even later time $t_{\mathrm{EC}} \approx  9 \times 10^9~\s$, elastic 
Compton scattering also freezes out.  Thus, photons injected at times 
$t \gtrsim t_{\mathrm{EC}}$ attain neither kinetic nor thermal equilibrium with
photons in the radiation bath.  Nevertheless, the injected photons continue to 
interact with electrons in the radiation bath at an appreciable rate until the 
time of last scattering $\tLS \approx 1.19 \times 10^{13}$~s.  During this window, 
photons in the radiation bath are 
up-scattered by electrons which acquire significant energy from these interactions.  
The ultimate result is a photon spectrum which is suppressed at low frequencies and 
enhanced at high frequencies.  This spectral distortion is characterized by a non-zero 
Compton $y$ parameter $y_C$.  Finally, after last scattering, any additional photons injected
from particle decay simply contribute to the diffuse photon background.    

In order to derive analytic expressions for the constraints on the parameters $\mu$ and 
$y_C$ in the presence of a decaying ensemble, we shall once again make use of the 
uniform-decay approximation as well as a linear approximation similar to the approximation
we invoked in calculating the $\delta Y_a$ in Sect.~\ref{sec:LightElemAbundances}.~
The validity of such an approximation follows from the fact that both $\mu$ and $y_C$ 
are tightly constrained by observation.  The most stringent limits on these parameters, 
which are currently those from COBE/FIRAS data~\cite{COBEFIRAS}, are  
\begin{eqnarray}
  |\mu| &<& 9.0 \times 10^{-5} \nonumber \\
  y_C &<& 1.2 \times 10^{-5}~.
  \label{eq:muandyClimits}
\end{eqnarray}
Moreover, proposed broad-spectrum experiments, such as 
PIXIE~\cite{PIXIE}, could potentially increase the sensitivity to which these 
distortions can be probed by more than an order of magnitude.  These stringent 
constraints on the shape of the CMB-photon spectrum imply that the contribution  
$\delta \rho_\gamma$ to the photon energy density from injection during the 
relevant epoch must be quite small, in the sense that $\delta \rho_\gamma \ll \rho_\gamma$.  
Thus, to a very good approximation, we may ignore feedback effects on the overall photon 
number density $n_\gamma$ and energy density $\rho_\gamma$ when analyzing distortions
in that spectrum.  This allows us to  replace these quantities by their equilibrium values.  

This linear approximation significantly simplifies the analysis of the resulting CMB 
spectrum~\cite{HuAndSilk,HuAndSilk2,ChlubaSunyaev,BeyondYAndMu}.  For example, 
in this approximation we may study the evolution of the CMB-photon 
spectrum using the method of Green's functions~\cite{ChlubaGreensFns1,BeyondYAndMu,ChlubaGreensFns2}.
In this approach, we may express a generic spectral distortion which constitutes a shift
$\Delta I_\nu(\tnow)$ in the present-day intensity of photons with frequency $\nu$ relative 
to the expected intensity in the standard cosmology in the form
\begin{equation}
  \Delta I_\nu (\tnow) ~=~ \int G(\nu,t,\tnow) 
    \,\frac{d}{dt}\left(\frac{Q}{\rho_\gamma}\right) dt~,
  \label{eq:GreensFuncSpectralDistortions}
\end{equation}
where $Q$ is the energy density injected at time $t$ and where 
$G(\nu,t,t')$ is a Green's function which relates the intensity of photons
with frequency $\nu$ injected at time $t$ to the photon intensity at that same 
frequency at a later time $t' > t$.  In general, $G(\nu,t,\tnow)$ is well approximated
by a sum of terms representing an overall temperature shift, a $\mu$-type 
distortion, and $y$-type distortion~\cite{ChlubaGreensFns1}.  This Green's-function
formalism provides a tool for numerically estimating the overall distortion to the 
CMB-photon spectrum from any arbitrary injection history, provided that the 
overall injected energy is small.  We shall draw on this formalism extensively in
deriving our analytic approximations for $\delta \mu$ and $\delta y_C$.

%%%%%%%%%%%%%%%%%%%%%%%%%%%%%%%%%%%%%%%%%%%%%%%%%%%%%%%%%%%%%%%%%%%%%%%%%%%%%%%%%%%%%%
\FloatBarrier
\subsection{The $\mu$ Parameter\label{sec:muParam}}
%%%%%%%%%%%%%%%%%%%%%%%%%%%%%%%%%%%%%%%%%%%%%%%%%%%%%%%%%%%%%%%%%%%%%%%%%%%%%%%%%%%%%%

The time-evolution of the pseudo-degeneracy parameter $\mu$ is governed by an 
equation of the form~\cite{HuAndSilk,HuAndSilk2}
\begin{equation}
  \frac{d\mu}{dt} ~=~ \frac{d\mu_{\mathrm{inj}}}{dt} 
    - \mu (\Gamma_{\mathrm{DC}} + \Gamma_{\mathrm{BR}})~.
  \label{eq:mu_diffeq}
\end{equation}
The first term in Eq.~(\ref{eq:mu_diffeq}) is a source term which arises due to 
the injection of energetic photons.  Within the linear approximation, this source term 
takes the generic form
\begin{eqnarray}
  \frac{d\mu_{\mathrm{inj}}}{dt} ~\approx ~ \frac{1}{2.143} \left[\frac{3}{\rho_\gamma} 
  \frac{d\rho_\gamma}{dt} - \frac{4}{n_\gamma} \frac{d n_\gamma}{dt} \right]~,
  \label{eq:muInj}
\end{eqnarray}
where $n_\gamma$ and $\rho_\gamma$ respectively denote the overall number density and 
energy density of photons.  We emphasize that this expression is completely general
and applies even in absence of any source of photon injection, in which 
case the two terms in Eq.~(\ref{eq:muInj}) cancel. 

The second term in Eq.~(\ref{eq:mu_diffeq}) is a sink term associated with processes
which serve to bring the population of photons in the radiation bath into thermal 
equilibrium and thereby erase $\mu$.  As discussed above, the dominant such 
processes are double-Compton scattering and bremsstrahlung, the respective thermalization 
rates for which are well approximated by power-law expressions of the 
form~\cite{HuAndSilk}
\begin{eqnarray}
  \Gamma_{\mathrm{DC}} &=&  \Gamma_{\mathrm{DC}}^{\mathrm{MRE}}
    \left(\frac{t}{\tMRE}\right)^{-9/4} \nonumber\\
  \Gamma_{\mathrm{BR}} &=& \Gamma_{\mathrm{BR}}^{\mathrm{MRE}}
    \left(\frac{t}{\tMRE}\right)^{-13/8}~,
  \label{eq:muParamDampingRates}
\end{eqnarray}
where $\tMRE$ once again denotes the time of matter-radiation equality.  The 
coefficients in this expression are given by
\begin{eqnarray}
  \Gamma_{\mathrm{DC}}^{\mathrm{MRE}} &\approx&  
    \left(5.73 \times 10^{-39} \gev\right) 
    \left( \Omega_B h^2 \right) \nonumber \\ & & \times
    \left( 1 - \frac{Y_p}{2}  \right)
    \left( \frac{\Tnow}{2.7\mbox{~K}} \right)^{3/2}  \nonumber\\
  \Gamma_{\mathrm{BR}}^{\mathrm{MRE}}~ &\approx&  
    \left(1.57 \times 10^{-36} \gev\right) 
    \left( \Omega_B h^2 \right)^{3/2} \nonumber \\ & & \times
    \left( 1 - \frac{Y_p}{2}  \right)
    \left( \frac{\Tnow}{2.7\mbox{~K}} \right)^{-5/4} ~,~~~
\end{eqnarray}
where $h$ is the Hubble constant, $\Omega_B$ is the present-day abundance of baryons,  
and $\Tnow \approx 2.7$~K is the present-day temperature of the CMB.
For completeness, we note that the expression for $\Gamma_{\mathrm{BR}}$ in Eq.~(\ref{eq:muParamDampingRates}) represents an overestimate of the true bremsstrahlung rate.  In particular, a full frequency-dependent treatment of the evolution of the photon spectrum reveals that bremsstrahlung is less efficient at early times~\cite{KhatriSunyaev}.  As a result, the true bremsstrahlung rate is actually smaller than the result in Eq.~(\ref{eq:muParamDampingRates})  would suggest.  However, we shall find below that the effect of bremsstrahlung on $\mu$ is subleading in comparison to the effect of double-Compton scattering, even with our overestimate for $\Gamma_{\mathrm{BR}}$.  Given this, 
we shall choose to ignore the effect of bremsstrahlung entirely at a later stage of our analysis. 
Thus, our use of the expression in 
Eq.~(\ref{eq:muParamDampingRates}) for the bremsstrahlung rate will ultimately 
have no effect on our overall results.

Up to this point, our expressions governing the evolution 
of $\mu$ are completely general and may be applied to any injection history, 
provided that the total energy density injected is sufficiently small that the
linear approximation is valid.  In order to derive an approximate analytic expression 
for the overall change $\delta \mu$ in $\mu$ due
to an ensemble of unstable particles, we proceed in essentially the same manner 
as we did in deriving expressions for the changes $\delta Y_a$ in the comoving 
number densities of light nuclei in Sect.~\ref{sec:LightElemAbundances}.~
We begin by deriving an expression for $\delta \mu$ in the presence
of a single decaying particle species $\chi$ with mass $m_\chi$ and lifetime 
$\tau_\chi$. 
When we take into account the full exponential nature 
of the decay, we find 
\begin{eqnarray}
  \frac{d\rho_\gamma}{dt} &=& - 4 H\rho_\gamma + 
    \epsilon_\chi \Gamma_\chi \rho_\chi \nonumber \\ 
  \frac{d n_\gamma}{dt} &=& - 3 H n_\gamma + 
    \frac{N_\chi^{(\gamma)}\Gamma_\chi\rho_\chi}{m_\chi}~,  
\end{eqnarray}
where $N_\chi^{(\gamma)}$ is the average number of photons produced per decay 
and where $\epsilon_\chi$ is the fraction of the energy released by $\chi$ decays which 
is transferred to photons.  By contrast, within the uniform-decay approximation 
in which $d\rho_\chi/dt$ is given by Eq.~(\ref{eq:drhodtinjUniformDec}), we have
\begin{eqnarray}
  \frac{d\rho_\gamma}{dt} &=& - 4 H\rho_\gamma + 
    \epsilon_\chi \rho_\chi(\tau_\chi) \delta(\tau_\chi - t) \nonumber \\ 
  \frac{d n_\gamma}{dt} &=& - 3 H n_\gamma + 
    \frac{N_\chi^{(\gamma)}\rho_\chi(\tau_\chi)}{m_\chi} \delta(\tau_\chi - t)~.  
\end{eqnarray}
In this approximation, a simple, analytic expression for $\delta \mu$ may 
be obtained by directly integrating Eq.~(\ref{eq:mu_diffeq}) from the time 
$t_e\approx 1.69 \times 10^3~\s$ at which electron-positron annihilation freezes out 
to the time $t_{\mathrm{EC}}$ at which elastic Compton scattering freezes out.  
In particular, we find that a particle species with a lifetime $\tau_\chi$ 
anywhere within this range gives rise to a $\mu$-type distortion of size  
\begin{eqnarray}
  \delta \mu &\approx & 
    \frac{\eta}{2.143} \left[
    \frac{90 \zeta (3)\epsilon_\chi}{\pi^4}  
    \left(\frac{\tau_\chi}{\s} \right)^{1/2} \!\!\!
    - 4N_\chi^{(\gamma)}\left(\frac{m_\chi}{\mev} \right)^{-1} \right] \nonumber\\
  &\,& \times \left(\frac{m_p}{\mev} \right) \frac{\Omega_\chi}{\Omega_B} 
    \exp \Bigg[  \frac{8}{5} \Gamma_{\mathrm{BR}}^{\mathrm{MRE}}\tMRE^{13/8}
      \left( t_{\mathrm{EC}}^{-5/8} -  \tau_\chi^{-5/8} \right) \nonumber \\ 
    && + \frac{4}{5} \Gamma_{\mathrm{DC}}^{\mathrm{MRE}}\tMRE^{9/4} 
      \left( t_{\mathrm{EC}}^{- 5/4} - \tau_\chi^{- 5/4} \right) \Bigg]~,
   \label{eq:deltamuNoSimplifying}
\end{eqnarray}
where $m_p$ is the proton mass, where $\Omega_B$ is the total present-day abundance 
of baryons, where $\eta \equiv n_B/n_\gamma \approx 6.2 \times 10^{-10}$ is the 
baryon-to-photon ratio of the universe, and where $\Omega_\chi$ once again refers to 
the extrapolated abundance of $\chi$.  For $\tau_\chi$ outside this range, 
$\delta \mu \approx 0$.

Since the bremsstrahlung term in the exponential in Eq.~(\ref{eq:deltamuNoSimplifying})
is generically negligible in comparison with the double-Compton-scattering term, we 
may safely neglect it in what follows.  Moreover, for $m_\chi$ and $\tau_\chi$ 
within our regimes of interest, the second term in the prefactor in 
Eq.~(\ref{eq:deltamuNoSimplifying}) is subleading and can likewise be neglected.   
Thus, we find that $\delta\mu$ is well approximated by
\begin{eqnarray}
\delta \mu &\approx &
  \big(4.8 \times 10^{-7}\big) 
    \, \frac{\Omega_\chi\epsilon_\chi}{\Omega_B} 
    \left(\frac{\tau_\chi}{\s} \right)^{1/2} \!\!
    e^{-(t_{\mu}/\tau_\chi)^{5/4}}~,~~~~~~~
  \label{eq:muSol}
\end{eqnarray}
where we have defined
\begin{equation}
  t_{\mu} ~\equiv~ \left(\frac{4}{5}\Gamma_{\mathrm{DC}}^{\mathrm{MRE}}\right)^{4/5}
  \tMRE^{9/5} ~ \approx ~ 6.6 \times 10^8\mbox{~s}~.
\end{equation}
This latter quantity represents the timescale at which the damping due to double-Compton 
scattering becomes inefficient and contributions to $\delta \mu$ 
become effectively unsuppressed.  

The dependence of $\delta \mu$ on $\tau_\chi$ has a straightforward physical
interpretation.  The exponential suppression factor reflects the washout of $\mu$ 
by double-Compton scattering at early times $\tau_\chi \ll t_{\mu}$.  The 
additional polynomial factor reflects the fact that the equilibrium photon density 
grows more rapidly with time than does the matter density at earlier times. 
This implies that particles decaying at an earlier time yield smaller perturbations
to the photon spectrum.

%%%%%%%%%%%%%%%%%%%%%%%%%%%%%%%%%%%%%%%%%%%%%%%%%%%%%%%%%%%%%%%%%%%%%%%%%%%%%%%%%%%%%%
\FloatBarrier
\subsection{The Compton $y$ Parameter\label{sec:yparam}}
%%%%%%%%%%%%%%%%%%%%%%%%%%%%%%%%%%%%%%%%%%%%%%%%%%%%%%%%%%%%%%%%%%%%%%%%%%%%%%%%%%%%%%

As discussed above, elastic Compton scattering serves to maintain kinetic 
equilibrium among photons in the radiation bath after double-Compton scattering 
and bremsstrahlung freeze out.  However, elastic Compton scattering eventually 
freezes out as well, at time $t \sim t_{\mathrm{EC}}$.  An injection of photons 
which occurs after $t_{\mathrm{EC}}$ but before the time $\tLS$ of last scattering
contributes to the development of a non-vanishing Compton $y$ parameter $y_C$.
Within the linear approximation, a reliable estimate for the rate of change of $y_C$ may  
be obtained from the relation~\cite{DaneseDeZotti}
\begin{equation}
  \frac{d y_C}{d\tinj} ~\approx~ 
  \frac{1}{4\rho_\gamma(\tinj)}\frac{d\rho(\tinj)}{d\tinj}~,
\end{equation}
where $d\rho(\tinj)/d\tinj$ denotes the rate at which energy density is injected
into the photon bath.  We note that in contrast with the corresponding evolution 
equation for $\mu$ in Eq.~(\ref{eq:mu_diffeq}), this equation contains no damping 
term.

In deriving our analytic approximation for the change $\delta y_C$ from
an ensemble of unstable particle species, we once again begin by considering the case of
a single such species $\chi$.  When we take into account the full 
exponential nature of the decay, we obtain 
\begin{eqnarray}
  \frac{dy_C}{d\tinj} &\approx &\frac{1}{4 \rho_\gamma(\tinj)} 
    \epsilon_\chi \Gamma_\chi \rho_\chi(\tinj)~.
\end{eqnarray}
The corresponding result within the uniform-decay approximation is 
\begin{eqnarray}
  \frac{dy_C}{d\tinj} &\approx &
    \frac{1}{4 \rho_\gamma(\tinj)}  \epsilon_\chi \rho_\chi(\tau_\chi)
    \delta(\tau_\chi - \tinj)~.
\end{eqnarray}
In integrating this latter expression over $t$ in order to obtain our 
approximate expression for $\delta y_C$, we must account for the fact that
the epoch during which injection contributes to $y_C$ straddles the time 
$\tMRE$ of matter-radiation equality.  Using the appropriate time-temperature 
relations 
\begin{equation}
T ~\approx~ T_{\mathrm{MRE}} \times \begin{cases}
   (\tMRE/t)^{1/2} & t \lesssim \tMRE \\
   (\tMRE/t)^{2/3} & t \gtrsim \tMRE   
\end{cases}
\end{equation}
before and after the transition to matter-domination,
where $T_{\mathrm{MRE}}$ is the temperature at matter-radiation equality,
we find 
\begin{equation}
  \delta y_C ~\approx ~ 
    \frac{15 \zeta(3)\eta m_p}{2\pi^4 T_{\mathrm{MRE}}}
    \frac{\Omega_\chi\epsilon_\chi}{\Omega_B} \times 
    \begin{cases} 
      \left(\frac{\tau_\chi}{\tMRE} \right)^{1/2} & \tau_\chi < \tMRE \\
      \left(\frac{\tau_\chi}{\tMRE} \right)^{2/3} & \tau_\chi > \tMRE~,
    \end{cases}
  \label{eq:deltayCExpressions}
\end{equation} 
where $\eta$ once again denotes the baryon-to-photon ratio of the universe.

%%%%%%%%%%%%%%%%%%%%%%%%%%%%%%%%%%%%%%%%%%%%%%%%%%%%%%%%%%%%%%%%%%%%%%%%%%%%%%%%%%%%%%
\FloatBarrier
\subsection{The Fruits of Linearization:\\~
  CMB-Distortion Constraints on Ensembles of Unstable Particles \label{sec:Superpose}}
%%%%%%%%%%%%%%%%%%%%%%%%%%%%%%%%%%%%%%%%%%%%%%%%%%%%%%%%%%%%%%%%%%%%%%%%%%%%%%%%%%%%%%

\begin{table*}[t!]
\centering
\smaller
\begin{tabular}{||c|c|c|c|c|c|c||}
\hline
\hline
  ~Distortion~  & $t_{0a} (s) $ & $t_{Ba} (s) $  & 
    $t_{1a} (s) $ & $\alpha_{a} $ & 
    $A_a $ & $B_a $ 
    \\
\hline
\hline
  $\mu$-type  & $ ~3.3 \times 10^{6}~$  & $ ~2.4 \times 10^{9}~$ & 
    $ ~1.2 \times 10^{10}~$ &  $ ~1.0  \times 10^{0}~ $ &
     $ ~5.4 \times 10^{-3}~$  & $~3.1 \times 10^{-1}~ $ 
     \\
\hline
  $y_c$-type &  $ 3.3  \times 10^{6} $  & $ 6.8 \times 10^{7}$ & 
    $ 5.4 \times 10^{9}$ & $  1.2 \times 10^{0} $ & 
    $3.7 \times 10^{-5} $  & $2.7 \times 10^{-1}$  
    \\
\hline
\hline
\end{tabular}
\caption{~Values for the parameters appearing in Eqs.~(\protect\ref{eq:specfitsmu}) and (\ref{eq:specfitsy})  which
  characterize the spectral distortions $\delta\mu$ and $\delta y_C$ due to
  the electromagnetic decays of unstable particles decaying after electron-positron 
  annihilation, with $a= \{\mu,y\}$.
\label{tab:spec}}
\end{table*}

Having derived analytic approximations for the spectral distortions $\delta\mu$ and 
$\delta y_C$ due to a single decaying particle species within the uniform-decay approximation, 
we now proceed to generalize these results to the case of an arbitrary injection history.

Within the linear approximation, the overall contribution to $\delta \mu$ from a 
decaying ensemble is simply the sum of the individual contributions from the 
constituent particles $\chi_i$ with lifetimes in the range 
$t_e \lesssim \tau_i \lesssim t_{\mathrm{EC}}$.  Likewise, the overall contribution 
to $y_C$ is a sum of individual contributions from $\chi_i$ with 
$t_{\mathrm{EC}} \lesssim \tau_i \lesssim \tLS$.
However, injection
from constituents with lifetimes $\tau_i \sim t_{\mathrm{EC}}$ gives rise to  
intermediate-type distortions which are distinct from the purely
$\mu$-type or purely $y$-type distortions discussed above.  While 
these intermediate-type distortions are not merely a superposition of $\mu$-type
and $y$-type distortions~\cite{BeyondYAndMu}, isolating and constraining such 
intermediate-type distortions would require a more precise measurement of the 
CMB-photon spectrum than COBE/FIRAS data currently provide.  We therefore find
that --- at least for the time being --- the net effect of injection at times 
$t \sim t_{\mathrm{EC}}$ may indeed be modeled through such a superposition,
with $\delta \mu$ and $\delta y_C$ given by modified versions of   
Eqs.~(\ref{eq:muSol}) and~(\ref{eq:deltayCExpressions}).
The appropriate modifications of these expressions can be inferred from
the approximate forms of the Green's functions in Ref.~\cite{ChlubaGreensFns1}.  In particular, 
we model the overall contributions to $\delta \mu$ 
from the decaying
ensemble by an expression of the form
\begin{eqnarray}
  \delta \mu  &=&
    \sum_{t_{e} < \tau_i  < t_{B \mu} }\!\!\!\!\!\!\!\! 
    A_\mu \Omega_i 
    \left(\frac{\tau_i}{t_{0 \mu}}\right)^{1/2}
    e^{-(t_{0 \mu}/\tau_i)^{5/4} }\nonumber\\
  &\,&+ \!\!\!\! \sum_{t_{B \mu} < \tau_i < \tMRE} \!\!\!\!\!\!\!\!
    B_\mu \Omega_i \left(\frac{\tau_i}{t_{1 \mu}}\right)^{1/2} 
    \left[ 1 - e^{-( t_{1 \mu}/\tau_i)^{\alpha_\mu}} \right] \nonumber\\
  &\,&+\!\!\!\! \sum_{\tMRE < \tau_i < \tLS}\!\!\!\!\!\!\!\! 
    B_\mu \Omega_i 
    \left(\frac{\tau_i}{t_{2 \mu}}\right)^{2/3} 
    \left[ 1 - e^{-( t_{2 \mu}/\tau_i)^{4\alpha_\mu/3}}\right] ~.~~\nonumber\\
\label{eq:specfitsmu}
\end{eqnarray}
Likewise, for $\delta y_C$ we have
\begin{eqnarray}
  \delta y_C &=& 
    \sum_{t_{e} < \tau_i  < t_{B y} }\!\!\!\!\!\!\!\! 
    A_y \Omega_i 
    \left(\frac{\tau_i}{t_{0 y}}\right)^{1/2}
    e^{-(t_{0 y}/\tau_i)^{5/4}}
    \nonumber\\
  &\,&+ \!\!\!\! \sum_{t_{B y} < \tau_i < \tMRE}\!\!\!\!\!\!\!\! 
    B_y \Omega_i 
    \frac{(\tau_i/t_{1 y})^{1/2}} 
    {1 + \left( t_{1 y} / \tau_i \right)^{\alpha_y}}
    \nonumber\\
  &\,&+ \!\!\!\! \sum_{\tMRE < \tau_i < \tLS} \!\!\!\!\!\!\!\! 
    B_y \Omega_i 
    \frac{(\tau_i/t_{2 y})^{2/3}} 
    {1 + \left( t_{2 y} / \tau_i \right)^{4\alpha_y/3}}~.~~~~
\label{eq:specfitsy}
\end{eqnarray}
In these equations we have introduced two sets of parameters $A_\mu$, $B_\mu$, $\alpha_\mu$, 
$t_{0\mu}$, \etc, and $A_y$, $B_y$, $\alpha_{y}$, $t_{1y}$, \etc, in order to
characterize $\delta \mu$ and $\delta y_C$, respectively.  We note, however,
that the timescales $t_{2\mu} \equiv \tMRE^{1/4} t_{1\mu}^{3/4}$ and 
$t_{2y} \equiv \tMRE^{1/4} t_{1y}^{3/4}$ are not independent model parameters
but merely shorthand notation for quantities which are completely determined by $t_{1\mu}$
and $t_{1y}$, respectively.  We also note that this parametrization 
accounts not only for $\mu$-type distortions at times $t \lesssim t_{\mathrm{EC}}$ and 
$y$-type distortions at times $t \gtrsim t_{\mathrm{EC}}$, but also for subdominant 
contributions to $\delta y_C$ prior to $t_{\mathrm{EC}}$ and to $\delta \mu$ after 
$t_{\mathrm{EC}}$.  Finally, we note that in formulating Eqs.~(\ref{eq:specfitsmu}) and (\ref{eq:specfitsy}) we 
have once again implicitly taken $\epsilon_i = 1$ for all $\chi_i$, just as we did in 
Eqs.~(\ref{eq:BBNFitsTerm1})--(\ref{eq:BBNFitsLi6}).   
The corresponding expressions for $\delta \mu$ and $\delta y_C$ may nevertheless be
obtained for the case in which $\epsilon_i$ differs from unity for one or more 
of the $\chi_i$ by once again replacing $\Omega_i$ for each species with the product 
$\Omega_i \epsilon_i$. 

\begin{center}
\begin{figure*}[t]
\includegraphics[width=0.45\textwidth]{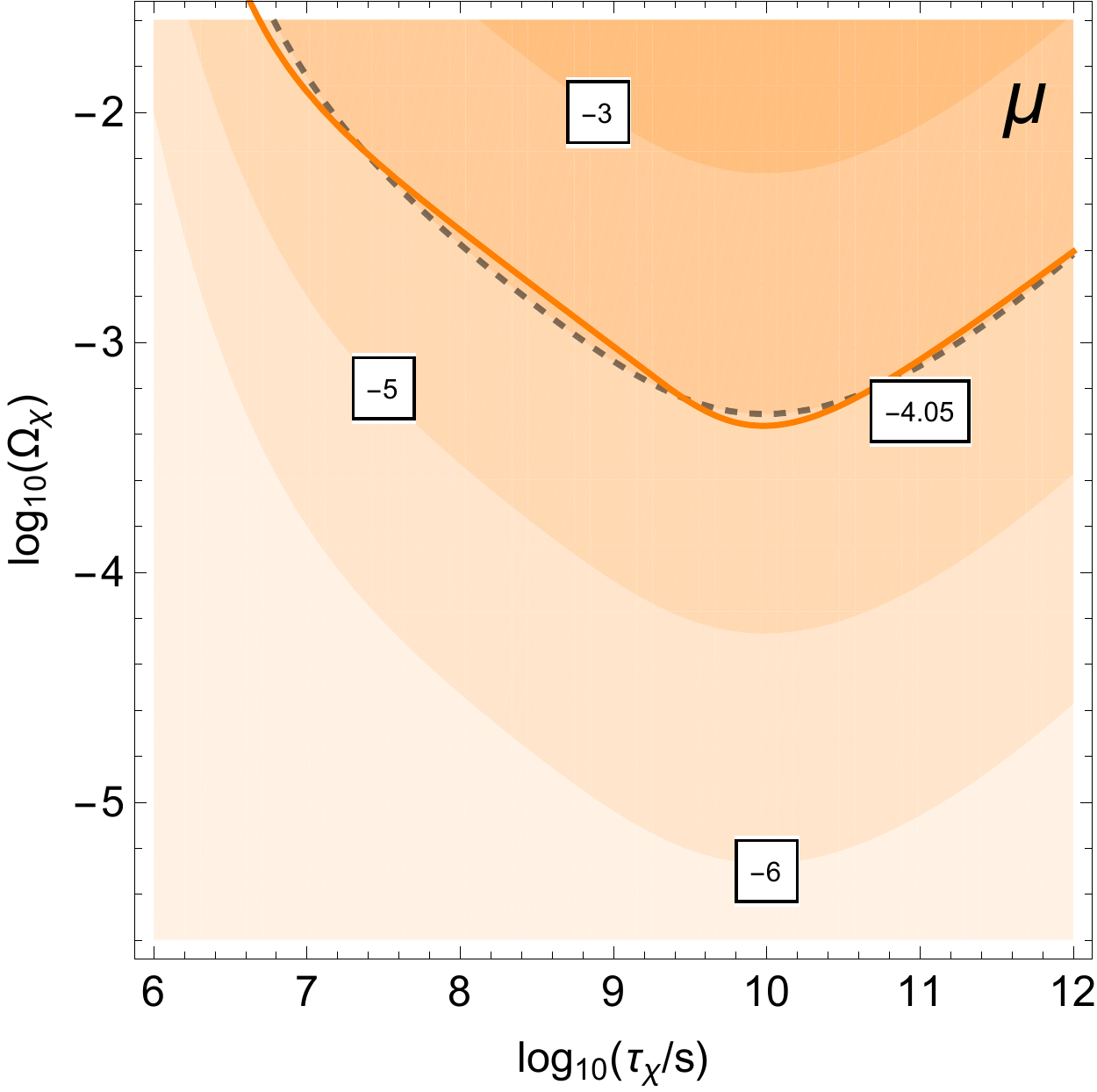}~~~
\includegraphics[width=0.45\textwidth]{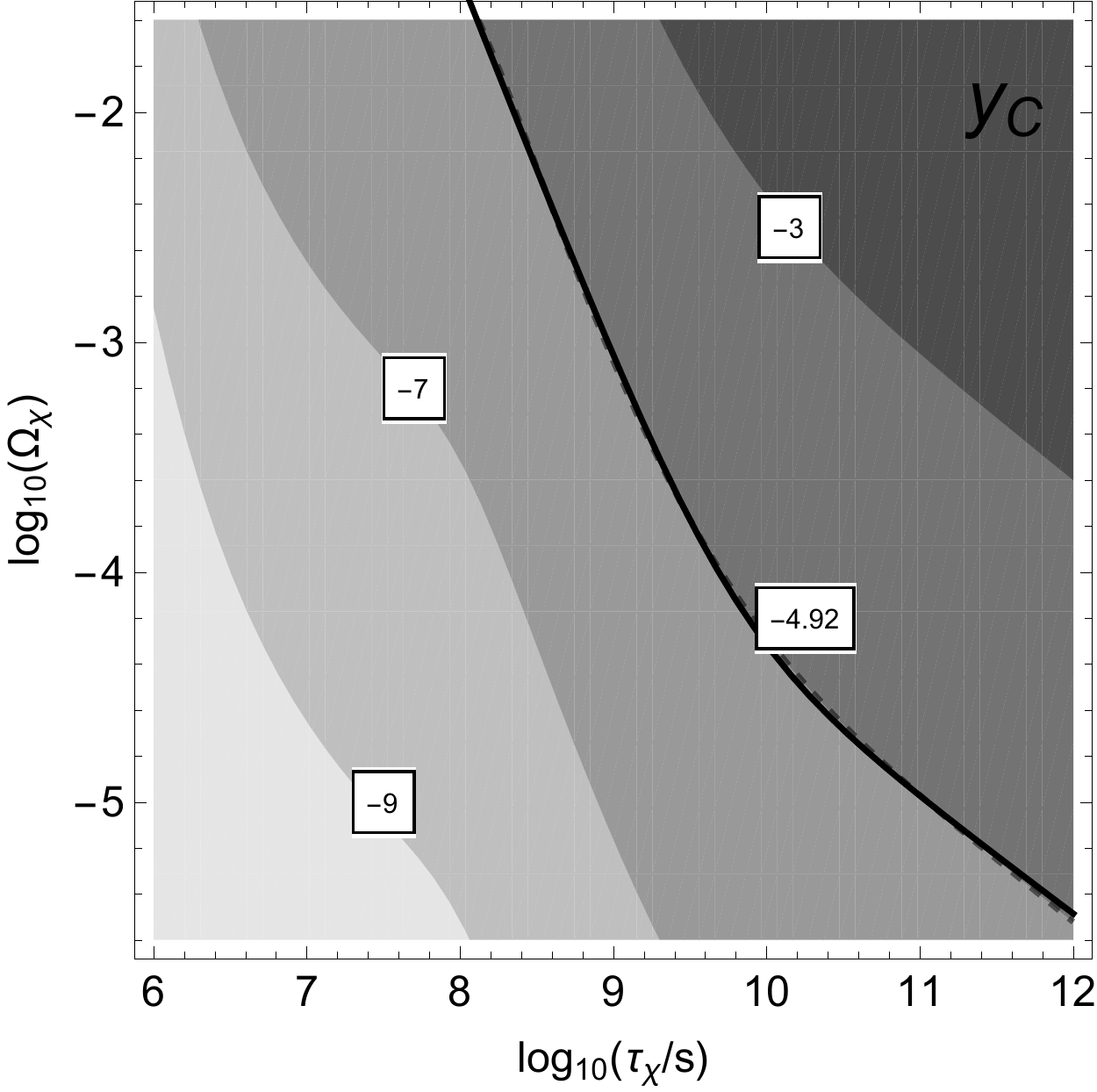}
\caption{Contours of $\log_{10}(\delta \mu)$ (left panel) and 
  $\log_{10} (\delta y_C)$ (right panel) obtained from 
  the analytic approximations in Eqs.~(\protect\ref{eq:specfitsmu}) and (\ref{eq:specfitsy}) in the presence of a 
  single electromagnetically-decaying particle with extrapolated abundance 
  $\Omega_\chi$ and lifetime $\tau_\chi$,
    plotted within the $(\Omega_\chi,\tau_\chi)$ plane.
   The gray dashed line in each panel demarcates the upper boundary of the region 
   within the $(\Omega_\chi,\tau_\chi)$ plane which is consistent with the observed limits 
   quoted in Eq.~(\ref{eq:muandyClimits}),  
    while the solid curve represents 
  the constraint contour obtained by fitting the corresponding analytic approximation 
  in Eqs.~(\protect\ref{eq:specfitsmu}) and (\ref{eq:specfitsy}) to these numerical results.
  \label{fig:MuAndYcConstraints}}
\end{figure*}
\end{center}

We now determine the values of the parameters in Eqs.~(\ref{eq:specfitsmu}) and (\ref{eq:specfitsy}) in 
a manner similar to that in which we determined the values of the parameters in 
Eqs.~(\ref{eq:BBNFitsTerm1})--(\ref{eq:BBNFitsLi6}). 
Specifically, we demand that our analytic 
approximations yield constraint contours consistent with those obtained from 
numerical analysis in the case in which the ensemble comprises a single decaying
particle with lifetime $\tau_\chi$ and abundance $\Omega_\chi$.
In particular, we survey $(\Omega_\chi, \tau_\chi)$ space and compute 
the generic spectral distortions numerically at each point using the Green's-function formalism of 
Refs.~\cite{ChlubaGreensFns1,ChlubaGreensFns2}.  
The generic spectra are then fit by a superposition of a $\mu$-type and a $y$-type distortion in order to determine a corresponding pair of values for $\delta \mu$ and $\delta y_C$.
The undetermined parameters in 
Eqs.~(\ref{eq:specfitsmu}) and (\ref{eq:specfitsy}) 
are then chosen such that our analytic expressions for
$\delta \mu$ and $\delta y_C$ provide good fits to the corresponding constraint 
contours.  Our best-fit values for all of these parameters are quoted in 
Table~\ref{tab:spec}.

In Fig.~\ref{fig:MuAndYcConstraints}, 
we display contours showing the 
results of our numerical calculations for $\delta \mu$ (left panel) and 
$\delta y_C$ (right panel) 
within the $(\Omega_\chi,\tau_\chi)$ plane.
The gray dashed line in each panel demarcates the upper boundary of the region 
within the $(\Omega_\chi,\tau_\chi)$ plane which is consistent with the limits on these
distortions quoted in Eq.~(\ref{eq:muandyClimits}).  
By contrast, the solid curve in each panel represents the corresponding 
constraint contour obtained through our analytic approximation in 
Eqs.~(\ref{eq:specfitsmu}) and (\ref{eq:specfitsy}).
As we see,
the results obtained through our 
analytic approximation represent a good fit to our numerical results for 
both $\delta \mu$ and $\delta y_C$.

The constraint contours obtained 
from our analytic approximations for both $\mu$- and $y$-type distortions are also 
included in Fig.~\ref{fig:comb} in order to facilitate comparison with the
constraint contours associated with the primordial abundances of light nuclei.
We note that the constraint on $\delta \mu$ is more stringent 
than the constraint on $y_C$ for $\tau_\chi \lesssim t_{\mathrm{EC}}$, as expected, while the 
opposite is true for $\tau_\chi \gtrsim t_{\mathrm{EC}}$.
We also see that 
both of these constraints 
are subdominant in comparison with the constraints on the primordial abundances of 
light nuclei,
except at late times $\tau_\chi \gtrsim 10^{10}$~s.

%%%%%%%%%%%%%%%%%%%%%%%%%%%%%%%%%%%%%%%%%%%%%%%%%%%%%%%%%%%%%%%%%%%%%%%%%%%%%%%%%%%%%%

\FloatBarrier
\section{Modifications to the Ionization History of the Universe\label{sec:Ionization}}

%%%%%%%%%%%%%%%%%%%%%%%%%%%%%%%%%%%%%%%%%%%%%%%%%%%%%%%%%%%%%%%%%%%%%%%%%%%%%%%%%%%%%%

As we have seen in Sect.~\ref{sec:CMBDistortions},
electromagnetic injection from late-decaying particles can give rise to 
certain distortions in the CMB-photon spectrum.
However, such injections can affect the CMB in other ways as well.  In particular, when such injection 
occurs during recombination, the resulting photons and electrons contribute to the 
heating and ionization of neutral hydrogen and helium as they cool, thereby expanding 
the surface of last scattering.  This in turn leads to alterations
in the pattern of CMB anisotropies, including  
a damping of correlations between temperature fluctuations as 
well as an enhancement of correlations
between polarization fluctuations at low multipole moments~\cite{ChenKamionkowski}.
These considerations therefore place additional constraints on decaying particle
ensembles.

In a nutshell, these effects arise because 
the rates for many of the Class-I processes discussed in 
Sect.~\ref{sec:InjectionConstraints} are still sizable at times $t\sim \tLS$. 
These processes therefore continue to rapidly redistribute  the energies of photons 
injected during the recombination epoch to lower energies, thereby allowing for 
efficient photoionization of neutral hydrogen and helium at a rate that exceeds 
the rate of cosmic expansion.  However, there is only a somewhat narrow time window around $\tLS$ during 
which an injection of photons can have a significant impact on the surface of last 
scattering.   At times $t \ll \tLS$, any additional ionization of 
particles in the plasma is effectively washed out.  By contrast, at times $t \gg \tLS$,
the relevant Class-I processes effectively shut off and photoionization is suppressed.  

To a good approximation, then, we may regard any modification of the ionization history
of the universe as being due to injection at $t\sim \tLS$.  Thus, the crucial 
quantity which is constrained is the overall injection rate of energy density in the form 
of photons or other electromagnetically-interacting particles at $\tLS$.  Since the
universe is already matter-dominated by the recombination epoch, the injection rate
associated with a single decaying particle species $\chi$ with lifetime $\tau_\chi$ and 
extrapolated abundance $\Omega_\chi$ at $\tLS$ is given by 
\begin{equation}
  \left.\frac{d\rho_\gamma}{dt}\right|_{t=\tLS} ~=~ \rhocrit(\tnow)
    \left(\frac{\tnow}{\tLS}\right)^2
    \epsilon_\chi \frac{\Omega_\chi}{\tau_\chi} e^{-\tLS/\tau_\chi}~, 
\end{equation} 
where $\epsilon_\chi$ once again denotes the fraction of the energy density released by 
$\chi$ decays which is transferred to photons.  Thus, we model the 
constraint on such a decaying particle from its effects on the ionization 
history of the universe by an inequality of the form 
\begin{equation}
  \epsilon_\chi \frac{\Omega_\chi}{\tau_\chi}e^{-t_{\mathrm{IH}}/\tau_\chi} ~\lesssim~ 
    \Gamma_{\mathrm{IH}}~,~~
  \label{eq:IonizationConstraintSinglePc}
\end{equation}
where $t_{\mathrm{IH}}$ and $\Gamma_{\mathrm{IH}}$ are taken to be free parameters.

We next determine the values of $\Gamma_{\mathrm{IH}}$ and $t_{\mathrm{IH}}$ 
by fitting these parameters to the constraint contours obtained in 
Refs.~\cite{SlatyerInjectionHistory,SlatyerWuIonizationUpdate} from a numerical
analysis of the injection and deposition dynamics.  The constraint contours obtained 
from subsequent numerical studies by other authors~\cite{PoulinIonization} are quite
similar.  In particular, we find 
\begin{eqnarray}
  \Gamma_{\mathrm{IH}} &\approx & 6.2 \times 10^{-26}\mbox{~s}^{-1} \nonumber \\
  t_{\mathrm{IH}} &\approx & 1.4\times 10^{13}\mbox{~s}~.
  \label{eq:GammaIHtIH}
\end{eqnarray} 
Note that recent results from the final Planck data release~\cite{PlanckFinal} will affect
the numerical values for $\Gamma_{\mathrm{IH}}$ and $t_{\mathrm{IH}}$ slightly
but not dramatically.    

\begin{center}
\begin{figure}[t]
\includegraphics[width=0.45\textwidth]{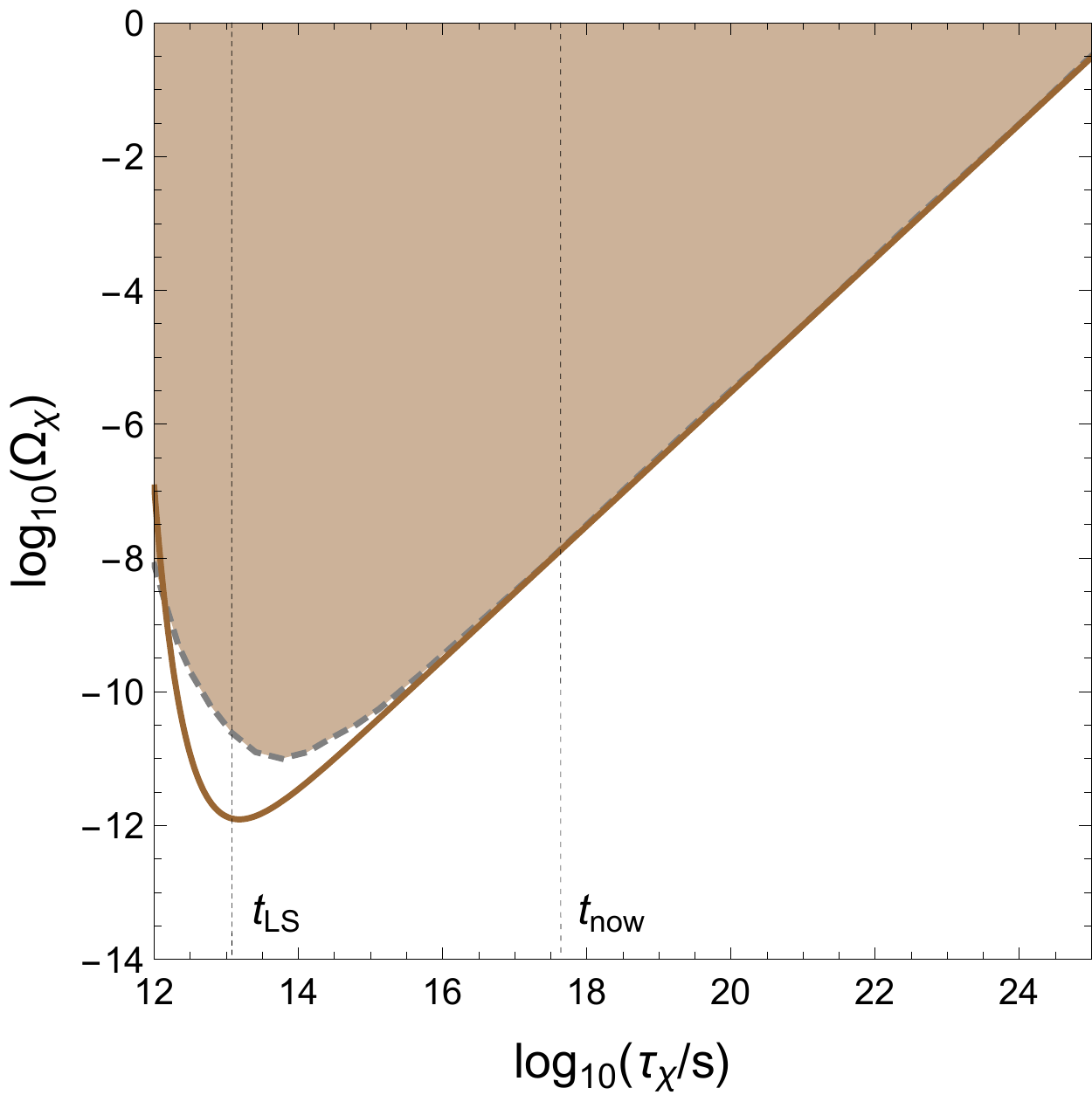}~~~
\caption{Constraints on a decaying particle with lifetime $\tau_\chi$ and 
  extrapolated abundance $\Omega_\chi$ from modifications of the ionization history of 
  the universe.  These constraints correspond to the case in which $\epsilon_\chi = 1$
  and the entirety of the energy density released by $\chi$ decays is transferred to 
  photons.  The gray dashed contour represents the bound obtained in 
  Refs.~\cite{SlatyerInjectionHistory,SlatyerWuIonizationUpdate} from numerical analysis.  
  The brown shaded region above the contour is excluded.  The solid curve represents the 
  constraint contour obtained by fitting the corresponding analytic approximation in 
  Eq.~(\protect\ref{eq:IonizationConstraintSinglePc}) to these results.  Dashed vertical 
  lines indicating the time $\tLS$ of last scattering and the present age of the universe 
  $\tnow$ have also been included for reference. 
  We see that 
  our approximation for the constraint contour is in good agreement with 
  these numerical results for $\tau_\chi \gtrsim 10^{15}$~s 
  and deviates significantly from this contour only when $\tau_\chi \sim \tLS$.
  \label{fig:IonizationFit}}
\end{figure}
\end{center}

In Fig.~\ref{fig:IonizationFit}, we 
compare the results of our analytic approximation with the numerical results of
Refs.~\cite{SlatyerInjectionHistory,SlatyerWuIonizationUpdate}.  We observe that
our approximation for the constraint contour is in good agreement with 
these numerical results for $\tau_\chi \gtrsim 10^{15}$~s 
and deviates significantly from this contour only when $\tau_\chi \sim \tLS$.
However, as discussed in Ref.~\cite{SlatyerInjectionHistory}, the results to which 
we perform our fit yield an artificially conservative bound on $\Omega_\chi$ 
for lifetimes $\tau_\chi \sim \tLS$, as the production of additional 
Lyman-$\alpha$ photons due to excitations from the ionizing particles was 
intentionally neglected.  Other analyses of the ionization 
history~\cite{ZhangChenKamionkowski} which include this effect yield
constraint contours which more closely resemble that obtained from our
analytic approximation.

The constraint in Eq.~(\ref{eq:IonizationConstraintSinglePc}) may be generalized to an 
ensemble of decaying particles in a straightforward manner.  Indeed, since this constraint 
is ultimately a bound on the overall injection rate at $t\sim \tLS$, the corresponding 
constraint on an ensemble of decaying particles $\chi_i$ takes the form
\begin{equation}
  \sum_i \epsilon_i \frac{\Omega_i}{\tau_i} e^{-t_{\mathrm{IH}}/\tau_i} 
    ~\lesssim~  \Gamma_{\mathrm{IH}}
  \label{eq:IonizationConstraint}
\end{equation}
with values of $\Gamma_{\mathrm{IH}}$ and $t_{\mathrm{IH}}$ once again given in 
Eq.~(\ref{eq:GammaIHtIH}).

We emphasize that the constraint in Eq.~(\ref{eq:IonizationConstraint}) stems solely
from the effect of electromagnetic injection from the decays of the ensemble constituents 
on the surface of last scattering.  Since this effect arises primarily from injection at 
$t \sim \tLS$, the constraints are most relevant for ensembles in which the constituents
have lifetimes roughly around this timescale.
However, electromagnetic injection can affect the ionization 
history of the universe in other ways as well.  For example, energy deposited in the 
intergalactic medium as a result of particle decays at timescales well after $\tLS$ serves 
both to elevate the gas temperature and to increase the ionization fraction of hydrogen.  
These effects can shift the onset of reionization to earlier 
times~\cite{SlatyerWuIonizationUpdate,SlatyerArbitraryInjections} and impact the 
21-cm absorption/emission signal arising from hyperfine transitions in neutral 
hydrogen~\cite{ClarkDutta,Liu21cm,Mitridate21cm}.  
These considerations are relevant primarily for 
particles with lifetimes $\tau_i \gg \tnow$ --- particles which would in principle 
contribute to the present-day dark-matter abundance.  Since our primary focus in
this paper is on ensembles whose constituents have lifetimes $\tau_i \lesssim \tnow$,
we do not investigate these additional considerations here.  However, we note that 
they can play an important role in constraining ensembles of particles with longer
lifetimes, such as those which arise within the context of the DDM framework.

%%%%%%%%%%%%%%%%%%%%%%%%%%%%%%%%%%%%%%%%%%%%%%%%%%%%%%%%%%%%%%%%%%%%%%%%%%%%%%%%%%%%%%

\FloatBarrier
\section{Contributions to the Diffuse Photon Background\label{sec:DiffusePhoton}}

%%%%%%%%%%%%%%%%%%%%%%%%%%%%%%%%%%%%%%%%%%%%%%%%%%%%%%%%%%%%%%%%%%%%%%%%%%%%%%%%%%%%%%

At times $t \gg \tLS$, the universe becomes transparent to photons with a 
broad range of energies $\mathcal{O}(\mathrm{keV}) \lesssim E_\gamma \lesssim 
\mathcal{O}(\mathrm{TeV})$~\cite{ChenKamionkowski}.   
Photons within this energy range which are injected on timescales $t \gg \tLS$ are 
not reprocessed by Class-I processes and do not interact with 
CMB photons at an appreciable rate.  Thus, as discussed in 
Sect.~\ref{sec:CMBDistortions}, such photons therefore simply free-stream until 
the present epoch and contribute to the diffuse photon background.  The total diffuse
background of photons observed from the location of Earth includes both a galactic 
contribution from processes occurring within the Milky-Way halo (at redshifts $z\approx 0$) 
and an extra-galactic contribution from processes which occurred within the more distant
past.  The decays of unstable particles with lifetimes $\tau_i \ll \tnow$ contribute 
essentially exclusively to the extra-galactic component of this total background.  By 
contrast, particles with lifetimes $\tau_i \gtrsim \tnow$ --- particles which would 
contribute non-negligibly to the present-day abundance of dark matter --- in 
principle contribute to both components.  Since our primary focus in this paper
is on particles with lifetimes in the regime $\tau_i \ll \tnow$, we focus 
on the extra-galactic component of the diffuse photon background.  
However, we also note that since the local dark-matter density within the Milky-Way halo 
is several orders of magnitude larger than the corresponding cosmological density,
the galactic component of the diffuse photon background is typically the more 
important of the two for constraining particles with lifetimes $\tau_i \gtrsim \tnow$. 

The overall diffuse extra-galactic contribution $d\Phi/dE_\gamma$ to the 
differential flux of photons per unit energy observed by 
a detector on Earth from the decays of an ensemble of unstable constituent particles
$\chi_i$ is simply the direct sum of the individual contributions $d\Phi_i/dE_\gamma$ from 
these constituents.  In order to evaluate these individual contributions, we begin by noting 
that the decays of a given species $\chi_i$, averaged over all possible decay channels, 
produce a characteristic injection spectrum of photons $d\mathcal{N}_i/dE_\gamma$.
The spectrum of photons arriving at present time at a detector on Earth is the
integrated total contribution from decays occurring at different redshifts.
A photon injected at redshift $z$ with energy $E_\gamma'$ is redshifted to 
energy $E_\gamma = (1+z)^{-1}E_\gamma'$ when it arrives at the location of Earth.
Thus, we find that the observed flux of photons observed at a detector on Earth
from the decay of an individual decaying species $\chi_i$ may be written in the form  
\begin{eqnarray}
  \frac{d\Phi_i}{dE_\gamma} &=& \frac{1}{4\pi} \int d\Omega
    \int_0^{\infty} dz  \, 
\Bigg\{ \frac{\rho_i(z)}{(1+z)^3} \frac{\Gamma_i}{m_i}\,
    \frac{d\ell}{dz} e^{-\kappa(z)} \nonumber \\ & & ~\times \int dE_\gamma'
    \delta \left[E_\gamma' - (1+z)E_\gamma\right]
    \frac{d\mathcal{N}_i}{dE_\gamma'}  \Bigg\}~,~~~~~~
  \label{eq:dPhiidEgammaRaw}   
\end{eqnarray} 
where $d\Omega$ is the differential solid angle of the detector, where $\rho_i(z)$
is the energy density of $\chi_i$ at redshift $z$, where $\kappa(z)$ is the 
optical depth of the universe to photons emitted at redshift $z$, and where $\ell(z)$ is 
the line-of-sight distance away from Earth, expressed as a function of $z$.

We focus here primarily on photons with energies in the range 
$\mathcal{O}(\mathrm{keV}) \lesssim E_\gamma \lesssim \mathcal{O}(\mathrm{TeV})$, for which
the universe is effectively transparent at all times $t \gtrsim \tLS$ and opaque at 
times $t \lesssim \tLS$.  Moreover, for simplicity, we shall approximate the optical 
depth for such photons with a function of the form
\begin{equation}
  \kappa(z) ~\approx~ 
  \begin{cases}
    0 & z < z_{\mathrm{LS}} \\ 
    \infty & z > z_{\mathrm{LS}}~,   
  \end{cases}  
\end{equation}
where $z_{\mathrm{LS}} \equiv z(\tLS) \approx  1100$ is the redshift at the time of last 
scattering.  In other words, we shall assume that the universe is completely 
transparent to all photons within this energy range after last scattering and 
completely opaque beforehand.  Moreover, since $\tLS \gg \tMRE$, we approximate the 
line-of-sight distance $\ell(z)$ in a flat universe dominated by matter and dark energy as 
\begin{equation}
  \ell(z) ~=~ \frac{c}{H_{\mathrm{now}}} \int_0^z \frac{dz}{\mathcal{F}^{1/2}(z)}~,  
\end{equation}
where $c$ is the speed of light, where $H_{\mathrm{now}}$ is the present-day value
of the Hubble parameter, and where we have defined
\begin{equation}
  \mathcal{F}(z) ~\equiv~ \Omega_m(1+z)^3 + \Omega_\Lambda~,
\end{equation}
where $\Omega_m$ and $\Omega_\Lambda$ respectively denote the present-day abundances 
of massive matter and dark energy,
the former including 
contributions from both dark and baryonic matter.
The energy density $\rho_i(z)$ of a decaying particle can be expressed in terms of 
its extrapolated abundance $\Omega_i$ as
\begin{equation}
  \rho_i(z) ~=~ \Omega_i \rho_{\mathrm{crit}}(\tnow)(1+z)^3e^{-\Gamma_i t(z)}~,
\end{equation}  
where the time $t(z)$ which corresponds to redshift $z$ is 
\begin{equation}
  t(z) ~=~ \frac{1}{H_{\mathrm{now}}}\int_0^z\frac{dz}{(1+z)\mathcal{F}^{1/2}(z)}~.
\end{equation} 
With these substitutions, Eq.~(\ref{eq:dPhiidEgammaRaw}) becomes
\begin{eqnarray}
  \frac{d\Phi_i}{dE_\gamma} &=& \frac{c\rho_{\mathrm{crit}}(\tnow)\Omega_i\Gamma_i}
    {4\pi H_{\mathrm{now}} m_i} \! \int\! d\Omega \!
    \int_0^{z_{\mathrm{LS}}}\!\! dz  \, \Bigg[ \frac{1}{\mathcal{F}^{1/2}(z)}
      \nonumber  \\ & & ~~~~~\times \,
         e^{-\Gamma_i t(z)}  
    \left.\frac{d\mathcal{N}_i}{dE_\gamma'}\right|_{E_\gamma' = (1+z)E_\gamma} 
      \Bigg]~.~~~  
  \label{eq:dPhiidEgamma}
\end{eqnarray} 

Finally, in assessing our prediction for $d\Phi/dE_\gamma$ for a decaying particle
ensemble, we must account for the fact that at a realistic detector, the photon energies 
recorded by the instrument differ from the actual energies of the incoming
photons due to the non-zero energy resolution of the detector.  We account for this 
effect by convolving the spectrum of incoming photons with a smearing function
$R_\epsilon(E_\gamma - E_\gamma')$, which represents the probability for 
the detector to register a photon energy $E_\gamma$, given an actual incoming 
photon energy $E_\gamma'$.
For concreteness, we consider a Gaussian smearing function of the form
\begin{eqnarray}
  R_\epsilon ( E_\gamma - E_\gamma') ~=~ \frac{1}{\sqrt{2 \pi}\epsilon E_\gamma'} 
    \exp\left[-\frac{(E_\gamma - E_\gamma')^2}{2 \epsilon^2 E_\gamma'^2 }\right]~,
  \label{eq:PhotonDetectorSmearing}~~~
\end{eqnarray}
where $\epsilon$ is a dimensionless parameter which sets the overall, energy-dependent
standard deviation $\sigma(E_\gamma') = \epsilon E_\gamma'$ of the distribution.  
Thus, we have our final result for the extra-galactic photon flux:
\begin{equation}
  \frac{d\Phi}{dE_\gamma} ~=~ \sum_i \int dE_\gamma' 
    \frac{d\Phi_i}{dE_\gamma'} R_\epsilon(E_\gamma - E_\gamma')~,
  \label{dPhidEgammaTot}
\end{equation}
with the individual fluxes $d\Phi_i/dE_\gamma'$ given by Eq.~(\ref{eq:dPhiidEgamma})
and with $R_\epsilon (E_\gamma - E_\gamma')$ given by Eq.~(\ref{eq:PhotonDetectorSmearing}).

Limits on additional contributions to the extra-galactic photon 
flux at energies $E_\gamma \gtrsim 800$~keV can be derived from measurements of 
the total gamma-ray flux by instruments such as COMPTEL, EGRET, and Fermi-LAT.~
For photon energies in the range $800\mbox{~keV} \lesssim E_\gamma \lesssim 30\mbox{~MeV}$, 
COMPTEL data on the diffuse extra-galactic background spectrum~\cite{COMPTELExGalPhotonBG}  
are well modeled by a power-law expression of the form   
\begin{eqnarray}
  \frac{d\Phi}{dE_\gamma d\Omega} &\approx & (49.9 \times 10^{-3}) \,
    \left(\frac{E_\gamma}{\mathrm{MeV}}\right)^{-2.40} \nonumber \\ & &
    ~~~~~~ \mbox{~MeV}^{-1}\mathrm{cm}^{-1}\mathrm{s}^{-1}\mathrm{sr}^{-1}~.
  \label{eq:COMPTELDiffFluxBound} 
\end{eqnarray}
Likewise, for energies in the range 
$30\mbox{~MeV} \lesssim E_\gamma \lesssim 1.41\mbox{~GeV}$,
EGRET data~\cite{EGRETExGalPhotonBG} are well modeled by the power-law expression
\begin{eqnarray}
  \frac{d\Phi}{dE_\gamma d\Omega} &\approx & (7.35 \times 10^{-3}) \,
    \left(\frac{E_\gamma}{\mathrm{MeV}}\right)^{-2.35} \nonumber \\ & &
    ~~~~~~ \mbox{~MeV}^{-1}\mathrm{cm}^{-1}\mathrm{s}^{-1}\mathrm{sr}^{-1}~.
\end{eqnarray}
At even higher energies, the spectrum of the extra-galactic diffuse photon background
can be inferred from Fermi-LAT data.  Within the energy range
$100\mbox{~MeV}\lesssim E_\gamma \lesssim 820\mbox{~GeV}$, these data are well 
described by a power-law expression with an exponential suppression at high 
energies~\cite{FermiExGalPhotonBG}:
\begin{eqnarray}
  \frac{d\Phi}{dE_\gamma d\Omega} &\approx & (4.15 \times 10^{-3}) \, 
    \left(\frac{E_\gamma}{\mathrm{MeV}}\right)^{\! -2.32} \!\!\!\!
    \exp\!\left(\frac{-E_\gamma}{279\mbox{~GeV}}\right) \nonumber \\ & &     
    ~~~~~~~~~~~~~ \mbox{~MeV}^{-1}\mathrm{cm}^{-1}\mathrm{s}^{-1}\mathrm{sr}^{-1}~.
  \label{eq:FermiDiffFluxBound}
\end{eqnarray}
Finally, we note that while we focus here primarily on photons with energies 
above $800$~keV, estimates of the extra-galactic photon background at lower energies 
have been computed from BAT~\cite{BATExGalPhotonBG} and 
INTEGRAL~\cite{INTEGRALExGalPhotonBG} data and from observations of 
Type-I supernovae~\cite{SupernovaExGalPhotonBG}.

The expression in Eq.~(\ref{dPhidEgammaTot}) is completely general and 
applicable to any ensemble of particles which decays to photons with energies 
within the transparency window
$\mathcal{O}(\mathrm{keV}) \lesssim E_\gamma \lesssim \mathcal{O}(\mathrm{TeV})$.
To illustrate how this expression can be evaluated in practice, we
consider a concrete example involving 
a specific set of injection spectra $d\mathcal{N}_i/dE_\gamma$ for the ensemble 
constituents.  In particular, we consider a scenario in which each constituent $\chi_i$ decays with a 
branching fraction of effectively unity to a pair of photons via the process 
$\chi_i \rightarrow \gamma + \gamma$.  Under the assumption that the $\chi_i$
are ``cold'' (\ie, non-relativistic) by the time $\tLS$, each of these photons has 
energy $E_\gamma \approx m_i/2$ in the comoving frame at the moment of injection.  
We thus have
\begin{equation}
  \frac{d{\cal N}_i}{dE_\gamma} ~\approx~ 
    2\delta\left(E_\gamma - \frac{m_i}{2}\right)~.
  \label{eq:ExampleDiphotonSpectrum}
\end{equation}
While higher-order processes will generically also give rise to a continuum spectrum 
which peaks around $E_\gamma \sim \mathcal{O}(10^{-3} m_i)$, we focus here on the
contribution from the photons with $E_\gamma \approx m_i/2$, since
line-like features arising from monochromatic photon emission are far easier to 
detect than continuum features.
Thus, for the injection spectrum in Eq.~(\ref{eq:ExampleDiphotonSpectrum}), the 
total contribution to the differential extra-galactic diffuse photon flux 
from the decays of the constituents within this ensemble
is given by
\begin{eqnarray}
  \frac{d\Phi_i}{dE_\gamma} &\approx & 
  \frac{c\rho_{\mathrm{crit}}(\tnow)}
    {2\pi H_{\mathrm{now}}} \int\! d\Omega 
    \int dE_\gamma' \Bigg\{ \frac{R_\epsilon(E_\gamma - E_\gamma')}{E_\gamma'} 
    \nonumber \\ & &
  \times \sum_i \frac{\Omega_i\Gamma_i}{m_i} 
      \mathcal{F}^{-1/2}\!\left(\frac{m_i}{2E_\gamma'} - 1\right)
      \nonumber \\ & & \times \,
         \exp\left[-\Gamma_i t\left(\frac{m_i}{2E_\gamma'} - 1\right)\right]  
      \Bigg\}~.  
  \label{eq:dPhiidEgammaSubbed}
\end{eqnarray}

%%%%%%%%%%%%%%%%%%%%%%%%%%%%%%%%%%%%%%%%%%%%%%%%%%%%%%%%%%%%%%%%%%%%%%%%%%%%%%%%%%%%%%

\FloatBarrier
\section{Applications: ~Two Examples\label{sec:MultiCompResults}}

%%%%%%%%%%%%%%%%%%%%%%%%%%%%%%%%%%%%%%%%%%%%%%%%%%%%%%%%%%%%%%%%%%%%%%%%%%%%%%%%%%%%%%

In Sects.~\ref{sec:LightElemAbundances}--\ref{sec:DiffusePhoton}, we 
examined the observational limits on electromagnetic injection from particle 
decays after the BBN epoch and formulated a set of constraints which
can be broadly applied to any generic ensemble of unstable particles.
In this section, we examine the implications of these constraints by applying
them to a pair of toy ensembles which 
exhibit 
a variety of scaling behaviors for the 
constituent masses $m_i$, extrapolated 
abundances $\Omega_i$, and lifetimes $\tau_i$.
For simplicity, we shall focus primarily on 
the ``early-universe'' regime in which $1\mbox{~s} \ll \tau_i \lesssim \tLS$ for all of the $\chi_i$.  
Within this regime, the leading constraints on injection are those from 
modifications of the abundances of light nuclei and from distortions in
the CMB.   

We consider two different classes of mass spectra for our decaying ensemble constituents.
The first consists of spectra in which the masses of these decaying particles are evenly 
distributed on a linear scale according to the scaling relation
\begin{equation}
  m_k ~=~ m_0 + k\, \Delta m~,
  \label{eq:linearmassscaling}
\end{equation}
where $k=0,1,...,N-1$ and 
where $m_0$ and $\Delta m$ are taken to be free parameters
which describe the mass 
of the lightest ensemble constituent and 
the subsequent mass splittings respectively. 
Mass spectra of this approximate linear form
are realized, for example, for the KK excitations of a particle propagating in 
a flat extra spacetime dimension whose total length is small compared with $1/m_0$. 
The second class of mass 
spectra we consider are spectra in which the particle masses are evenly distributed on a 
logarithmic scale according to the scaling relation     
\begin{equation}
  m_k ~=~ m_0 \,\xi^k~,
  \label{eq:logmassscaling}
\end{equation}
where $m_0$ once again denotes the mass of the lightest ensemble constituent, and
where $\xi$ is a dimensionless free parameter.  A mass spectrum of this 
sort is expected, for example, for axion-like particles in axiverse
constructions~\cite{Axiverse}.  

A variety of scaling relations for the decay widths $\Gamma_i$ of 
the $\chi_i$ across the ensemble can likewise be realized in different scenarios.
For simplicity, in this paper we focus on the case in which each $\chi_i$
decays with a branching fraction of effectively
unity to a pair of photons via the process $\chi_i\rightarrow \gamma +\gamma$.  
Moreover, we shall assume that the scaling relation for $\Gamma_i$ takes the form
\begin{equation}
  \Gamma_i ~=~ C_i \frac{m_i^3}{\Lambda^2}~,
  \label{eq:GammaScaling}
\end{equation}
where $\Lambda$ is a parameter with dimensions of mass and where $C_i$ is a 
dimensionless coefficient --- a coefficient which in principle can also have a
non-trivial dependence on $m_i$.  Such a scaling relation arises naturally,
for example, in situations in which each of the $\chi_i$ decays via a  
non-renormalizable Lagrangian operator of dimension $d=5$ in an effective theory 
for which $\Lambda$ is the cutoff scale.  For purposes of illustration, we focus on 
the case in which the coefficient $C_i = C_\chi$ is identical for all $C_i$.  In
this case, the $\Gamma_i$ are specified by a single free parameter: the ratio 
$M_\ast \equiv \Lambda/\sqrt{C_\chi}$.  We emphasize that it is possible --- and 
indeed in most situations even expected --- that $C_\chi \ll 1$.  Thus, unlike 
$\Lambda$ itself, the parameter $M_\ast$ may exceed the reduced Planck mass $M_P$
without necessarily rendering the theory inconsistent.  However, we remark that for 
values of $M_\ast$ in this regime, the Weak Gravity 
Conjecture~\cite{WeakGravityConjecture} imposes restrictions on the manner in 
which the decay operator in Eq.~(\ref{eq:GammaScaling}) can arise. 
 
Finally, we consider a range of different possible scaling relations for the
extrapolated  abundances $\Omega_i$ across the ensemble.  In particular, we 
consider a family of scaling relations for the $\Omega_i$ of the form
\begin{equation}
  \Omega_i ~=~ \Omega_0 \left(\frac{m_i}{m_0}\right)^\gamma~,
  \label{eq:AbundanceScaling}
\end{equation}
in which the abundance $\Omega_0$ of the lightest ensemble constituent and the 
scaling exponent $\gamma$ are both taken to be free parameters.  In
principle, $\gamma$ can be either positive or negative.  However, since 
$\tau_i \propto m_i^{-3}$ and since the leading constraints on injection at 
timescales $\tinj \lesssim \tLS$ are typically less stringent for earlier $\tinj$,  
the most interesting regime turns out to be that in which $\gamma \geq 0$ and the
initial energy density associated with each $\chi_i$ is either uniform across the
ensemble or decreases with increasing $\tau_i$. 
 
In summary, then, each of our toy ensembles is characterized by a set of six parameters:
$\{N,m_0,\Delta m,M_\ast,\Omega_0,\gamma\}$ for the mass spectrum in 
Eq.~(\ref{eq:linearmassscaling}) and $\{N,m_0,\xi,M_\ast,\Omega_0,\gamma\}$ for the 
mass spectrum in Eq.~(\ref{eq:logmassscaling}).  Note that for any particular choice of these 
parameters, the lifetimes of the individual ensemble constituents span a range 
$\tau_{N-1} \leq \tau_i \leq \tau_0$.

%%%%%%%%%%%%%%%%%%%%%%%%%%%%%%%%%%%%%%%%%%%%%%%%%%%%%%%%%%%%%%%%%%%%%%%%%%%%%%%%%%%%%%
\FloatBarrier
\subsection{Results for Uniform Mass Splittings\label{sec:LinearScalingEnsembles}}
%%%%%%%%%%%%%%%%%%%%%%%%%%%%%%%%%%%%%%%%%%%%%%%%%%%%%%%%%%%%%%%%%%%%%%%%%%%%%%%%%%%%%%

We begin by examining the results for a toy ensemble with a mass spectrum given
by Eq.~(\ref{eq:linearmassscaling}).  For such a mass spectrum, the density $n_m$ of states 
per unit mass is uniform across the ensemble; however, it is also useful to consider
the density $n_\tau$ of states per unit lifetime, which is more directly
related to the injection history.  In the continuum limit in which the difference 
between the lifetimes of sequential states $\chi_i$ and $\chi_{i+1}$ in the ensemble 
is small and the lifetime $\tau$ may be approximated as a continuous variable, we find 
that
\begin{equation}
  n_\tau ~=~ \frac{M_\ast^{2/3}}{3\,\Delta m\,\tau^{4/3}}~.
\end{equation}
The density of states per unit lifetime is therefore greatest within the most 
unstable regions of the ensemble.  Moreover, we also note that the density of 
states per unit $\ln(\tau)$ also decreases with $\ln(\tau)$.  These considerations 
turn out to have important implications for the bounds on such ensembles, as shall 
become apparent below.   

\begin{figure*}[t]
\begin{center}
\includegraphics*[width=0.325\textwidth]{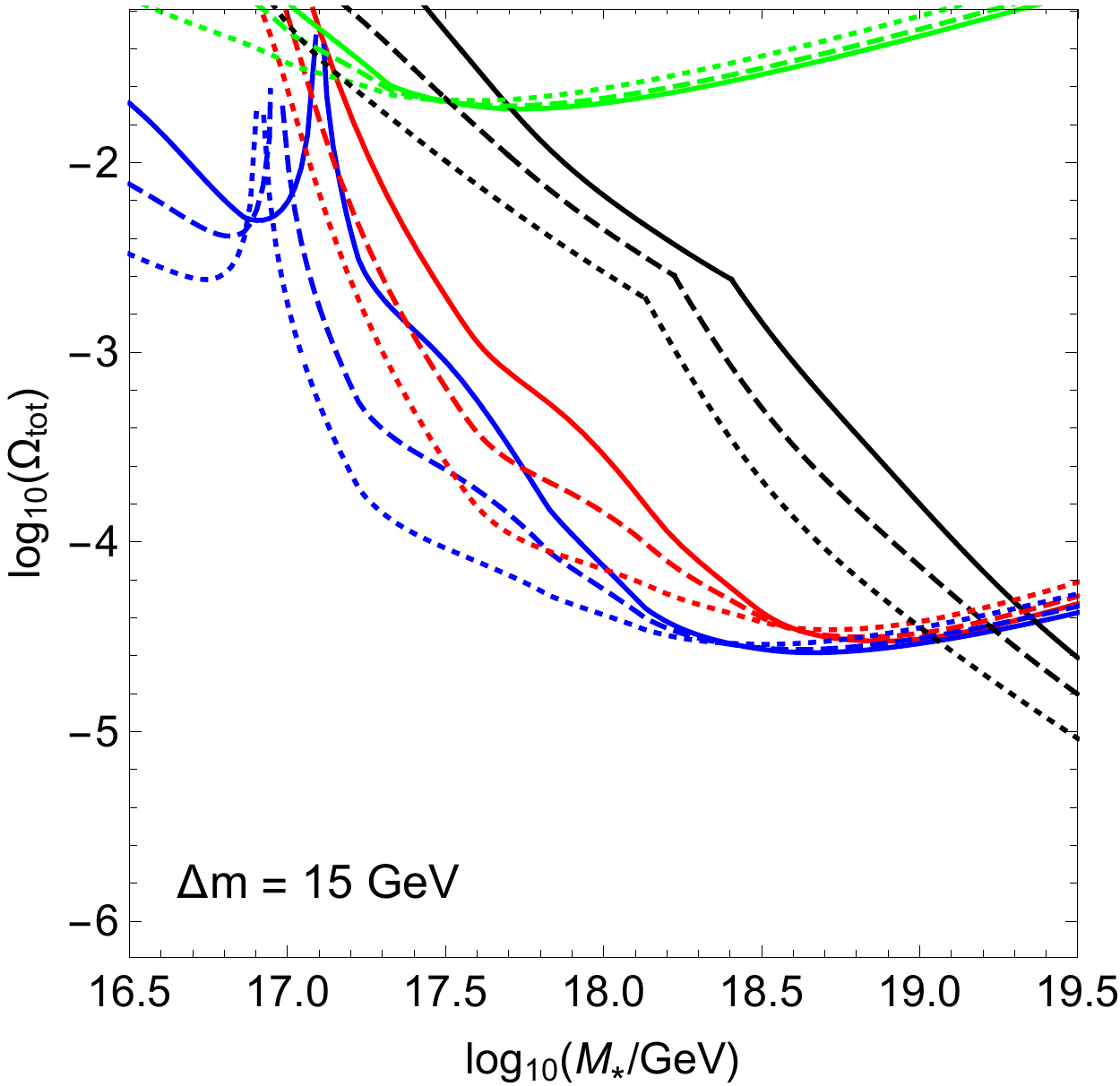}~
\includegraphics*[width=0.325\textwidth]{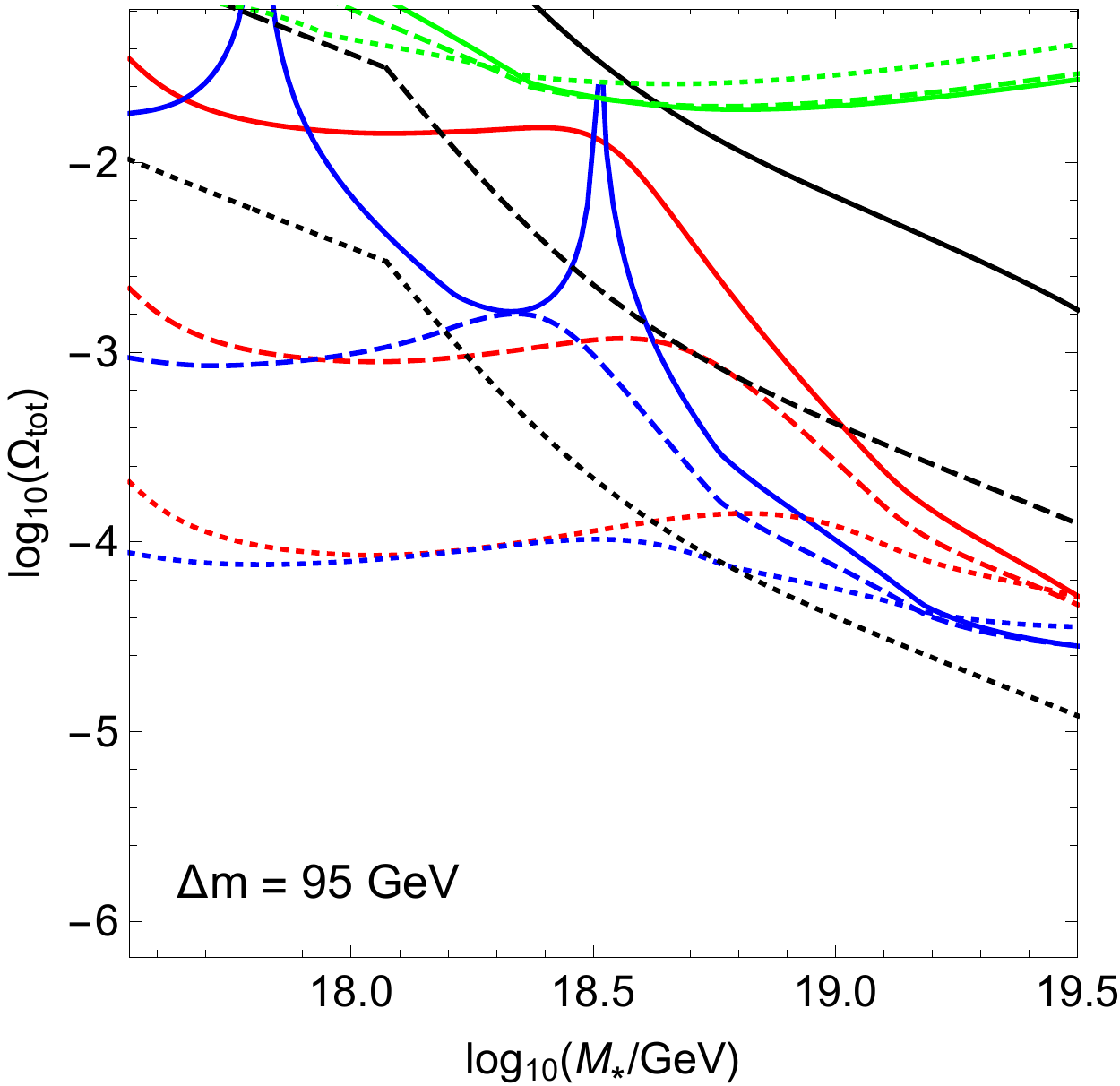}~
\includegraphics*[width=0.315\textwidth]{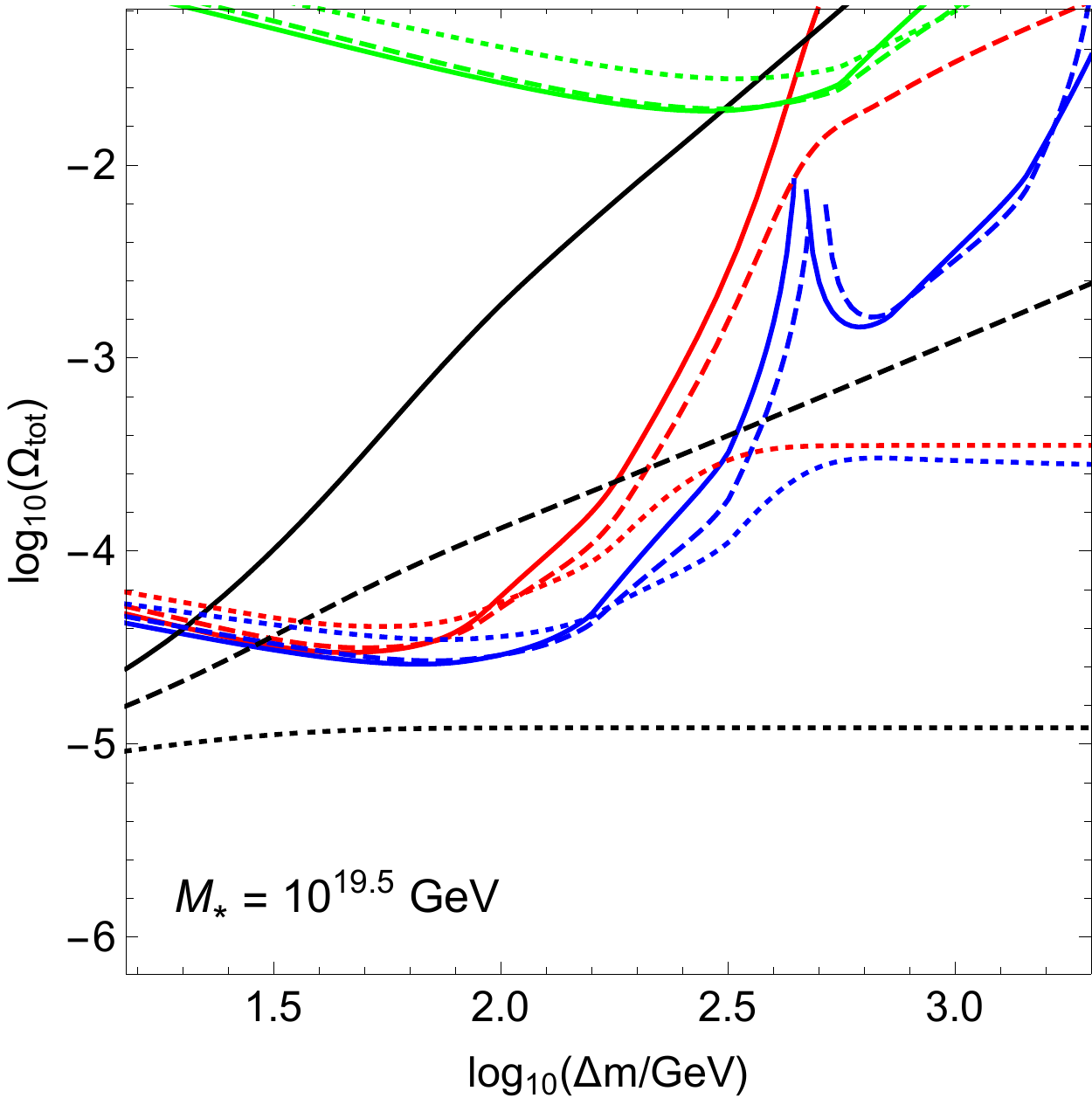}
\end{center}
\caption{Compilation of constraints on electromagnetic injection for ``low-density'' ensembles
  with mass spectra of the form given in Eq.~(\ref{eq:linearmassscaling}), in
  which the mass splitting $\Delta m$ between successive $\chi_i$ is constant.
  The contours shown correspond to the choice of $N = 3$ with $m_0 = 10$~GeV.~ 
  The curves shown in each panel indicate the upper bound on the total abundance 
  $\Omegatot$ of the ensemble from constraints on the primordial 
  abundances of $\ce{D}$ (blue curve), $\ce{^{7}Li}$ (green curve), and $\ce{^{6}Li}$ 
  (red curve) and from limits on distortions of the CMB-photon spectrum (black curve).
  The dotted, dashed, and solid curves correspond respectively to the choices 
  $\gamma = \{0,1,2\}$ for the scaling exponent in Eq.~(\ref{eq:AbundanceScaling}).
  In the left and center panels, we plot this upper bound as a function of the 
  coupling-suppression scale $M_\ast$ with fixed $\Delta m = 15$~GeV and with fixed
  $\Delta m = 95$~GeV, respectively.  In the right panel, we plot the
  bound as a function of $\Delta m$ with fixed $M_\ast = 10^{19.5}$~GeV. 
  Comparing the results shown in the center panel with those of Fig.~\protect\ref{fig:comb},
    we see that the bounds on a decaying ensemble 
   differ not only quantitatively but even qualitatively from the bounds on its individual constituents.
\label{fig:LimitsMLinearLoDens}}
\begin{center}
\includegraphics*[width=0.325\textwidth]{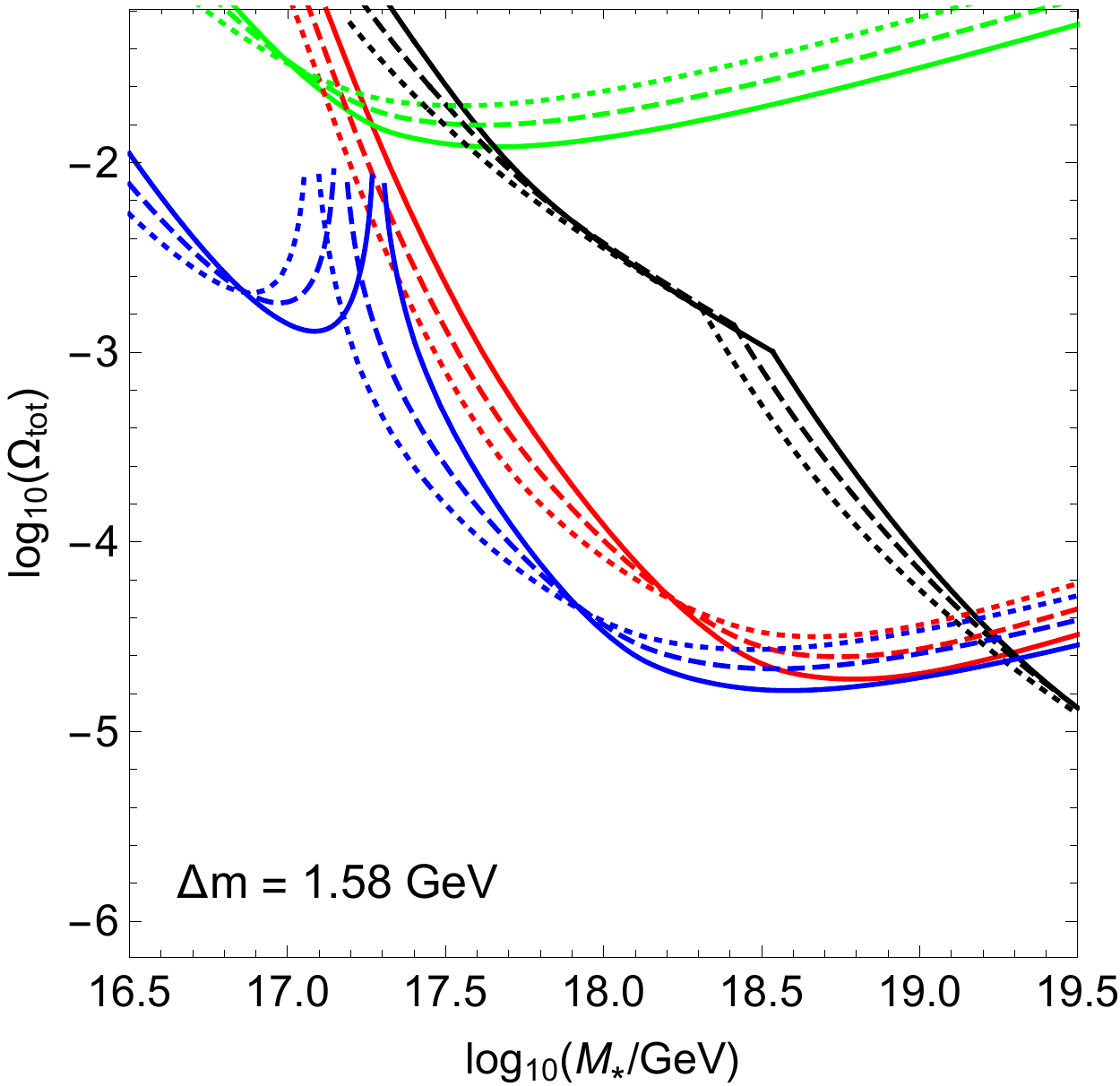}~
\includegraphics*[width=0.325\textwidth]{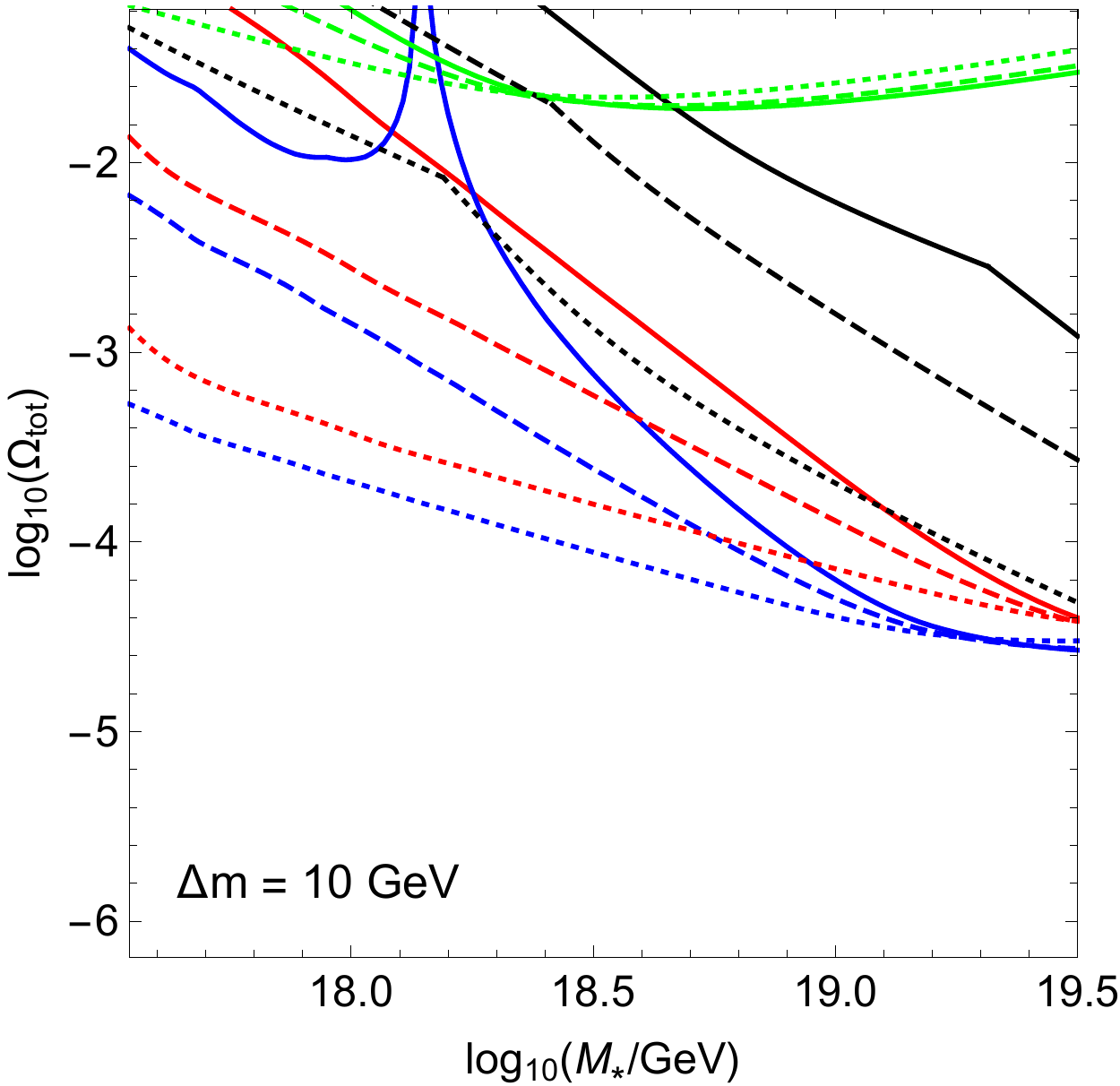}~
\includegraphics*[width=0.315\textwidth]{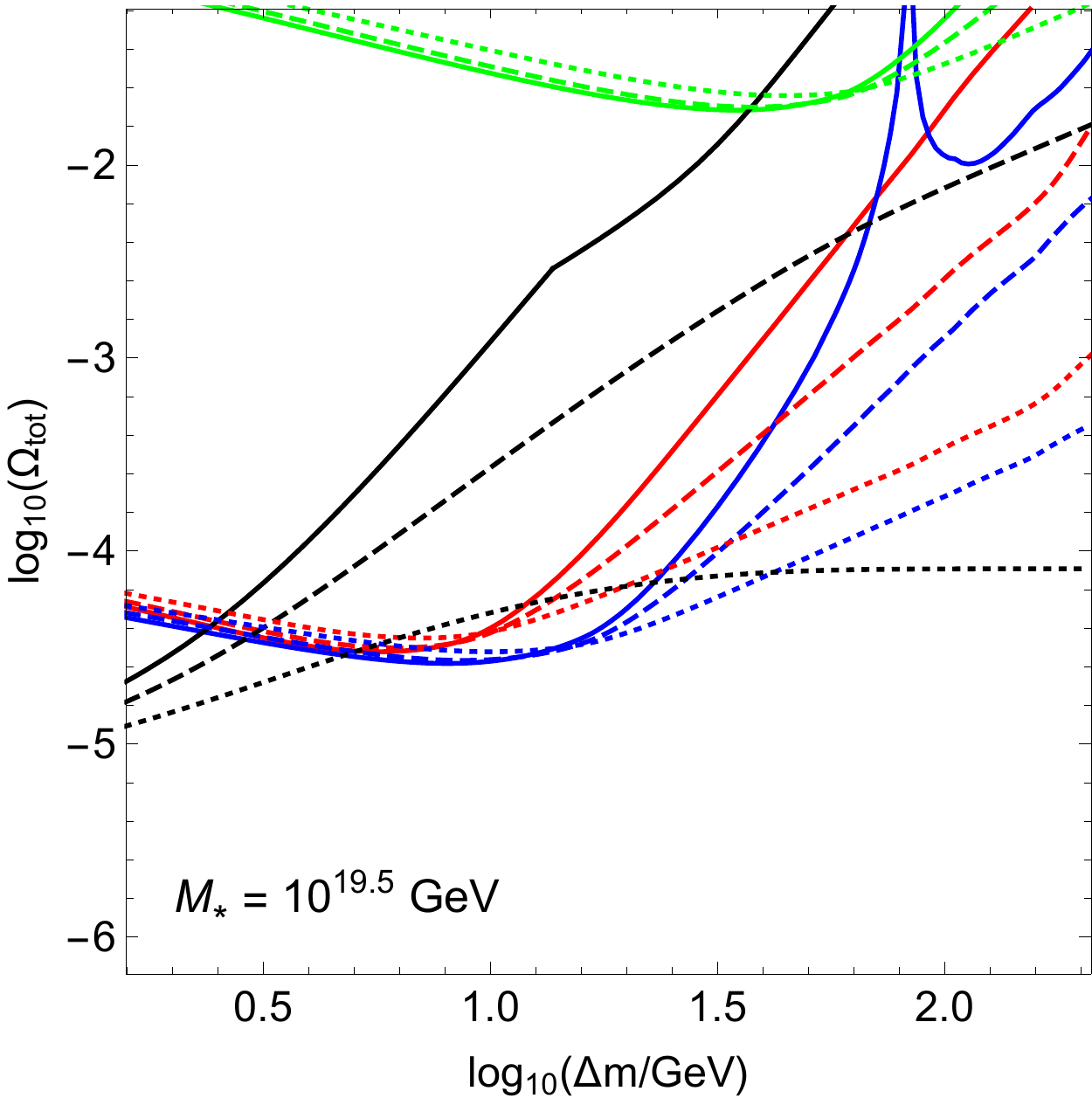}
\end{center}
\caption{Same as in Fig.~\protect\ref{fig:LimitsMLinearLoDens}, but for 
   ``high-density'' ensembles with $N = 20$.  In the left and center panels, we plot the 
   upper bound on $\Omegatot$ as a function of the coupling-suppression scale $M_\ast$ 
   with fixed $\Delta m = 1.58$ and with fixed $\Delta m = 10$~GeV, respectively.  
   In the right panel, we plot the bound as a function of $\Delta m$ with 
   fixed $M_\ast = 10^{19.5}$~GeV.
\label{fig:LimitsMLinearHiDens}}
\end{figure*}

In Fig.~\ref{fig:LimitsMLinearLoDens},
we show the constraints on an ensemble comprising $N = 3$ unstable particles
in which the lightest ensemble constituent has a mass $m_0 = 10$~GeV.~  
We consider this ensemble
to be ``low-density'' in that it has 
a density of states per unit $\tau$ which is small throughout the
range of parameters shown.
(This will later be contrasted with analogous results in Fig.~\ref{fig:LimitsMLinearHiDens}
for a ``high-density'' ensemble.)
The contours shown in the left and center panels 
of Fig.~\ref{fig:LimitsMLinearLoDens}
represent upper bounds on 
the total extrapolated abundance $\Omegatot$ as functions of $M_\ast$ for fixed
$\Delta m$.  The contours shown in the left panel correspond to the choice of
$\Delta m = 15$~GeV; those in the center panel correspond to the 
choice of $\Delta m = 95$~GeV.~ In the right panel, we show the corresponding 
contours as functions of $\Delta m$ for fixed $M_\ast = 10^{19.5}$~GeV.~     
The blue, green, and red curves shown in each panel represent the bounds on
$\Omegatot$ from constraints on the primordial abundances of $\ce{D}$, $\ce{^{7}Li}$, 
and $\ce{^{6}Li}$, respectively.  We use a uniform color for the $\ce{D}$ 
contour --- a contour which reflects the sum of contributions to $\delta Y_{\ce{D}}$ 
from both production and destruction processes. 
The dotted, dashed, and solid curves of each color 
correspond respectively to the choices $\gamma = \{0,1,2\}$ for the scaling exponent 
in Eq.~(\ref{eq:AbundanceScaling}).  The corresponding black curves shown in each
panel represent the bounds arising from limits on distortions of the CMB-photon 
spectrum obtained by taking whichever bound (either that from $\delta \mu$ or that from $\delta y_C$)
is more stringent at every point.

In interpreting the results shown in Fig.~\ref{fig:LimitsMLinearLoDens}, 
we note that each choice of parameters specifies a particular range of lifetimes 
$\tau_{N - 1} \leq \tau_i \leq \tau_0$ for the ensemble constituents.  This range of 
lifetimes varies across the range of $M_\ast$ shown in the left panel, 
from $1.03\times 10^4\mathrm{~s} \leq \tau_i \leq 6.58 \times 10^5\mbox{~s}$
for $M_\ast = 10^{16.5}$~GeV to  
$1.03\times 10^{10}\mathrm{~s} \leq \tau_i \leq 6.58 \times 10^{11}\mathrm{~s}$
for $M_\ast = 10^{19.5}$~GeV.~  Across this entire range of $M_\ast$ values,
the range of $\tau_i$ values for the ensemble is reasonably narrow, spanning only roughly 
a single order of magnitude.  As a result, the bounds on $\Omegatot$ for a decaying 
ensemble shown in the left panel of Fig.~\ref{fig:LimitsMLinearLoDens} as 
functions of $M_\ast$ closely resemble the bounds on $\Omega_\chi$ for a single decaying 
particle species shown in Fig.~\ref{fig:comb} as a function of $\tau_\chi$.
Moreover, since the $\tau_i$ for all particles in the ensemble are roughly comparable,
the results are not particularly sensitive to $\gamma$, which determines how $\Omegatot$ 
is distributed across the ensemble.
The most stringent constraint on $\Omegatot$ is the bound associated with the  
primordial $\ce{D}$ abundance over almost the entire range of $M_\ast$ shown.  The
only exception occurs in the regime where $M_\ast$ is large and the 
lifetimes of all of the ensemble constituents are extremely long, in which case
the constraint associated with CMB distortions supersedes the $\ce{D}$ constraint.
We note that for values of $\Omegatot$ which lie above the $\ce{D}$ contour, 
$\ce{D}$ may be either overabundant or underabundant. 
We also note that a ``funnel'' similar to 
that appearing
in Fig.~\ref{fig:comb} is visible in the $\ce{D}$ constraint contours for all 
values of $\gamma$ shown in the figure, and that this funnel occurs where 
$M_\ast$ is such that the lifetimes of the ensemble constituents are approximately
$\tau_i \approx 10^{6.5}$~s.  

By contrast, the results shown in the center panel of 
Fig.~\ref{fig:LimitsMLinearLoDens} correspond to the regime in which the range of 
lifetimes spanned by the ensemble is quite broad, extending over several orders of 
magnitude in $\tau_i$.  In particular, the range of lifetimes varies across the 
range of $M_\ast$ shown in this panel from 
$1.03\times 10^4\mathrm{~s} \leq \tau_i \leq 8.22 \times 10^7\mbox{~s}$
for $M_\ast = 10^{17.5}$~GeV to  
$8.22\times 10^{7}\mathrm{~s} \leq \tau_i \leq 6.58 \times 10^{11}\mathrm{~s}$
for $M_\ast = 10^{19.5}$~GeV.~  We note that the range of $M_\ast$ shown in this
panel has been truncated relative to the left panel in order to ensure that the
characteristic decay timescales for all ensemble constituents occur well after
the BBN epoch.  For an ensemble with such a broad range of lifetimes, 
the effect of distributing $\Omegatot$ over multiple particle
species is readily apparent.  Indeed, the shapes of the constraint contours in 
the center panel of Fig.~\ref{fig:LimitsMLinearLoDens} bear little 
resemblance to the shapes of the contours in Fig.~\ref{fig:comb}.
Thus, comparing the results shown in the center panel of Fig.~\ref{fig:LimitsMLinearLoDens} 
with those in Fig.~\ref{fig:comb},
we see that the bounds on a decaying ensemble 
indeed differ {\it not only quantitatively but even qualitatively}\/ from the bounds on its individual constituents.

For example, two separate funnels arise in the $\ce{D}$ constraint contour 
for $\gamma = 2$.  Each of these funnels represents not merely a cancellation 
between the production and destruction contributions to $Y_{\ce{D}}$ from a single 
particle, but rather a cancellation among the individual contributions from all 
constituent particle species in the ensemble.  The funnel on the left arises
due to a cancellation between the net negative contributions to
$Y_{\ce{D}}$ from $\chi_1$ and $\chi_2$ and the net positive contribution
from $\chi_0$.  By contrast, the funnel on the right arises due to a 
cancellation between a net negative contribution to
$Y_{\ce{D}}$ from $\chi_2$ and positive contributions from $\chi_0$ and 
$\chi_1$, with the latter contribution suppressed because 
$\tau_1 \approx 10^{6.5}$~s.  The positions of these funnels, then, 
reflect the collective nature of the ensemble --- and they have important
consequences.  Indeed, within certain regions of parameter space within which 
the $\ce{D}$ constraints are weakened by cancellations among 
$\delta Y_{\ce{D}}$ contributions from different ensemble constituents, 
the constraint associated with the $\ce{^{6}Li}$ abundance actually represents 
the leading bound on $\Omegatot$ for the ensemble.           

Finally, in the right panel of Fig.~\ref{fig:LimitsMLinearLoDens}, we show 
how the bounds on $\Omegatot$ vary as a function of $\Delta m$ for fixed 
$M_\ast = 10^{19.5}$~GeV.~  We choose this benchmark for $M_\ast$ because 
the CMB-distortion constraints and the leading constraints from light nuclei 
are comparable for this choice of $M_\ast$.  In interpreting the results displayed
in this panel, it is important to note that $\tau_0$ is independent of $\Delta m$.
By contrast, the $\tau_i$ for all other $\chi_i$ decrease with increasing $\Delta m$.
Since, broadly speaking, the impact that a decaying particle has on both CMB distortions
and on the primordial abundances of light nuclei tends to decrease as $\tau_i$
decreases, the bounds on decaying ensembles generally grow weaker as $\Delta m$
increases.  This weakening of the constraints generally grows more pronounced as 
$\gamma$ increases and as a larger fraction of the abundance is carried by the
heavier ensemble constituents.  For $\gamma = 0$, on the other hand, the fraction of 
$\Omegatot$ carried by the lightest ensemble constituent $\chi_0$ is independent
of $\Delta m$.  As a result, many of the constraint contours shown in 
the right panel of Fig.~\ref{fig:LimitsMLinearLoDens} become essentially flat for 
sufficiently large $\Delta m$ as the lifetimes of the remaining $\chi_i$ 
become so short that the impact of their decays on the corresponding cosmological 
observables is negligible in comparison with those of $\chi_0$.    

Thus far, we have focused on the regime in which the spacing between the lifetimes 
$\tau_i$ of the ensemble constituents is large --- in other words, the regime in 
which the density of states per unit $\tau$ is small across the ensemble.
However, it is also interesting to consider the opposite regime, 
in which the density of states per unit $\tau$ is sufficiently large over the 
entire range $\tau_{N-1} \leq \tau \leq \tau_0$ that the ensemble effectively
acts as a continuous source of electromagnetic injection over this interval.
This range of lifetimes is completely determined in our toy model by $M_\ast$, $m_0$, 
and the quantity $m_{N-1}= (N-1)\Delta m+ m_0$.
Thus, we may make a direct comparison between the bounds 
on such ``high-density ensembles'' and the bounds on ``low-density'' ensembles with 
the same $\tau_i$ range by varying $N$ and $\Delta m$ oppositely while holding 
$M_\ast$, $m_0$, and $m_{N-1}$ fixed.  Towards this end, in
Fig.~\ref{fig:LimitsMLinearHiDens}, we show the upper bounds on $\Omegatot$ for
an ensemble with a higher density of states per unit lifetime.  The constraint 
contours in the left and center panels 
of the figure represent the same choices of $m_0$ and $m_{N-1}$ as in the 
corresponding panels of Fig.~\ref{fig:LimitsMLinearLoDens}, but for $N=20$ rather
than $N=3$.  For these parameter choices, the ensemble is effectively in the 
``high-density'' regime over the entire range of $M_\ast$ shown in each panel.  
In the right panel of  Fig.~\ref{fig:LimitsMLinearHiDens}, we once again show 
constraint contours as functions of $\Delta m$ for fixed $M_\ast$.

As one might expect, the primary effect of distributing the abundance of 
the ensemble more evenly across the same range of lifetimes is that features
in the constraint contours associated with particular $\chi_i$ are in large part 
smoothed out.  While the $\ce{D}$ funnels in the left panel are still present, 
this reflects the fact that the range of $\tau_i$ for the ensemble is sufficiently 
narrow that distributing the abundance more democratically across this range
has little impact on the constraint contours.
A single $\ce{D}$ funnel also appears in the constraint contour for $\gamma = 2$
in the center panel of Fig.~\ref{fig:LimitsMLinearHiDens}.  The presence of
this feature reflects the fact that for $\gamma = 2$, the vast majority of 
$\Omegatot$ is carried by the most massive --- and therefore most 
unstable --- ensemble constituents.  Moreover, the density of states per unit 
lifetime is also much higher for these shorter-lived states than it is for the 
rest of the ensemble.  The funnel can be understood as corresponding to the 
value of $M_\ast$ for which the lifetimes of these states, which are 
smaller than but roughly similar to $\tau_{N-1}$, satisfy 
$\tau_i \sim \tau_{N-1} \approx 10^{6.5}$~s.  Once again, however, we note that
the location of the funnel is shifted slightly away from this value of $M_\ast$ 
due to the collective contribution to $Y_D$ from the longer-lived constituents 
in the ensemble.  A similar feature is also apparent in the constraint contour for 
$\gamma = 2$ in the right panel of Fig.~\ref{fig:LimitsMLinearHiDens}.

We can gain further insight into the results displayed in 
Fig.~\ref{fig:LimitsMLinearHiDens} by examining the continuum limit in which 
$\Delta m \rightarrow 0$.
In this limit, the sums appearing in Eqs.~(\ref{eq:BBNFitsTerm1})--(\ref{eq:BBNFitsLi6}) 
and in Eqs.~(\ref{eq:specfitsmu}) and (\ref{eq:specfitsy}) 
can be recast as integrals over the continuous lifetime variable $\tau$ ---
integrals which may be evaluated in a straightforward manner, yielding 
analytic expressions for $\delta Y_a$, $\delta \mu$, and $\delta y_C$.
For example, in the case in which the $\tau_i$ span the entire range from 
$t_{Aa}$ to $t_{fa}$ for each relevant nucleus, we find that for any $\gamma \geq 0$ 
the term $\delta Y_a^{(1)}$ in Eq.~(\ref{eq:BBNFits}) takes the form
\begin{eqnarray}
  \delta Y_a^{(1)} &=&
    A_a \Xi(t_{Ca}) \Gamma\left(\frac{\gamma}{3}-\frac{2}{3},x^{-1}\right)
      \Bigg|_{x^{\mathrm{min}}_{Aa}}^{x^{\mathrm{max}}_{Aa}}
    \nonumber \\ 
   + && \!\!\!\!\!
    B_a \Xi(t_{Xa}) \Bigg[  
      \frac{6\beta x^{-\frac{5+2\gamma}{6}} }{5+2\gamma}  
      -\frac{3 x^{-\frac{1+\gamma}{3}} }{1+\gamma}  \Bigg]
      \Bigg|_{x^{\mathrm{min}}_{Ba}}^{x^{\mathrm{max}}_{Ba}}
      \nonumber\\  
    + && \!\!\!\!\! B_a \Xi(t_{Xa}) \Bigg[  
      \frac{6(1+\beta) x^{-\frac{5+2\gamma}{6}} }{5+2\gamma}  
      -\frac{24 x^{-\frac{7+4\gamma}{12}} }{7+4\gamma}  \Bigg]
      \Bigg|_{x^{\mathrm{min}}_{Ca}}^{x^{\mathrm{max}}_{Ca}}~,
    \nonumber \\
  \label{eq:ContLimitBBNLinear}
\end{eqnarray}
where $\Gamma(n,z)$ is the incomplete gamma function,
where we have defined 
\begin{equation}
  \Xi(t) ~\equiv~ \frac{(\gamma + 1) \Omegatot M_\ast^{2(1+\gamma)/3}}
    {3\big(m_{N-1}^{\gamma+1}-m_0^{\gamma+1}\big) t^{(1+\gamma)/3}}~,
  \label{eq:XifnDef}
\end{equation}
where $A_a$ and $B_a$ are defined in Table~\ref{tab:BBN} and where the limits 
of integration are 
\begin{equation}
\begin{array}{ll}
  x^{\mathrm{min}}_{Aa} ~=~ t_{Aa}/t_{Ca} ~~~~&~~~~
  x^{\mathrm{max}}_{Aa} ~=~ t_{Ba}/t_{Ca} \\
  x^{\mathrm{min}}_{Ba} ~=~ t_{Ba}/t_{Xa} ~~~~&~~~~
  x^{\mathrm{max}}_{Ba} ~=~ 1 \\
  x^{\mathrm{min}}_{Ca} ~=~ 1 ~~~~&~~~~
  x^{\mathrm{max}}_{Ca} ~=~ t_{fa}/t_{Xa}~.
\end{array}
  \label{eq:ContinuumLimitsOfIntYa}    
\end{equation}
In cases in which $\tau_i$ does not span the entire range from $t_{Aa}$ to $t_{fa}$,
the values of the quantities in Eq.~(\ref{eq:ContinuumLimitsOfIntYa})
should be replaced by values which restrict the overall range of lifetimes to 
$\max[\tau_{N-1},t_{Aa}] \leq \tau_i \leq \min[\tau_0,t_{fa}]$.

For $\ce{D}$ and $\ce{^{6}Li}$, the term $\delta Y_a^{(2)}$ in Eq.~(\ref{eq:BBNFits})
is non-vanishing and must be evaluated separately.
For $\ce{D}$, the term $\delta Y_a^{(2)}$ in Eq.~(\ref{eq:BBNFits}) has exactly
the same form as Eq.~(\ref{eq:ContLimitBBNLinear}), but with $A^\ast_a$ in place
of $A_a$, $B^\ast_a$ in place of $B_a$, \etc~  For $\ce{^{6}Li}$, we find 
for all $\gamma \geq 0$ that the $\delta Y_a^{(2)}$ term takes the form
\begin{widetext}
\begin{eqnarray}
  \delta Y_a^{(2)} &=&
    A^*_a \Xi(t^*_{Ca}) \Gamma\left(\frac{\gamma}{3}-\frac{2}{3},x^{-1}\right)
      \Bigg|_{x^{\ast\mathrm{min}}_{Aa}}^{x^{\ast\mathrm{max}}_{Aa}}
      \nonumber \\ 
+&&\!\!\!\!\!\!\!
    B^*_a \Xi(t^*_{Xa}) \Bigg[   
      \frac{72\beta^5\theta x^{-\frac{17+2\gamma}{6}} }{17+2\gamma}
      -\frac{45\beta^4 (1 + \theta) x^{-\frac{7+\gamma}{3}} }{7+\gamma}
      +\frac{120\beta^3 x^{-\frac{11+2\gamma}{6}} }{11+2\gamma} 
      -\frac{3(5-3\theta)x^{-\frac{1+\gamma}{3}} }{1+\gamma} \Bigg]
      \Bigg|_{x^{\ast\mathrm{min}}_{Ba}}^{x^{\ast\mathrm{max}}_{Ba}}
      \nonumber\\ 
+&&\!\!\!\!\!\!\!
    B^*_a \Xi(t^*_{Xa}) \Bigg[  
      \frac{8\theta(1+9\beta^5)x^{-\frac{17+2\gamma}{6}}}{17+2\gamma}
      -\frac{45(1+\theta)(1+7\beta^4)x^{-\frac{7+\gamma}{3}}}{7(7+\gamma)} 
      +\frac{24(1+5\beta^3)x^{-\frac{11+2\gamma}{6}}}{11+2\gamma} 
      -\frac{64(9-5\theta)x^{-\frac{7+4\gamma}{12}}}{7(7+4\gamma)} \Bigg]
      \Bigg|_{x^{\ast\mathrm{min}}_{Ca}}^{x^{\ast\mathrm{max}}_{Ca}}~,
      \nonumber \\
  \label{eq:ContLimitBBNLinearLi6}
\end{eqnarray}
where the limits of integration are given by expressions analogous to 
those appearing in Eq.~(\ref{eq:ContinuumLimitsOfIntYa}). 
  
Expressions for the CMB-distortion parameters $\delta \mu$ and $\delta y_C$ may 
be obtained in a similar fashion.  In particular, for the case in which
the $\tau_i$ span the entire range from $t_e$ to $\tLS$, we find that $\delta \mu$
takes the form
\begin{eqnarray}
  \delta\mu &=&
    \frac{4A_\mu}{5} \Xi(t_{0\mu}) 
      \Gamma\left(\frac{4\gamma - 2}{15},x^{-5/4}\right)
      \Bigg|_{x^{\mathrm{min}}_{0\mu}}^{x^{\mathrm{max}}_{0\mu}}
    \nonumber \\ 
    &&~+ B_\mu\Xi(t_{1\mu})
      \Bigg[ \frac{6}{1-2\gamma}x^{\frac{1-2\gamma}{6}} 
      - \frac{1}{\alpha_\mu}
      \Gamma\left(\frac{2\gamma-1}{6\alpha_\mu},x^{-\alpha_\mu}\right)\Bigg]
      \Bigg|_{x^{\mathrm{min}}_{1\mu}}^{x^{\mathrm{max}}_{1\mu}} 
    \nonumber \\ 
    &&~+ B_\mu \Xi(t_{2\mu})
      \Bigg[ \frac{3}{1-\gamma}x^{\frac{1-\gamma}{3}} 
      - \frac{3}{4\alpha_\mu}
      \Gamma\left(\frac{\gamma-1}{4\alpha_\mu},x^{-\frac{4\alpha_\mu}{3}}\right)\Bigg]
      \Bigg|_{x^{\mathrm{min}}_{2\mu}}^{x^{\mathrm{max}}_{2\mu}}
  \label{eq:ContLimitmuLinear}
\end{eqnarray}
for all $\gamma \geq 0$, except for the special case in which $\gamma = 1$.
For this special case, the replacement
\begin{equation}
  -\frac{3}{1-\gamma}x^{\frac{1-\gamma}{3}} ~\longrightarrow~ \ln x
\end{equation} 
should be made in the fourth line of Eq.~(\ref{eq:ContLimitmuLinear}).
Likewise, we find that $\delta y_C$ takes the form 
\begin{eqnarray}
  \delta y_C &=&
    \frac{4A_y}{5} \Xi(t_{0y}) \left.
      \Gamma\left(\frac{4\gamma - 2}{15},x^{-5/4}\right)
      \right|_{x^{\mathrm{min}}_{0y}}^{x^{\mathrm{max}}_{0y}}    
      \nonumber \\ 
    &&~+ B_y \Xi(t_{1y}) 
      \frac{6x^{\alpha+\frac{1-2\gamma}{6}}}{6\alpha_\mu - 2\gamma +1} 
      \left.
      _2F_1\left(1,1+\frac{1-2\gamma}{6\alpha_y};
      2+ \frac{1-2\gamma}{6\alpha_y};-x^{\alpha_y}\right) 
      \right|_{x^{\mathrm{min}}_{1y}}^{x^{\mathrm{max}}_{1y}}      
      \nonumber \\
    &&~+ B_y \Xi(t_{2y})
      \frac{3x^{\frac{1-\gamma}{3}}}{1-\gamma}  \left.
      _{2}F_1\left(1,\frac{\gamma -1}{4\alpha_y};
      1+\frac{\gamma-1}{4\alpha_y};-x^{-\frac{4\alpha_y}{3}}\right) 
      \right|_{x^{\mathrm{min}}_{2y}}^{x^{\mathrm{max}}_{2y}}~,
  \label{eq:ContLimityLinear}
\end{eqnarray}
\end{widetext}
where $_{2}F_1(a,b;c;z)$ denotes the ordinary hypergeometric function,
where $A_a$, $B_a$, and $\alpha_a$ are defined in Table~\ref{tab:spec},
and where the limits of integration are
\begin{equation}
\begin{array}{ll}
  x^{\mathrm{min}}_{0a} ~=~ t_e/t_{0a}  ~~~~&~~~~
  x^{\mathrm{max}}_{0a} ~=~ t_{Ba}/t_{0a}  \\
  x^{\mathrm{min}}_{1a} ~=~ t_{Ba}/t_{1a} ~~~~&~~~~
  x^{\mathrm{max}}_{1a} ~=~ \tMRE/t_{1a}  \\
  x^{\mathrm{min}}_{2a} ~=~ \tMRE/t_{2a} ~~~~&~~~~
  x^{\mathrm{max}}_{2a} ~=~ \tLS/t_{2a}~. 
\end{array}
  \label{eq:ContinuumLimitsOfIntmuy}   
\end{equation}
Once again, in cases in which $\tau_i$ does not span the entire 
range from $t_e$ to $\tLS$, the 
values of the quantities in Eq.~(\ref{eq:ContinuumLimitsOfIntmuy})   
should be replaced by values which restrict the overall range of lifetimes to 
$\max[\tau_{N-1},t_{e}] \leq \tau_i \leq \min[\tau_0,\tLS]$.

%%%%%%%%%%%%%%%%%%%%%%%%%%%%%%%%%%%%%%%%%%%%%%%%%%%%%%%%%%%%%%%%%%%%%%%%%%%%%%%%%%%%%%
\FloatBarrier
\subsection{Results for Exponentially Rising Mass Splittings\label{sec:LogScalingEnsembles}}
%%%%%%%%%%%%%%%%%%%%%%%%%%%%%%%%%%%%%%%%%%%%%%%%%%%%%%%%%%%%%%%%%%%%%%%%%%%%%%%%%%%%%%

We now consider the results for a toy ensemble with a mass spectrum given
by Eq.~(\ref{eq:logmassscaling}).  The density of states per unit mass 
is inversely proportional to $m$ in this case, and the corresponding density of states 
per unit lifetime is 
\begin{equation}
  n_\tau ~=~ \frac{1}{3\ln(\xi)\tau}~,
\end{equation}
which likewise decreases with $\tau$.  However, we note that the density of states 
per unit $\ln(\tau)$ in this case remains constant across the ensemble.

\begin{figure*}[t]
\begin{center}
\includegraphics*[width=0.325\textwidth]{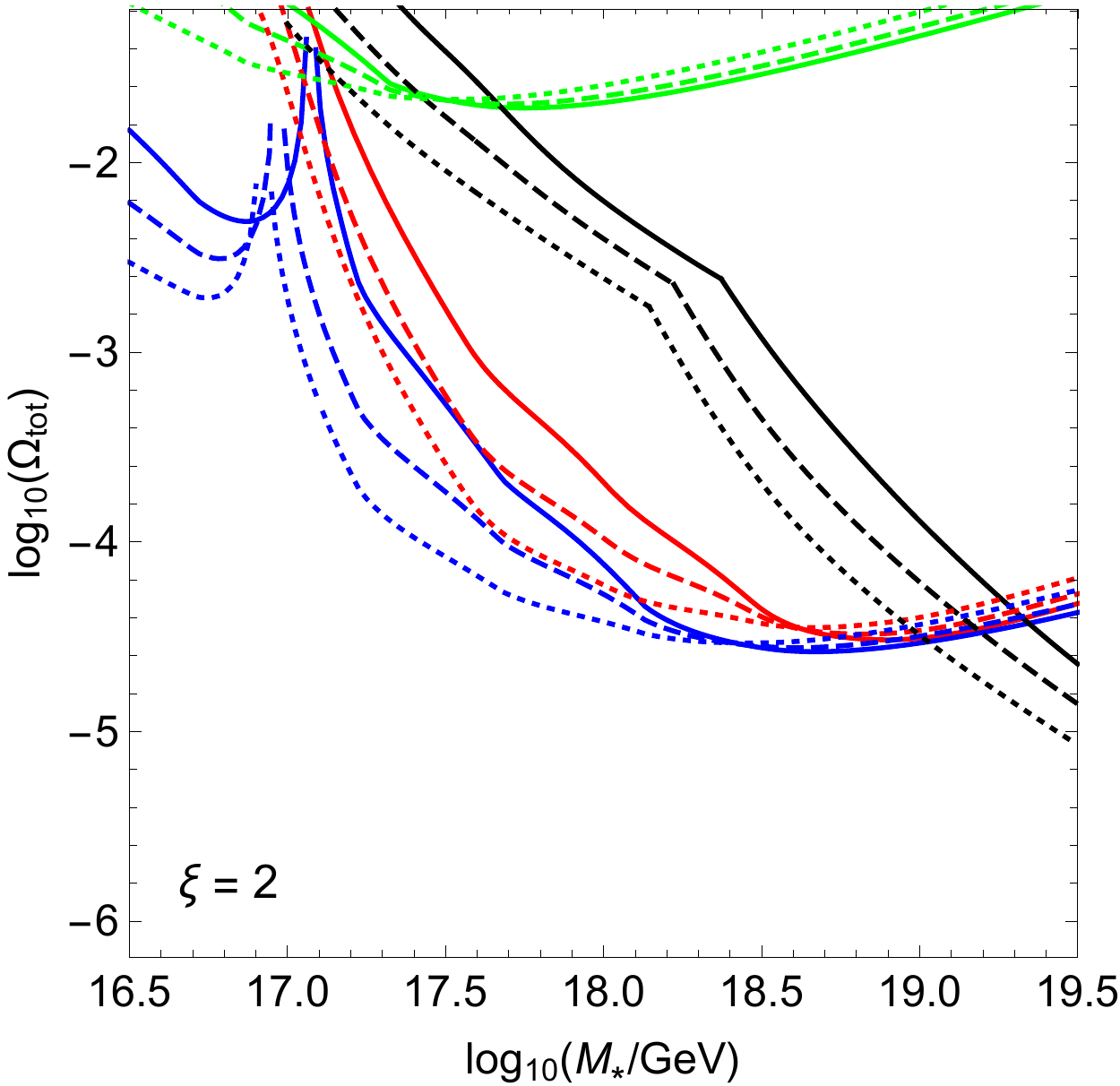}~
\includegraphics*[width=0.325\textwidth]{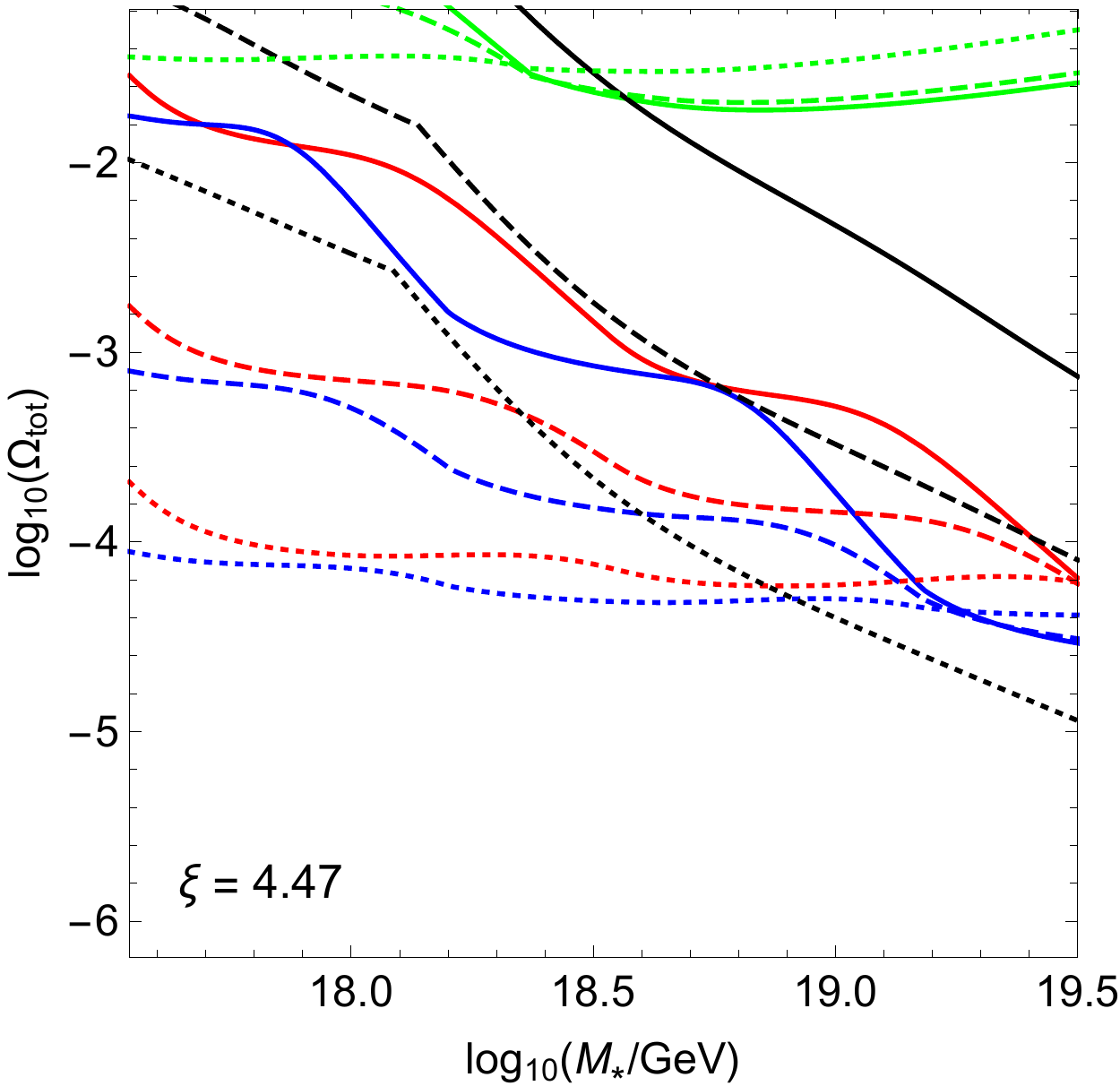}~
\includegraphics*[width=0.315\textwidth]{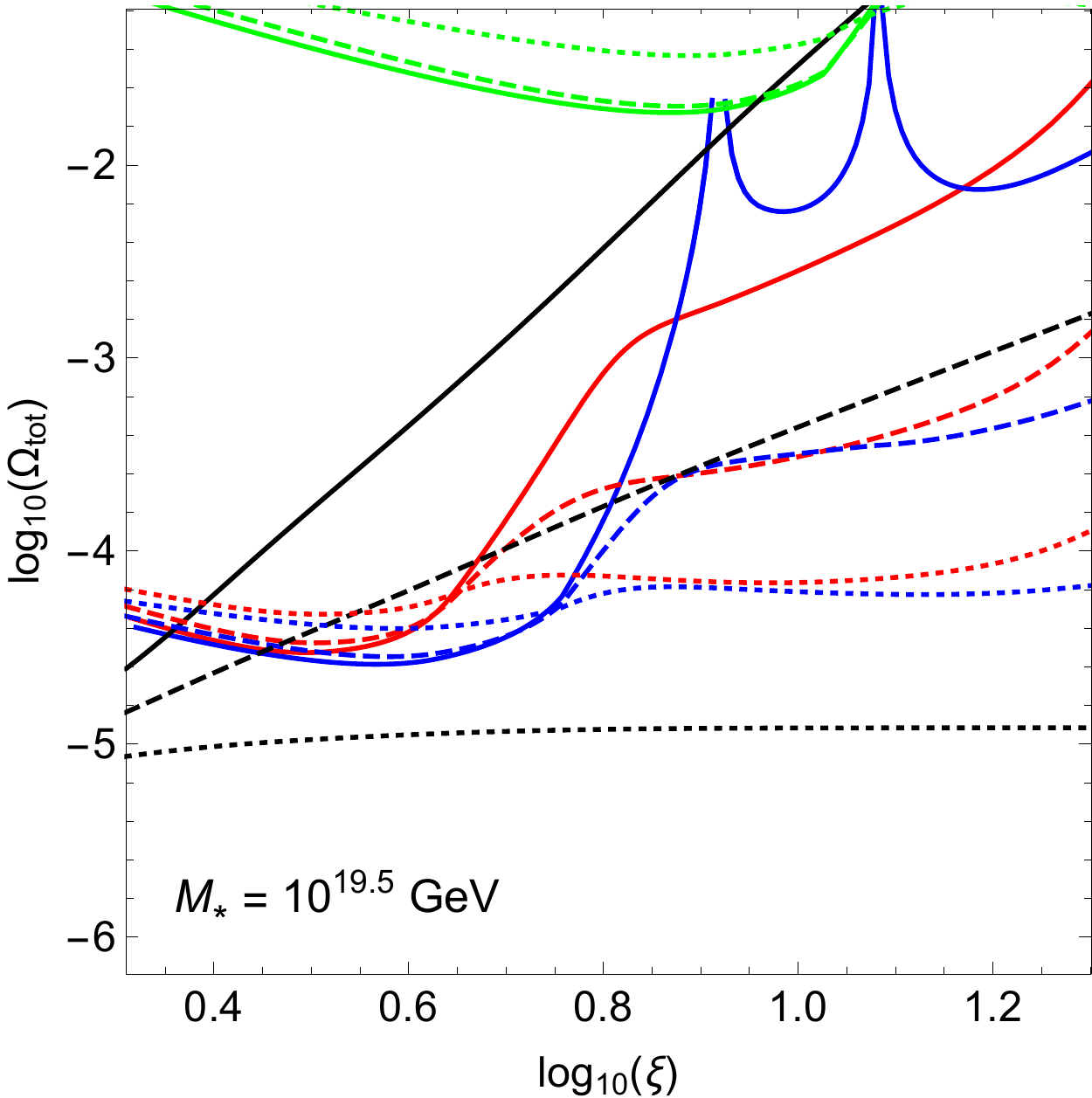}
\end{center}
\caption{Same as in Fig.~\protect\ref{fig:LimitsMLinearLoDens}, but for ``low-density'' ensembles
   with mass spectra of the form given in Eq.~(\ref{eq:logmassscaling}), in
   which the masses of the $\chi_i$ are evenly distributed on a
   logarithmic scale.  In the left and center panels, we plot the upper bound on 
   $\Omegatot$ as a function of the coupling-suppression scale $M_\ast$ with fixed 
   $\xi = 2$ and with fixed $\xi = 4.47$, respectively.  In the right panel, 
   we plot the bound as a function of $\xi$ with fixed 
   $M_\ast = 10^{19.5}$~GeV.   
\label{fig:LimitsMLogLoDens}}
\begin{center}
\includegraphics*[width=0.325\textwidth]{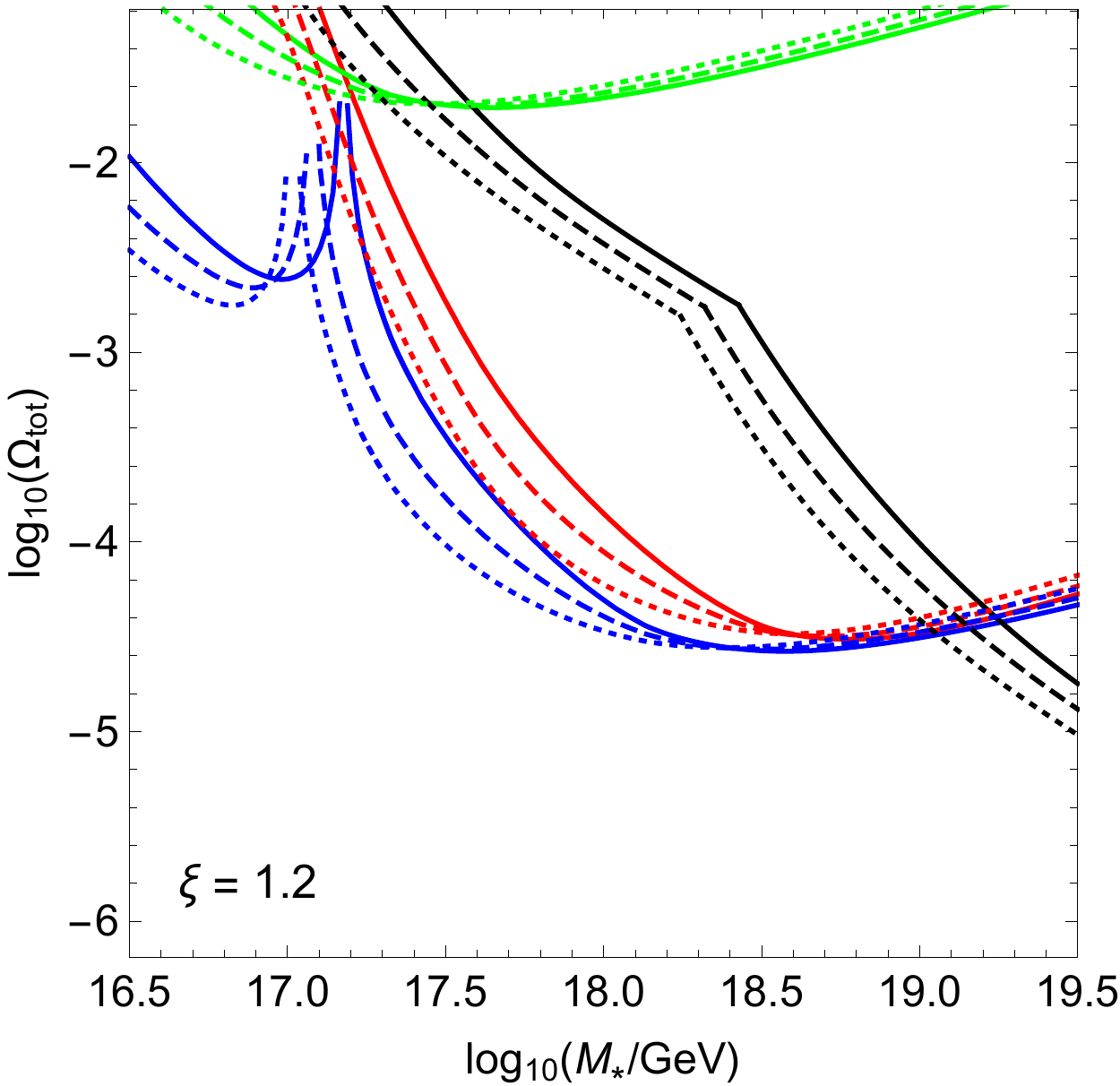}~
\includegraphics*[width=0.325\textwidth]{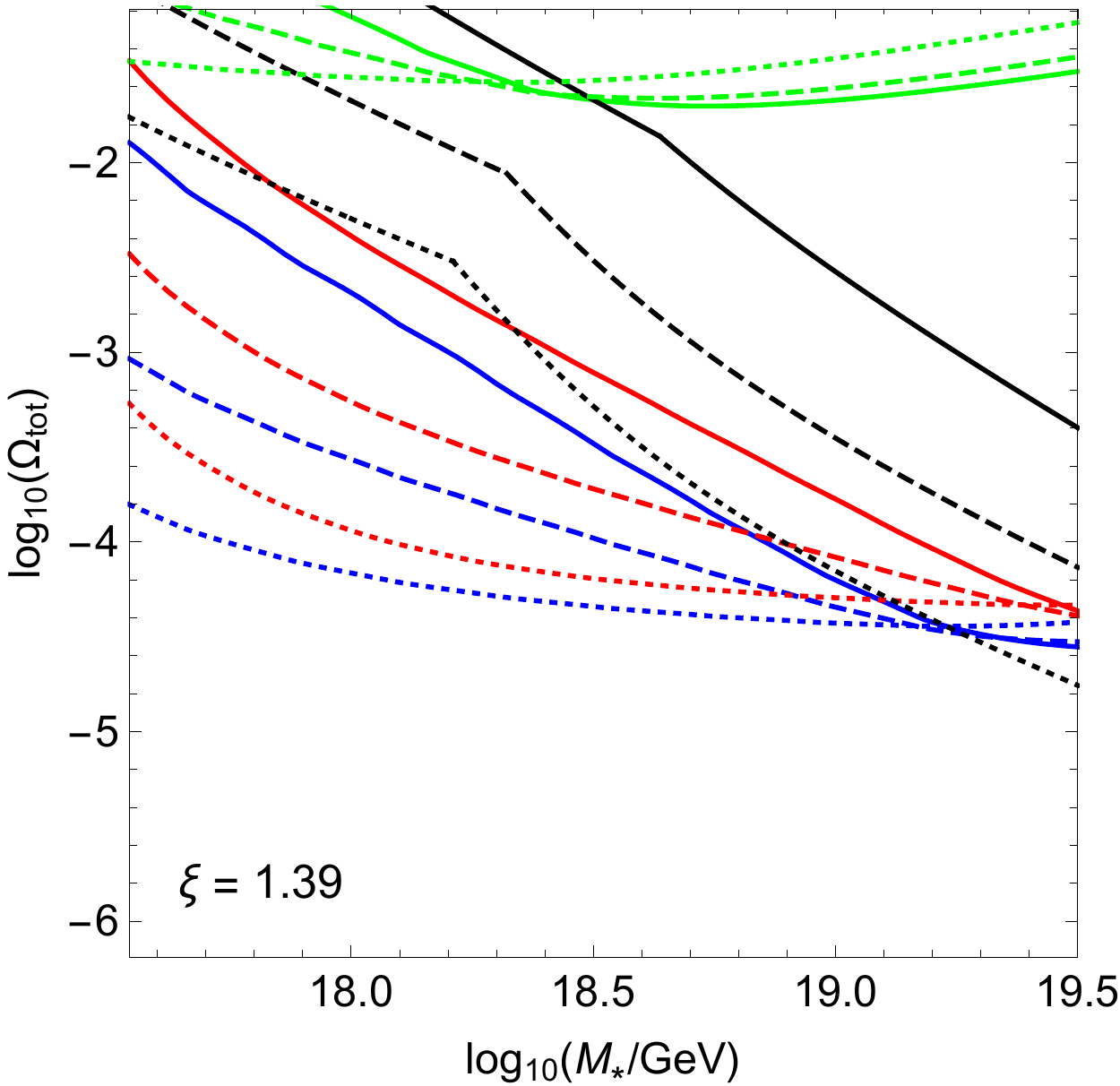}~
\includegraphics*[width=0.315\textwidth]{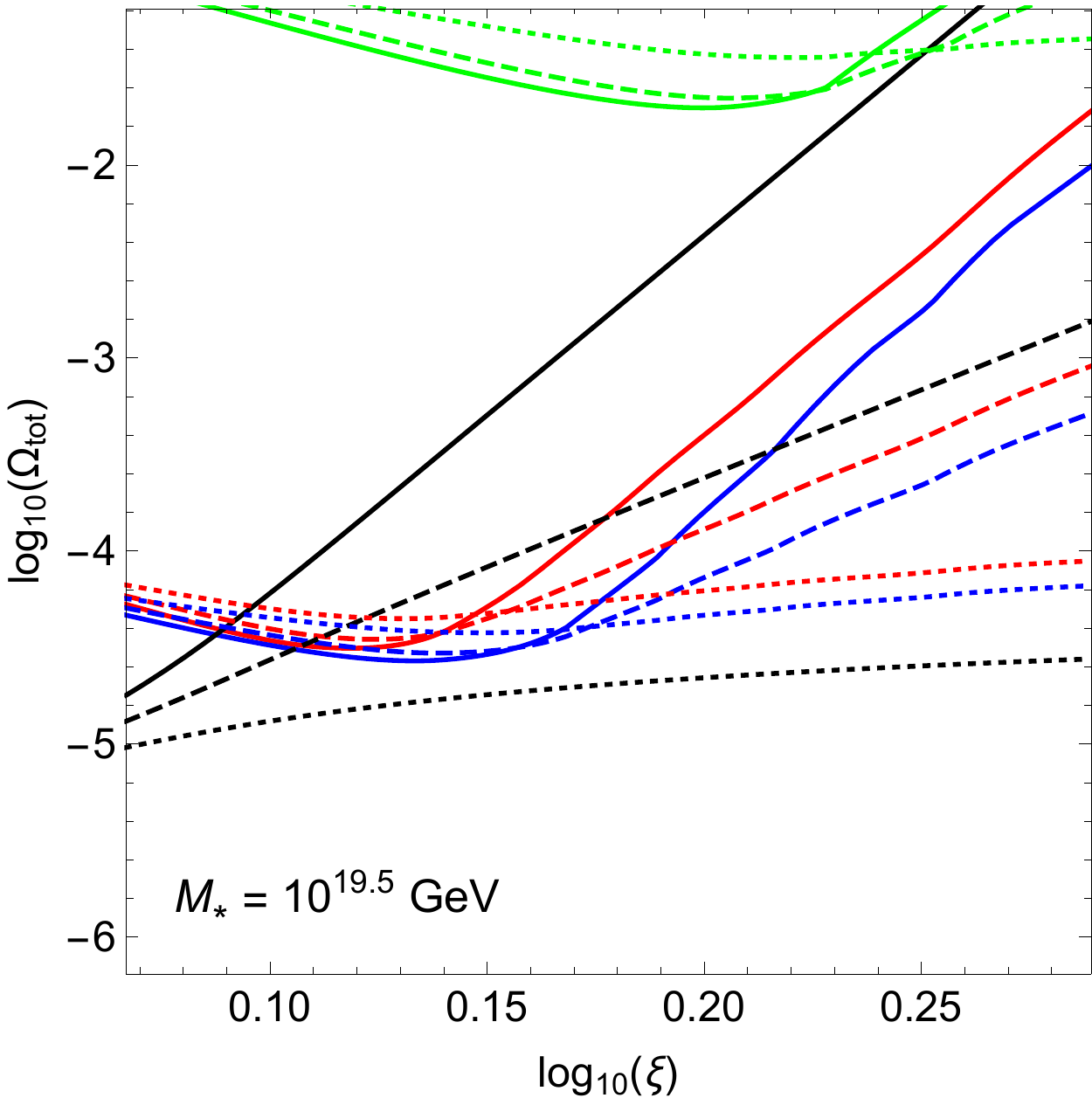}
\end{center}
\caption{Same as in Fig.~\protect\ref{fig:LimitsMLogLoDens}, but for 
   ``high-density'' ensembles with $N = 10$.
   In the left and center panels, we plot the upper 
   bound on $\Omegatot$ as a function of the coupling-suppression scale $M_\ast$ with 
   fixed $\xi = 1.2$ and with fixed $\xi = 1.39$, respectively.  In the right 
   panel, we plot the bound as a function of $\xi$ with fixed 
   $M_\ast = 10^{19.5}$~GeV.
\label{fig:LimitsMLogHiDens}}
\end{figure*}

In Fig.~\ref{fig:LimitsMLogLoDens} we show the bounds on $\Omegatot$ for a ``low-density'' 
ensemble with a mass spectrum given by Eq.~(\ref{eq:logmassscaling}) and the same benchmark 
parameter choices $N = 3$ and $m_0 = 10$~GeV as in Fig.~\ref{fig:LimitsMLinearLoDens}.
In the left and center panels, 
the value of $\xi$ has been chosen such that the range
of $\tau_i$ for the ensemble constituents is the same as in Fig.~\ref{fig:LimitsMLinearLoDens}
at both ends of the range of $M_\ast$ shown.  For the mass spectrum considered here, this
range of $\tau_i$ depends on $m_0$, on $M_\ast$, and on the quantity $\xi^{N-1}$.  
In the right panel of Fig.~\ref{fig:LimitsMLogLoDens},
we show how the bounds on $\Omegatot$ vary as a 
function of $\xi$ for fixed $M_\ast = 10^{19.5}$~GeV.~  The contours shown in 
Fig.~\ref{fig:LimitsMLogLoDens} exhibit the same overall scaling behaviors as those in 
Fig.~\ref{fig:LimitsMLinearLoDens}.  However, there are salient differences which 
reflect the fact that the density of states per unit $\ln(\tau)$ is uniform in this case
across the ensemble.

In Fig.~\ref{fig:LimitsMLogHiDens}, we show the corresponding results for an
ensemble with $m_0 = 10$~GeV and a mass spectrum again given by 
Eq.~(\ref{eq:logmassscaling}), but with a sufficiently high $n_\tau$ 
throughout the range of $M_\ast$ or $\xi$ shown in each panel that the 
ensemble is effectively always within the ``high-density'' regime.  We find that 
taking $N = 10$ is sufficient to achieve this.  The left and center panels of
Fig.~\ref{fig:LimitsMLogHiDens} correspond to the same ranges of $M_\ast$ and the 
same value of $\xi^{N-1}$ as the left and center panels of Fig.~\ref{fig:LimitsMLogLoDens}. 
The right panel once again shows how the bounds on $\Omegatot$ vary as a function of 
$\xi$ for fixed $M_\ast = 10^{19.5}$~GeV.~  Once again, we see that the principal 
consequence of increasing $n_\tau$ is that constraint contours become increasingly 
smooth and featureless.

Once again, as we did for the case of an ensemble with a uniform mass splitting, 
we may derive the corresponding expressions for 
$\delta Y_a$, $\delta \mu$, and $\delta y_C$ in the case of an exponentially rising
mass splitting in the continuum limit by direct 
integration of the expressions in Eqs.~(\ref{eq:BBNFitsTerm1})--(\ref{eq:BBNFitsLi6}) 
and in Eqs.~(\ref{eq:specfitsmu}) and (\ref{eq:specfitsy}).
For the mass spectrum in Eq.~(\ref{eq:logmassscaling}), this limit corresponds to 
taking $\xi\rightarrow 1$.  For the case in which the $\tau_i$ span the entire 
range from $t_{Aa}$ to $t_{fa}$ for each relevant nucleus, we find that for 
$\gamma > 0$, the term $\delta Y_a^{(1)}$ in Eq.~(\ref{eq:BBNFits}) takes the form
\begin{eqnarray}
  \delta Y_a^{(1)} &=&
    A_a \widetilde{\Xi}(t_{Ca}) 
      \Gamma\left(\frac{\gamma}{3}-1,x^{-1}\right)
      \Bigg|_{x^{\mathrm{min}}_{Aa}}^{x^{\mathrm{max}}_{Aa}}
      \nonumber \\ 
 +&& \!\!\!\!\!\!\! 
    B_a \widetilde{\Xi}(t_{Xa}) 
      \Bigg[ 
      \frac{6\beta x^{-\frac{3+2\gamma}{6}} }{3+2\gamma} 
      - \frac{3  x^{-\frac{\gamma}{3}}  }{\gamma}\Bigg]
      \Bigg|_{x^{\mathrm{min}}_{Ba}}^{x^{\mathrm{max}}_{Ba}}
      \nonumber \\ 
 +&& \!\!\!\!\!\!\! 
    B_a \widetilde{\Xi}(t_{Xa}) 
      \Bigg[ 
      \frac{6(1+\beta) x^{-\frac{3+2\gamma}{6}} }{3+2\gamma}  
      - \frac{24 x^{-\frac{3+4\gamma}{12}} }{3+4\gamma}  \Bigg]
      \Bigg|_{x^{\mathrm{min}}_{Ca}}^{x^{\mathrm{max}}_{Ca}} ~,
      \nonumber \\  
\label{eq:ContLimitBBNLog}
\end{eqnarray}
where the limits of integration are defined as in
Eq.~(\ref{eq:ContinuumLimitsOfIntYa}) and where we have defined 
\begin{equation}
  \widetilde{\Xi}(t) ~\equiv~ \frac{\gamma \Omegatot M_\ast^{2\gamma/3}}
    {3\big(m_{N-1}^\gamma - m_0^\gamma\big) t^{\gamma/3}}~.
  \label{eq:XiTildefnDef}
\end{equation}
By contrast, for the special case in which $\gamma = 0$, the replacement
\begin{equation}
  -\frac{3 x^{-\frac{\gamma}{3}} }{\gamma} ~\longrightarrow~ \ln x
  \label{eq:ReplacementLogCase}
\end{equation} 
should be made in the second term in square brackets appearing in the 
second line of Eq.~(\ref{eq:ContLimitBBNLog}), while 
$\widetilde{\Xi}(t)$ is given not by Eq.~(\ref{eq:XiTildefnDef}) but rather by  
\begin{equation}
  \widetilde{\Xi}(t) ~\equiv~ \frac{\Omegatot}{3\ln(m_{N-1}/m_0)}~.
  \label{eq:XiTildefnDefgammaEq0}
\end{equation}

For $\ce{D}$ and $\ce{^{6}Li}$, the term $\delta Y_a^{(2)}$ in Eq.~(\ref{eq:BBNFits})
is non-vanishing and once again must be evaluated separately.
For $\ce{D}$, the term $\delta Y_a^{(2)}$ in Eq.~(\ref{eq:BBNFits}) has exactly
that same form as Eq.~(\ref{eq:ContLimitBBNLog}), but with $A^\ast_a$ in place
of $A_a$, $B^\ast_a$ in place of $B_a$, \etc~  For $\ce{^{6}Li}$, we find that
the $\delta Y_a^{(2)}$ term takes the form
\begin{widetext}
\begin{eqnarray}
  \delta Y_a^{(2)} &=&
    A_a^* \widetilde{\Xi}(t^\ast_{Ca}) 
      \Gamma\left(\frac{\gamma}{3}-1,x^{-1}\right)
      \Bigg|_{x^{\ast\mathrm{min}}_{Aa}}^{x^{\ast\mathrm{max}}_{Aa}}
      \nonumber \\ 
+&&\!\!\!\!\!\!\!\! 
    B_a^* \widetilde{\Xi}(t^\ast_{Xa}) 
      \Bigg[ 
      \frac{72\beta^5\theta x^{-\frac{15+2\gamma}{6}} }{15+2\gamma}
      -\frac{45\beta^4(1 + \theta) x^{-\frac{6+\gamma}{3}} }{6+\gamma}
      +\frac{120\beta^3 x^{-\frac{9+2\gamma}{6}} }{9+2\gamma}
      -\frac{3(5-3\theta)x^{-\frac{\gamma}{3}} }{\gamma} \Bigg]
      \Bigg|_{x^{\ast\mathrm{min}}_{Ba}}^{x^{\ast\mathrm{max}}_{Ba}}
      \nonumber\\ 
+&&\!\!\!\!\!\!\!\! 
    B_a^* \widetilde{\Xi}(t^\ast_{Xa}) 
      \Bigg[
      \frac{8\theta(1+9\beta^5)x^{-\frac{15+2\gamma}{6}}}{15+2\gamma}
      -\frac{45(1+\theta)(1+7\beta^4)x^{-\frac{6+\gamma}{3}}}{7(6+\gamma)} 
      +\frac{24(1+5\beta^3)x^{-\frac{9+2\gamma}{6}}}{9+2\gamma} 
      -\frac{64(9-5\theta)x^{-\frac{3+4\gamma}{12}}}{7(3+4\gamma)} \Bigg]
      \Bigg|_{x^{\ast\mathrm{min}}_{Ca}}^{x^{\ast\mathrm{max}}_{Ca}}~,
      \nonumber \\
\label{eq:ContLimitBBNLogLi6}    
\end{eqnarray}
where the limits of integration are given by expressions analogous to 
those appearing in Eq.~(\ref{eq:ContinuumLimitsOfIntYa}).
Once again, as was the case with $\delta Y_a^{(1)}$, the expression for 
$\delta Y_a^{(2)}$ in Eq.~(\ref{eq:ContLimitBBNLogLi6}) requires modification 
for the special case in which $\gamma = 0$.  In particular, the replacement specified in
Eq.~(\ref{eq:ReplacementLogCase}) should likewise be made in the last term in 
square brackets on the second line of 
Eq.~(\ref{eq:ContLimitBBNLogLi6}).    
Moreover, for $\gamma = 0$,
the quantity $\widetilde{\Xi}(t)$ is given by Eq.~(\ref{eq:XiTildefnDefgammaEq0})
rather than by Eq.~(\ref{eq:XiTildefnDef}). 

The corresponding expressions for $\delta \mu$ and $\delta y_C$ for the case in 
which the $\tau_i$ span the entire range from $t_e$ to $\tLS$ can likewise be 
computed by direct integration.  In particular, we find that $\delta\mu$ is given by  
\begin{eqnarray}
  \delta\mu &=&
    \frac{4A_\mu}{5} \widetilde{\Xi}(t_{0\mu}) 
      \Gamma\left(\frac{4\gamma - 6}{15},x^{-5/4}\right)
      \Bigg|_{x^{\mathrm{min}}_{0\mu}}^{x^{\mathrm{max}}_{0\mu}}
    \nonumber \\ 
    &&~+ B_\mu\widetilde{\Xi}(t_{1\mu})
      \Bigg[\frac{6}{3-2\gamma}x^{\frac{3-2\gamma}{6}} 
      - \frac{1}{\alpha_\mu}
      \Gamma\left(\frac{2\gamma-3}{6\alpha_\mu},x^{-\alpha_\mu}\right)\Bigg]
      \Bigg|_{x^{\mathrm{min}}_{1\mu}}^{x^{\mathrm{max}}_{1\mu}} 
    \nonumber \\ 
    &&~+ B_\mu \widetilde{\Xi}(t_{2\mu})
      \Bigg[\frac{3}{2-\gamma}x^{\frac{2-\gamma}{3}}  
      - \frac{3}{4\alpha_\mu}
      \Gamma\left(\frac{\gamma-2}{4\alpha_\mu},x^{-\frac{4\alpha_\mu}{3}}\right)\Bigg]
      \Bigg|_{x^{\mathrm{min}}_{2\mu}}^{x^{\mathrm{max}}_{2\mu}}
  \label{eq:ContLimitmuLog}
\end{eqnarray}
for all $\gamma \geq 0$, except for the special case in which $\gamma = 2$.
For this special case, the replacement
\begin{equation}
  -\frac{3}{2-\gamma}x^{\frac{2-\gamma}{3}} ~\longrightarrow~ \ln x
\end{equation} 
should be made in the fourth line of Eq.~(\ref{eq:ContLimitmuLog}).
Likewise, we find that $\delta y_C$ is given by
\begin{eqnarray}
  \delta y_C &=&
    \frac{4A_y}{5} \widetilde{\Xi}(t_{0y}) \left.
      \Gamma\left(\frac{4\gamma - 6}{15},x^{-5/4}\right)
      \right|_{x^{\mathrm{min}}_{0y}}^{x^{\mathrm{max}}_{0y}}
      \nonumber \\
    &&~+ B_y \widetilde{\Xi}(t_{1y})
      \frac{6x^{\alpha+\frac{3-2\gamma}{6}}}{6\alpha_\mu - 2\gamma +3}
      \left.
      _2F_1\left(1,1+\frac{3-2\gamma}{6\alpha_y};
      2+ \frac{3-2\gamma}{6\alpha_y};-x^{\alpha_y}\right)
      \right|_{x^{\mathrm{min}}_{1y}}^{x^{\mathrm{max}}_{1y}}
      \nonumber \\
    &&~+ B_y \widetilde{\Xi}(t_{2y})
      \frac{3x^{\frac{2-\gamma}{3}}}{2-\gamma} \left.
      _{2}F_1\left(1,\frac{\gamma -2}{4\alpha_y};
      1+\frac{\gamma-2}{4\alpha_y};-x^{-\frac{4\alpha_y}{3}}\right)
      \right|_{x^{\mathrm{min}}_{2y}}^{x^{\mathrm{max}}_{2y}}~.
  \label{eq:ContLimityLog}
\end{eqnarray}
\end{widetext}
The limits of integration in Eqs.~(\ref{eq:ContLimitmuLog}) 
and (\ref{eq:ContLimityLog}) are defined as in  
Eq.~(\ref{eq:ContinuumLimitsOfIntYa}), and $\widetilde{\Xi}(t)$ is 
defined as in Eq.~(\ref{eq:XiTildefnDef}) for $\gamma > 0$ and as in 
Eq.~(\ref{eq:XiTildefnDefgammaEq0}) for $\gamma = 0$.  Once again, we note that in 
cases in which the $\tau_i$ do not span the relevant range of timescales during
which particle decays can affect one of these cosmological observables,
the limits of integration in Eqs.~(\ref{eq:ContLimitBBNLog})--(\ref{eq:ContLimityLog}) 
should be replaced by the values which restrict the overall range of lifetimes 
appropriately.

In summary, the results shown in Figs.~\ref{fig:LimitsMLinearLoDens}--\ref{fig:LimitsMLogHiDens} 
illustrate some of the ways in
which the constraints on injection from unstable-particle decays can be
modified in scenarios in which the injected energy density is distributed 
across an ensemble of particles with a range of lifetimes.   
These results also illustrate that a non-trivial interplay between 
the contributions from different decaying particle species within the ensemble
can have unexpected and potentially dramatic effects on the upper 
bound on $\Omegatot$ --- as when cancellations among positive and negative 
contributions to $\delta Y_{\ce{D}}$ from a broad range of particles within the 
ensemble result in a significant weakening of this upper bound.
{\it Thus, we see that the bounds on a decaying ensemble 
can exhibit collective properties and behaviors that transcend those
associated with the decays of its individual constituents.}

%%%%%%%%%%%%%%%%%%%%%%%%%%%%%%%%%%%%%%%%%%%%%%%%%%%%%%%%%%%%%%%%%%%%%%%%%%%%%%%%%%%%%%

\FloatBarrier
\section{Conclusions: ~Discussion and Summary of Results\label{sec:Conclusions}}

%%%%%%%%%%%%%%%%%%%%%%%%%%%%%%%%%%%%%%%%%%%%%%%%%%%%%%%%%%%%%%%%%%%%%%%%%%%%%%%%%%%%%%

\begin{table*}[t]
\centering
\smaller
\begin{tabular}{||m{2.25in}|m{2.5in}||}
\hline
\hline
  ~~~~~~~~~~~~~~~~~Constraint & ~~~~~~~~~~~~~~~ Analytic formulation \\ 
\hline
\hline
  ~Primordial abundances of light nuclei & Analytic approximations for
  $\delta Y_a$ given in Eq.~(\ref{eq:BBNFits}), with the corresponding limits 
  and parameter values given in Table~\ref{tab:BBN}. \\
\hline
  ~Spectral distortions in the CMB & 
  Analytic approximation for $\delta \mu$ and $\delta y_C$ given 
  in Eqs.~(\ref{eq:specfitsmu}) and (\ref{eq:specfitsy}), with the parameter values given in Table~\ref{tab:spec} 
  and limits given in Eq.~(\ref{eq:muandyClimits}). \\
\hline
  ~Ionization history of the universe & 
  Constraint specified by Eq.~(\ref{eq:IonizationConstraint}), with the values of 
  $\Gamma_{\mathrm{IH}}$ and $t_{\mathrm{IH}}$ given in Eq.~(\ref{eq:GammaIHtIH}). \\
\hline
  ~Diffuse extra-galactic photon background &
  Differential signal flux given by Eq.~(\ref{dPhidEgammaTot}), with differential background 
  fluxes given in Eqs.~(\ref{eq:COMPTELDiffFluxBound})--(\ref{eq:FermiDiffFluxBound}). \\
\hline
\hline
\end{tabular}
\caption{~Summary/compilation of our main results.
\label{tab:SummaryTable}}
\end{table*}

In this paper, we have considered ensembles of unstable particle species 
and investigated the cosmological constraints 
which may be placed on such ensembles due to limits on electromagnetic injection 
since the conclusion of the BBN epoch.
Indeed, as we have discussed, such injection has the potential
to modify the primordial abundances of light nuclei established during BBN.~ 
Such injection can also 
give rise to spectral distortions in the CMB, alter the ionization history of
the universe, and leave observable imprints in the diffuse photon background. 
For each of these individual considerations, we have presented an approximate 
analytic formulation for the corresponding constraint which may be applied
to generic ensembles of particles with lifetimes spanning a broad range of 
timescales $10^{2}\mathrm{~s} \lesssim \tau_i \lesssim \tnow$.  For ease of reference,
these analytic approximations, along with the corresponding equation numbers, 
are compiled in Table~\ref{tab:SummaryTable}.  In deriving these 
results, we have taken advantage of certain linear and uniform-decay approximations.
We have also demonstrated how these results can be applied within the context of
toy scenarios in which the mass spectrum for the decaying ensemble takes one of 
two characteristic forms realized in certain commonly-studied extensions of the SM.

Several comments are in order.  First, it is worth noting that while the values of the
parameters in Table~\ref{tab:BBN} were derived assuming a particular set of 
initial comoving number densities $Y_a$ for the relevant nuclei at the end of the 
BBN epoch, the corresponding parameter values for different sets of initial comoving number 
densities $Y_a'$ may be obtained in a straightforward manner from the values in 
Table~\ref{tab:BBN}.~  Provided that $Y_a'$ and $Y_a$ are not significantly 
different, the characteristic timescales $t_{Ba}$, $t_{Ca}$, and $t_{Xa}$ for 
each process are to a good approximation unchanged, and a shift in the 
initial comoving number densities can be compensated by an appropriate rescaling of 
the relevant normalization parameters $A_a$ and $B_a$.  
For example, a shift in the initial helium 
mass fraction at the end of BBN can be compensated by a rescaling of 
the fit parameters $A_{\ce{D}}^\ast$ and $B_{\ce{D}}^\ast$ associated with 
$\ce{D}$ production and of the parameters $A_{\ce{^{4}He}}$ and $B_{\ce{^{4}He}}$ 
associated with $\ce{^{4}He}$ destruction by $Y_p'/Y_p$, as well as a rescaling of the 
parameters $A_{\ce{^{6}Li}}^\ast$ and $B_{\ce{^{6}Li}}^\ast$ associated with
secondary $\ce{^{6}Li}$ production by $(Y_p'/Y_p)^2$.  The quadratic dependence 
on $Y_p'/Y_p$ exhibited by the last of these rescaling factors reflects the fact that 
the reactions through which the non-thermal populations of $\ce{^{3}He}$ and 
$\ce{T}$ nuclei are initially produced and the reactions through which these 
nuclei in turn contribute to $\ce{^{6}Li}$ production both involve $\ce{^{4}He}$ in the
initial state.  Similarly, a shift in the initial 
$\ce{^{7}Li}$ abundance can be compensated by a rescaling of the fit parameters 
$A_{\ce{^{7}Li}}$ and $B_{\ce{^{7}Li}}$ associated with $\ce{^{7}Li}$ destruction 
and of the parameters $A_{\ce{^{6}Li}}$ and $B_{\ce{^{6}Li}}$ associated with primary 
$\ce{^{6}Li}$ production by $Y_{\ce{^{7}Li}}'/Y_{\ce{^{7}Li}}$, while a shift in 
the initial $\ce{D}$ abundance can be compensated by a rescaling of the parameters 
$A_{\ce{D}}$ and $B_{\ce{D}}$ associated with $\ce{D}$ destruction by 
$Y_{\ce{D}}'/Y_{\ce{D}}$.     

Second, we remark that in this paper we have focused primarily on constraints related 
to electromagnetic injection.  In situations in which a significant fraction of 
the energy liberated during unstable particle decays is released in the form of
hadrons, the limits obtained from bounds on the primordial abundances of light nuclei
may differ significantly from those we have derived here.  Detailed analyses of 
the limits on hadronic injection in the case of a single decaying particle species 
have been performed by a number of authors~\cite{KawasakiHadronicCascades1,
KawasakiHadronicCascades2,CyburtBoundState,CyburtEllisGravitino}.  It would be interesting 
to generalize these results to the case of a decaying ensemble as well.

Third, as we have seen, there are two fundamentally different timescale regimes 
in which the corresponding physics is subject to very different leading bounds:
an ``early'' regime 
$10^{2}~{\rm s} \lsim  \tinj \lsim 10^{12}~{\rm s}$,
and a ``late'' regime
$\tinj \gsim 10^{12}~{\rm s}$.
This distinction is particularly important when considering 
the implications of our results within the context of
the Dynamical Dark Matter framework~\cite{DDM1,DDM2}.
Within this framework, the more stable particle species within the ensemble provide the dominant contribution to
the dark-matter abundance at present time, while the species with shorter lifetimes
have largely decayed away.  However, the masses, decay widths, abundances, \etc,
of the constituents of a realistic DDM ensemble are governed by the same underlying 
set of scaling relations.  Thus, in principle, one might hope to establish bounds on
the ensemble as a whole within the DDM framework by constraining the properties of those ensemble 
constituents which decay on timescales $\tau \lesssim \tLS$ --- \ie, ``early'' timescales on which 
the stringent limits on CMB distortions and the primordial abundances of light nuclei can be
brought to bear.  

In practice, however, it turns out that these considerations have little power to
constrain most realistic DDM scenarios.  On the one hand, both the 
lifetimes $\tau_i$ and extrapolated abundances $\Omega_i$ of the individual ensemble 
constituents typically decrease monotonically with $i$ in such scenarios --- either 
exponentially~\cite{DDMHadrons} or as a power law~\cite{DDM2,DDMAxion,ThermalDDM}.
On the other hand, as indicated in Fig.~\ref{fig:IonizationFit}, the constraint associated 
with the broadening of the surface of last scattering due to ionization from particle 
decays becomes increasingly stringent as the lifetime of the particle decreases down to 
around $\tau_i \sim \tLS$.  This implies that for an ensemble of unstable particles
in which the collective abundance of the cosmologically stable constituents is around 
$\OmegaDM \approx 0.26$, the $\Omega_i$ for those constituents with lifetimes 
$\tau_i \lesssim \tLS$ are constrained to be extremely small.  Thus, the constraints 
associated with CMB distortions and with the primordial abundances of light nuclei in the ``early'' regime
are not 
particularly relevant for constraining the properties of DDM ensembles in which the
abundances scale with lifetimes in this way.  Nevertheless, we note that in ensembles 
in which $\tau_i$ and $\Omega_i$ do {\it not}\/ increase monotonically 
with $i$ (such as those in Ref.~\cite{ThermalDDM}), 
the constraints associated with these considerations could indeed be relevant.
This is currently under study~\cite{toappear}.

%%%%%%%%%%%%%%%%%%%%%%%%%%%%%%%%%%%%%%%%%%%%%%%%%%%%%%%%%%%%%%%%%%%%%%%%%%%%%%%%%%%%%%

\FloatBarrier
\begin{acknowledgments}

%%%%%%%%%%%%%%%%%%%%%%%%%%%%%%%%%%%%%%%%%%%%%%%%%%%%%%%%%%%%%%%%%%%%%%%%%%%%%%%%%%%%%%

The research activities of KRD are supported in part by the Department of Energy 
under Grant DE-FG02-13ER41976 (DE-SC0009913) and by the National Science Foundation
through its employee IR/D program.  The research activities of JK are supported in 
part by National Science Foundation CAREER Grant PHY-1250573.  The research activities 
of PS are supported in part by the Department of Energy under Grant DE-SC0010504, in part 
by the Vetenskapsr{\aa}det (Swedish Research Council) through contract No.\ 
638-2013-8993 and the Oskar Klein Centre for Cosmoparticle Physics, and in 
part by the Department of Energy under Grant DE-SC007859 and the LCTP at the University 
of Michigan.  The research activities of BT are supported in part by National Science 
Foundation Grant PHY-1720430.  The opinions and conclusions expressed herein are those 
of the authors, and do not represent any funding agencies.
\end{acknowledgments}

%%%%%%%%%%%%%%%%%%%%%%%%%%%%%%%%%%%%%%%%%%%%%%%%%%%%%%%%%%%%%%%%%%%%%%%%%%%%%%%%%%%%%

\end{document}